\documentclass[manuscript, table, screen]{acmart}

\AtBeginDocument{%
  }

\setcopyright{acmcopyright}
\copyrightyear{2023}
\acmYear{2023}

\acmJournal{TOSEM}

\usepackage{adjustbox}
\usepackage{multirow}
\usepackage{graphicx}
\usepackage{stfloats}
\usepackage{caption}
\usepackage{subcaption}
\usepackage{enumitem}
\usepackage{makecell}

\makeatletter
\newcommand{\thickhline}{%
    \noalign {\ifnum 0=`}\fi \hrule height 1pt
    \futurelet \reserved@a \@xhline
}
\newcolumntype{"}{@{\hskip\tabcolsep\vrule width 1pt\hskip\tabcolsep}}
\makeatother
\definecolor{lightapricot}{rgb}{0.93, 0.92, 0.84}

\colorlet{lightgrey}{lightgray}
\usepackage{color,soul}

\newcommand{\secref}[1]{Section~\ref{sec:#1}}
\newcommand{\tabref}[1]{Table~\ref{table:#1}}

\newcommand{\figref}[1]{Figure~\ref{fig:#1}}

\begin{document}

\title{Can GitHub Issues Help in App Review Classifications?}

\author{Yasaman Abedini}
\orcid{0009-0005-8168-6116}
\affiliation{%
  \institution{Department
of Computer Engineering, Sharif University of Technology}
   \country{Iran}
}
\email{y.abedini14@sharif.edu}

\author{Abbas Heydarnoori}
\orcid{0000-0001-9785-2880}
\affiliation{%
    \institution{Department
of Computer Engineering, Sharif University of Technology}
    \country{Iran}}
\email{heydarnoori@sharif.edu}
\authornote{Abbas Heydarnoori is currently affiliated with the Department of Computer Science at Bowling Green State University, USA.}

\renewcommand{\shortauthors}{Abedini et al.}

\begin{abstract}
App reviews reflect various user requirements that can aid in planning maintenance tasks. Recently, proposed approaches for automatically classifying user reviews rely on machine learning algorithms. A previous study demonstrated that models trained on existing labeled datasets exhibit poor performance when predicting new ones. Therefore, a comprehensive labeled dataset is essential to train a more precise model. In this paper, we propose a novel approach that assists in augmenting labeled datasets by utilizing information extracted from an additional source, GitHub issues, that contains valuable information about user requirements. First, we identify issues concerning review intentions (bug reports, feature requests, and others) by examining the issue labels. Then, we analyze issue bodies and define 19 language patterns for extracting targeted information. Finally, we augment the manually labeled review dataset with a subset of processed issues through the \emph{Within-App}, \emph{Within-Context}, and \emph{Between-App Analysis} methods. We conducted several experiments to evaluate the proposed approach. Our results demonstrate that using labeled issues for data augmentation can improve the F1-score to 6.3 in bug reports and 7.2 in feature requests. Furthermore, we identify an effective range of 0.3 to 0.7 for the auxiliary volume, which provides better performance improvements.
\end{abstract}

\ccsdesc[500]{Software and its engineering~Software libraries and repositories}
\ccsdesc[500]{Information systems~Information extraction}

\keywords{App Review Classification, GitHub Issues, Intention Mining, Machine learning}

\received{18 September 2023}
\received[revised]{06 December 2023}
\received[revised]{31 May 2024}
\received[accepted]{24 June 2024}

\maketitle

\section{Introduction}
\label{sec:introduction}
App stores contain a large volume of mobile apps related to various fields, each offering different features~\cite{Maalej_Journal, Aslam, DATAR}. Users can express their feedback and experiences within these stores by posting reviews~\cite{Villarroel, Maalej_Journal}. User reviews contain valuable information, such as bug reports, feature suggestions, requests for improving current features, etc.,  which can assist developers in future maintenance and development plans~\cite{Maalej_Journal, Chen, Villarroel, Aslam}. 
Due to the substantial number of user reviews, particularly in popular apps, various approaches have been presented to automatically classify valuable reviews from the app developer's point of view~\cite{Guzman, Aslam, Villarroel, Devin}.
Previous studies often focus on providing novel approaches based on Natural Language Processing (NLP) and Machine Learning (ML) techniques to increase the accuracy of review classification~\cite{Dabrowski, Devin} and utilize manually labeled datasets to train and evaluate classification models~\cite{Devin}. 
They usually prepare these datasets by following a systematic process to minimize human errors in labeling reviews, but this can be both expensive and time-consuming~\cite{Guzman, Dabrowski}. As reported by Guzman et al.~\cite{Guzman}, labelers spend an average of 22 hours analyzing 1820 reviews manually. 
Therefore, it is challenging to prepare a precisely labeled dataset, and thus, they have a relatively low volume and are selected from a small number of apps~\cite{Dabrowski}. Dabrowski et al.~\cite{Dabrowski} identified this problem as a \emph{``significant threat''} for assessing the generalizability of existing approaches. 
Due to this limitation, although models trained on a specific dataset perform well during testing on the same dataset, a recent study demonstrates that these models perform poorly when applied to unseen datasets~\cite{Devin}. Devine et al.~\cite{Devin} showed that the performance of transformer-based models, trained separately on seven datasets, decreases when tested on each other. Therefore, the trained models are less generalizable than expected. They raise an important question: \emph{``How informative are the predictions of these models when used in the real world?''} To address this, they demonstrate that better results can be achieved by combining multiple labeled datasets. Although this approach enhances the performance of the classification models, their results indicate that further improvements in generalizability are still necessary.

However, the challenges of preparing labeled datasets have received comparatively less attention. In some studies, researchers have explored the use of semi-supervised~\cite{Deocadez} or active learning~\cite{Dhinakaran} methods or combined dataset~\cite{Devin} to address this critical challenge~\cite{Dabrowski}. The main problem with these approaches is their reliance on either insufficient existing datasets or human-dependent methods for collecting new data, both of which present significant difficulties as mentioned earlier. So, despite these efforts, the trained models still exhibit low accuracy and require further improvement. To answer this problem, we propose an approach to enhance the generalizability of trained models by augmenting the manually labeled review dataset.

In addition to app stores, some platforms such as Twitter, forums (e.g. VLC and Firefox forums~\cite{Tizard}), issue-tracking systems (e.g. GitHub~\cite{Khalajzadeh}), etc., receive user feedback~\cite{Devin, Nayebi}. 
Issue-tracking systems are not exclusive to developers, and users can also provide feedback, such as bug reports, feature requests, etc., by creating issues in app repositories~\cite{Rostami}. Developers usually assign one or more labels to an issue to present its purpose, topic, or priority~\cite{Izadi}. Also, issue bodies have similarities with app store user reviews~\cite{Al-Safoury, Zhang}. Al-Safoury et al.~\cite{Al-Safoury} leverage the similarity between issue reports and user reviews to complete mentioned change requests. Also, Zhang et al.~\cite{Zhang} utilize this similarity to assist in classifying unlabeled GitHub issues. In comparison, we use issue reports to improve review classifiers by applying linguistic patterns to adapt them to user reviews. Although issues contain more technical details than reviews, they can enrich review-based labeled datasets.
In this paper, we augment the manually labeled dataset by integrating processed labeled issues with them to train more generalizable models for review classification. To the best of our knowledge, this paper is the first study that uses labeled issues in GitHub to enhance the generalizability of app review classifiers.

For this purpose, we first identify GitHub labeled issues relevant to our task. To increase the confidence of the augmented dataset, we target repositories that have a minimum of two contributors and contain more than 30 labeled issues (\secref{data_collect}). In addition, we investigate issue labels to identify those relevant to our intended user intentions, including \emph{Bug Report}, \emph{Feature Request}, and \emph{Other} (e.g., User's Question). As a result, we collect 217K labeled issues that are suitable for augmenting the review dataset.

Since the issue tracking system serves as a platform for developing projects, developers are more inclined to contribute to repositories rather than users. Therefore, issue bodies can provide information that may not be visible in app store user reviews, such as technical details, code snippets, error messages, etc. We extract the target information from issue bodies to adapt them to user reviews. We define 19 linguistic patterns, derived from manual analysis of 577 related issue templates and 2,098 main constituent section titles of issue bodies that do not follow defined templates (\secref{extract_info}). By processing the issue bodies this way, we bring them closer to the user reviews posted in app stores.
Next, we integrate the labeled review datasets with a subset of processed issues selected through three methods \emph{Within-App} (using processed issues from the same app), \emph{Within-Context} (using processed issues from similar apps), and \emph{Between-App} (using randomly selected processed issues) \emph{Analysis} (\secref{data_aug}). 
Finally, we define this problem as two binary classification tasks: determining whether a review is a Bug Report and whether a review is a Feature Request. It is possible for a review to fall into both categories, resulting in positive outputs from both classifiers. We fine-tune transformer-based classifiers, which include pre-trained BERT~\cite{bert_model}, RoBERTa~\cite{roberta_model}, DistilBERT~\cite{distilbert_model}, and ALBERT~\cite{albert_model} models. These models have demonstrated state-of-the-art performance in previous studies~\cite{Hadi, Devin} for review classification tasks (\secref{rev_classify}). 

We performed several experiments to evaluate the effectiveness of our proposed approach. We analyze the impact of each augmentation method on classifying unseen datasets. For this reason, we created a truth dataset through a standard process. This involved manually labeling 1000 reviews randomly selected from five apps that met the necessary conditions for applying augmentation methods. The labeling process involved four developers to ensure accuracy and reliability. 

The results of our experiments demonstrate that augmented datasets can lead to a significant improvement in model performance. The F1-score increased to 6.3 and 7.2 for bug reports and feature requests, respectively. In augmented models related to bug reports, we observe that recall increases up to 9.2 compared to the baseline model, along with an improvement in precision. Notably, recall is particularly crucial for developers in review classifiers, especially when predicting bug reports~\cite{Maalej_Journal}. A higher recall ensures that less target information is lost. So, our approach enhances the reliability of review classifiers. As expected, the Within-App and Within-Context methods, which are used to train app-specific classifiers, exhibit more improvement compared to the Between-App method, which is used to train the general-purpose classifiers. This outcome can be attributed to the fact that Within-App and Within-Context methods leverage more relevant and app-specific information, leading to better performance. 

In addition, the augmentation of available public datasets also enhances the generalizability of models trained based on them, as evidenced by improvements in F1-score and recall. On the other hand, our experiments highlight the significance of the volume of the auxiliary dataset in achieving better results. It appears that an effective range for the volume, typically between 0.3 to 0.7, yields the best outcomes for data augmentation. Properly balancing the size of the auxiliary dataset seems to be crucial in ensuring optimal performance and avoiding overfitting or underfitting issues.

In general, the main contributions in this paper include the following:
\begin{enumerate}
    \item We propose a novel approach that augments review datasets by incorporating labeled GitHub issues. As a result, we enhance the generalizability of review classifiers and achieve significant improvements in F1-score and recall for review classification models. 
    
    \item We use three methods: Within-App, Within-Context, and Between-App Analysis, to apply our proposed approach. The first two methods make tuning classifiers specific to an app more accessible. As far as we know, previous approaches have often not focused on training app-specific classifiers.  
        
    \item We conduct several experiments to show how using labeled issues can improve the performance of review classifiers and discuss the effective volume of auxiliary dataset.
    
    \item The processed datasets, implementation and evaluation codes are publicly available to help future research and replicate the results \footnote{\url{https://github.com/ISE-Research/App-Reviews-Augmentation}}.

\end{enumerate}

The structure of this paper is organized as follows. In \secref{Motiv_Example}, we describe our motivation for this study with an example. In \secref{approach}, the proposed approach is described in detail. In \secref{experiment_setting}, we introduce the used dataset and experimental setup. In \secref{eval}, we analyze the experimental results. In \secref{rel_work}, we survey related work. In \secref{threat}, we discuss threats to the validity of our results. Finally, we conclude our paper in \secref{concusion}.

\section{A Motivating Example}
\label{sec:Motiv_Example}

\begin{figure*}[!t]
    \centering
    \captionsetup[subfloat]{labelfont=footnotesize ,textfont=footnotesize}
    \subfloat[An Example of a GitHub Issue~\cite{AtennaPod_Git}]{\includegraphics[width=3.2in]{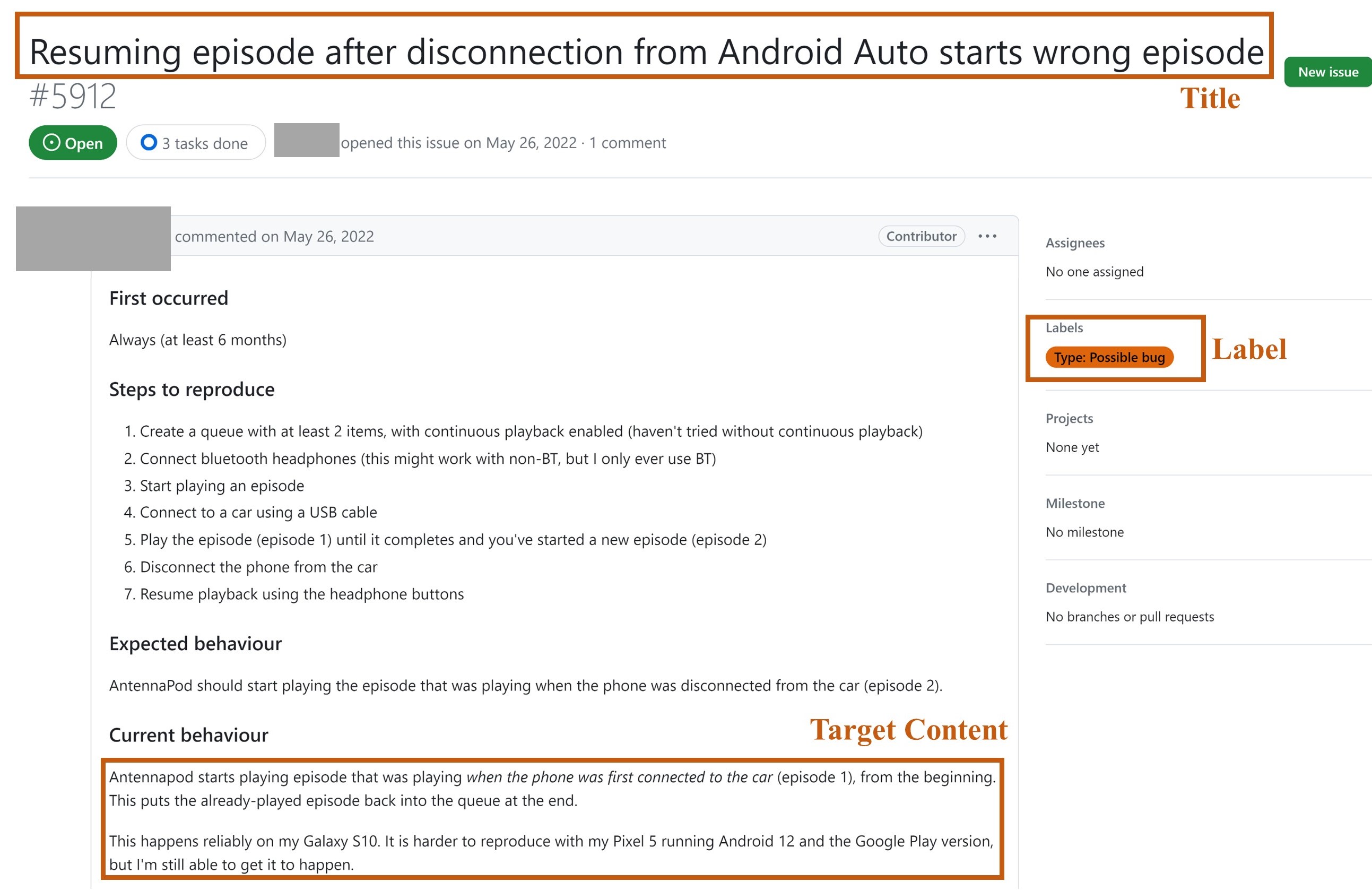}
    \label{fig:issue_example}}
    \hfil
    \subfloat[An Example of a Goggle Play Review~\cite{AtennaPod}]{\includegraphics[width=2.6in]{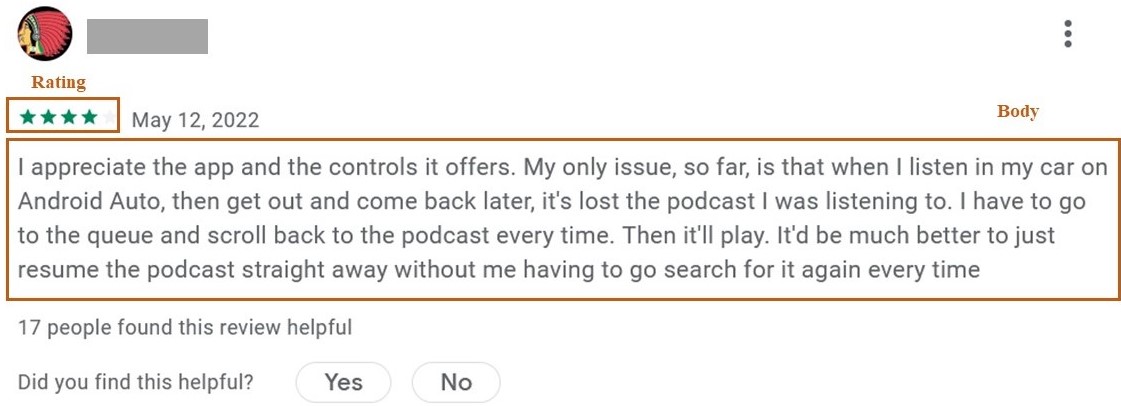}
    \label{fig:review_example}}
    \caption{Similarity Between Issues and Reviews}
\label{fig_sim}
\end{figure*}

\figref{review_example} shows a Google Play user review~\cite{Al-Safoury} for the AtennaPod~\cite{AtennaPod} open-source app, a podcast manager. The main parts of a user review are the \emph{body} and \emph{rating}. The review body contains a textual description of the user feedback about a bug encountered, a new feature requested, etc., expressed without a formal structure~\cite{Guzman}. For example, in \figref{review_example}, a user describes the problem of resuming a podcast after the app disconnects from the car. 

A user in this review reported a bug informally. However, this bug has been reported more formally in GitHub issues of the AtennaPod repository~\cite{AtennaPod_Git}. \figref{issue_example} shows an issue related to the same bug with the label \emph{``type: possible bug''}. The main parts of an issue include the \emph{title}, \emph{body}, and \emph{label}. As we can see,  a description of the same bug mentioned in the user review (\figref{review_example}) is provided in the \emph{Current Behavior} section of the body with slightly different constituent words. 
The issue title also briefly describes this problem. In this issue, other sections provide details for reproducing the bug, which is unrelated to our purpose. Therefore, although GitHub issues more technically report encountered bugs or describe requested features, they can be used for detecting the review's intention by extracting target information.

\section{Proposed Approach}
\label{sec:approach}

\begin{figure*}[!t]
\centering
  \includegraphics[width=6in]{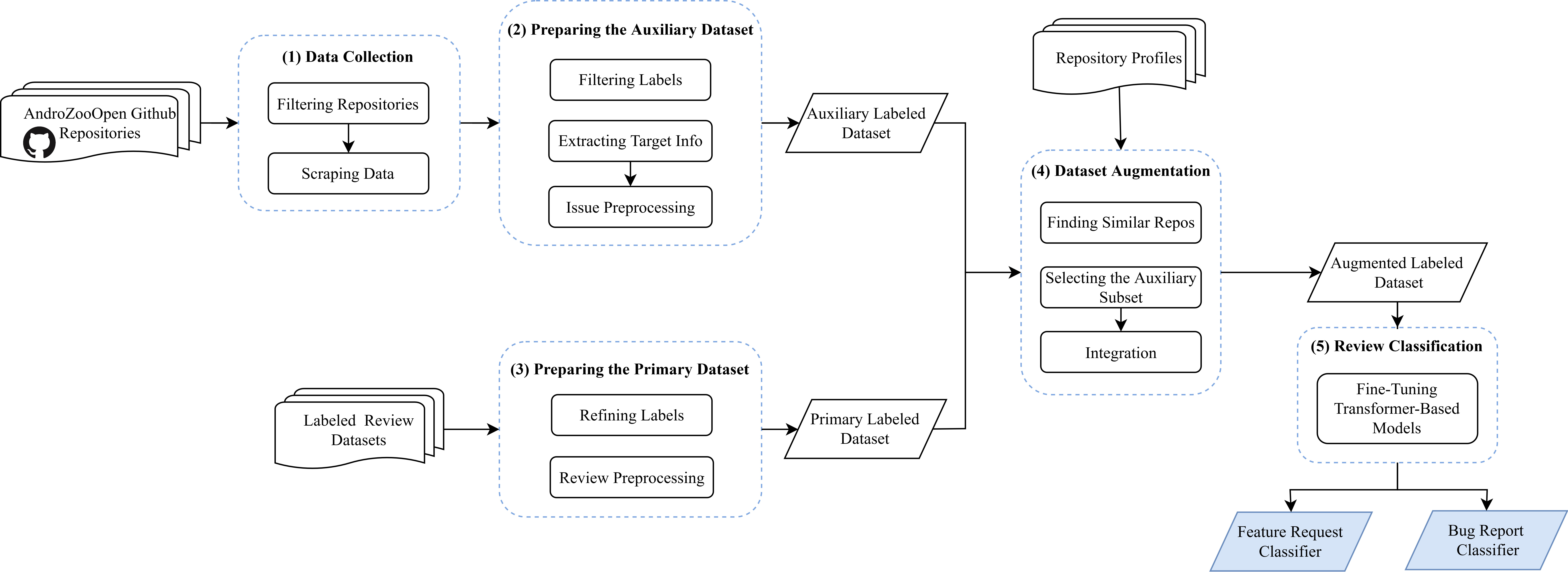}
\caption{An Approach Overview}
\label{fig:overview}  
\end{figure*}

In this research, we aim to increase the accuracy of classifying unseen reviews by augmenting the labeled review dataset used to train the predictive model. We enrich the training dataset by merging the review-based labeled dataset (\emph{primary dataset}) and the proposed dataset derived from labeled issues (\emph{auxiliary dataset}). So, we first create the auxiliary dataset from the available information and then merge it with the primary dataset in various methods.
To this end, we propose a 5-phase approach consisting of (1) collecting the required data, (2) preparing the auxiliary dataset using labeled issues, (3) preparing the primary dataset using five datasets of manually labeled reviews, (4) dataset augmentation (5) classifying unseen reviews. \figref{overview} provides an overview of our proposed approach.

In the first phase, we collect the required data to create an auxiliary dataset based on GitHub issues. We investigate the available Android repositories on GitHub and select those that provide us with more reliable information, considering their activity and popularity. 
In the second phase, we process the collected data. We refine the labels and define 19 language patterns for extracting textual information from issue bodies to prepare a compatible auxiliary dataset. 
Next, to prepare the primary dataset based on user reviews, we use five datasets that have been manually labeled in previous studies. Since they do not focus on a homogeneous set of labels, we adapt them to fit our intentions. Also, we preprocessed the review text by applying some common techniques in NLP.
In the fourth phase, we select a subset of the auxiliary dataset to augment the primary dataset based on the three methods, including \emph{Within-App} (using same app issues), \emph{Within-Context} (using similar app issues), and \emph{Between-App} (using random issues) \emph{Analysis}. In the second one, we need to find repositories related to apps similar to a target app. 
Lastly, we use the augmented dataset obtained from merging the primary and auxiliary datasets to train review classifiers according to fine-tuning a  transformer-based model (e.g., DistilBERT). In the rest of this section, we describe each phase in detail.

\subsection{Data Collection}
\label{sec:data_collect}

We first collect data needed for subsequent steps of our approach. Therefore, we investigate the Android repositories available on GitHub to identify appropriate ones that (i) contain adequate labelled issues and (ii) provide relevant and accurate data for creating the auxiliary dataset. Liu et al.~\cite{Liu_1} provided the AndroZooOpen dataset, which includes information about more than 46K Android repositories. To our best knowledge, this dataset is the largest available collection containing GitHub repositories related to open-source Android apps. This paper uses the AndroZooOpen dataset to find appropriate repositories by examining extracted information.

\begin{figure}[!t]
\centering
  \includegraphics[width=2.5in]{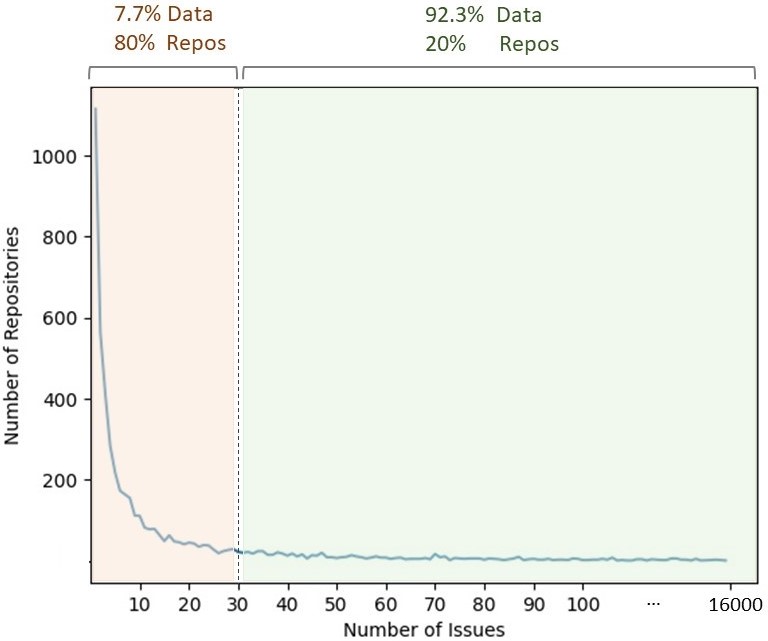}
\caption{Repository Counts per Each Number of Labelled Issues (Repos=Repositories)}
\label{fig:issu_repo_count}  
\end{figure}

In January 2023, approximately 43K repositories of this dataset were available on GitHub. We updated important fields, such as the contributor and star counts, to make our investigation more accurate and get labeled issues using the GitHub REST API~\cite{RESTAPI}.  Most available repositories had no labeled issue, and only 5,258 (12.1\%) repositories contained at least one. \figref{issu_repo_count} shows the repository counts for each number of labeled issues in these 5,258 repositories. As can be seen, most of the repositories have a few labeled issues. Although 20\% of them contain more than 30 labeled issues, they constitute 92.3\% of collected issues. Therefore, repositories with less than 30 labeled issues provide little information. In order to increase the quality and performance of the auxiliary dataset, we eliminate them for two reasons.

\begin{figure}[!t]
\centering
\captionsetup[subfloat]{labelfont=footnotesize ,textfont=footnotesize}
\subfloat[Distribution of Stars]{\includegraphics[width=1.6in]{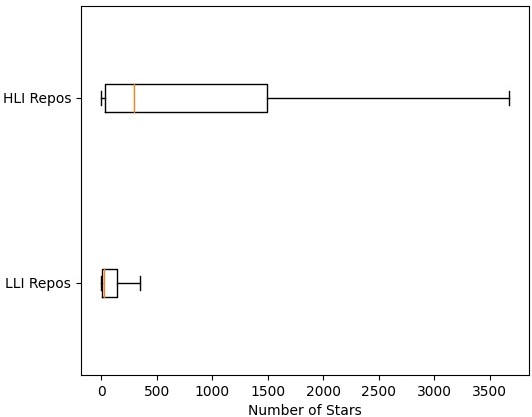}
\label{fig:issue_star}}
\hfil
\subfloat[Distribution of Contributors]{\includegraphics[width=1.6in]{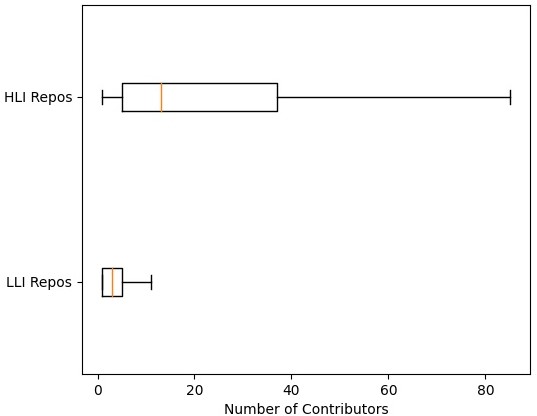}
\label{fig:issue_cont}}
\caption{The Comparison of HLI and LLI Repositories}
\label{fig:two_box_plots}
\end{figure}

First, in preparing the auxiliary dataset, we ignore unrelated labels to our review intentions (\secref{filter_label}) and then identify and extract the target information from selected issues (\secref{extract_info}). As explained, more than two-thirds of the collected issues, which are irrelevant to our intentions or do not provide sufficient information, are eliminated in these steps. The presence of many repositories with few labeled issues that ultimately provide little helpful information can reduce the performance of finding similar apps (\secref{find_similars}).
Second, \figref{two_box_plots} shows the distribution of the number of contributors and stars for repositories with at least one labeled issue. HLI (High-Labeled Issue) and LLI (Low-Labeled Issue) denote repositories with a minimum and a maximum of 30 labeled issues, respectively. As can be seen, more developers are involved in HLI repositories, which causes more monitoring. Also, developers have given HLI repositories more stars. Considering stars as a measure of repository popularity, following Borges et al.~\cite{Borges}, we can conclude that HLI repositories have garnered more interest from developers. Therefore, data provided by HLI repositories with more monitoring and popularity can be more reliable.

On the other hand, a developer may choose the incorrect label for an issue due to inexperience or human mistake. The presence of at least two contributors makes it possible for assigned labels by one developer to be reviewed by another. This possibility of double-checking can reduce the mistakes in labelling. We choose repositories with at least two contributors to create a more accurate auxiliary dataset. Therefore, we select repositories that (1) contain more than 30 labeled issues and (2) have at least two contributors. 

After selecting the appropriate Android repositories, we collect the required information. We need four types of information related to GitHub repositories and issues to apply the next phases of our approach, which include the following:

\begin{enumerate}
    \item Title, body and label of the issues to prepare the auxiliary dataset (\secref{aux_dataset})
    \item Names and descriptions of user-defined labels to identify related ones to bug report, feature request, and other review intentions (\secref{filter_label})
    \item Issues templates to identify helpful sections of an issue (\secref{extract_info})
    \item ReadMe file and About section to prepare repository profiles (\secref{find_similars})
\end{enumerate}

\tabref{dataset_info} shows the key specifications of the collected dataset in this step. 

\begin{table}
    \caption{Preliminary Collected Data}
    \renewcommand{\arraystretch}{1.3}
    \centering
    \begin{tabular}{ll}
        \hline
        Number of Repositories & 999 \\
        \hline
        Number of Issues & 318,685 \\
        \hline
        Number of Label Description & 7862\\
        \hline
        Number of Issue Templates & 644 \\
        \hline
        Number of Descriptions (ReadMe, About) & 820 \\
        \hline
    \end{tabular}
\label{table:dataset_info}
\end{table}

\subsection{Preparing the Auxiliary Dataset}
\label{sec:aux_dataset}
In this phase of our approach, we process the collected data to prepare an auxiliary labeled dataset. First, we filter the issue labels and adapt them to our desired intentions. Next, we extract the target information from the title and body of issues and preprocess them to remove noisy textual data, such as links, error messages, code fragments, etc. In the subsections of this section, we explain these steps in detail.

Previous studies have classified user reviews according to their intentions~\cite{Dabrowski}. They have focused on different perspectives and granularity to assist developers in maintenance tasks~\cite{Villarroel, Sorbo_1}. \tabref{Orig_Labels} displays the target intention classes of reviews in several studies with available labeled datasets. As seen, two intentions, namely \emph{Bug Report} and \emph{Feature Request}, have been repeatedly used with slight differences in their naming. Therefore, given the available data required to create the primary dataset, we follow Devine et al.~\cite{Devin} and consider the three most frequent classes of \textbf{Bug Report}, \textbf{Feature Request}, and \textbf{Other} as our target review intentions.

\begin{table}
    \caption{Target Intention Classes in Previous Studies (*For studies that presented multi-level taxonomy, we only mentioned the first level.)}
    \renewcommand{\arraystretch}{1.3}
    \begin{tabular}{lp{.60\linewidth}}
        \hline
        \textbf{Research} & \textbf{Intention Classes}\\
        \hline
        Pagano et al.~\cite{Pagano} & community, requirement, rating, user experience *\\
        \hline
        Maalej et al.~\cite{Maalej_Journal} & \textbf{bug report}, \textbf{feature request}, user experience, rating\\
        \hline
        Panichella et al.~\cite{Panichella} & information giving, information seeking, \textbf{feature request}, \textbf{problem discovery}, other\\
        \hline
        Gu et al.~\cite{Gu} & aspect evaluation, \textbf{bug reports}, \textbf{feature requests}, praise and others\\
        \hline
        Guzman et al.~\cite{Guzman} & \textbf{bug report}, feature strength, feature shortcoming, \textbf{user request}, praise, complaint, usage scenario\\
        \hline
        Ciurumelea et al.~\cite{Ciurumelea} & compatibility, usage, resources, pricing, and protection*\\
        \hline
        Jha et al.~\cite{Jha} & \textbf{bug report}, \textbf{feature request}, other\\
        \hline
        Scalabrino et al.~\cite{Scalabrino} & \textbf{functional bug report}, \textbf{suggestion for new feature}, report of performance problems, report of security issues, report of excessive energy consumption,  request for usability improvements and other\\
        \hline
    \end{tabular}
\label{table:Orig_Labels}
\end{table}

\subsubsection{Filtering the Issue Labels}
\label{sec:filter_label}
\begin{figure}
\centering
  \includegraphics[width=0.6\textwidth]{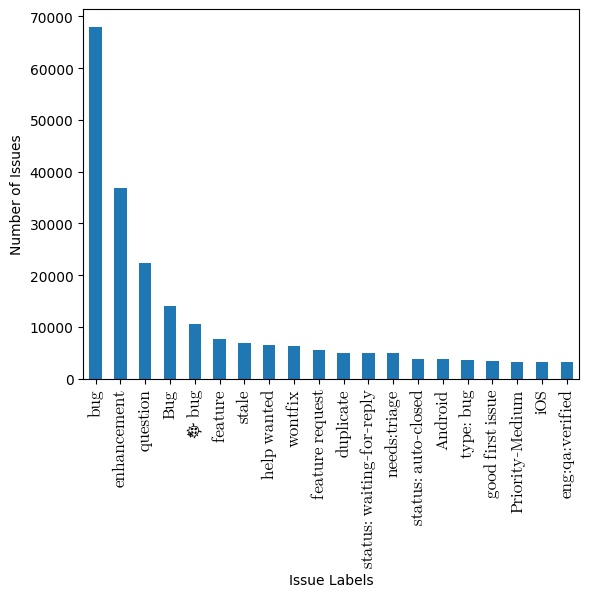}
\caption{The Most Frequent Labels in Preliminary Dataset}
\label{fig:TopLabels}  
\end{figure}

In GitHub, developers assign labels to an issue based on various criteria, such as intentions (e.g., bug, enhancement), states (e.g., duplicate, invalid), development stages (e.g., implementation, design), app components or features (e.g., component: client, feature: history), priorities (e.g., low priority, high priority), etc. These labels are helpful in managing and organizing issues more efficiently. By default, GitHub provides seven labels: \emph{Bug}, \emph{Enhancement}, \emph{Documentation}, \emph{Duplicate}, \emph{Good First Issue}, \emph{Help Wanted}, \emph{Invalid}, \emph{Question}, and \emph{Wontfix}. 
Developers also have the option to create new labels for categorizing their issues, which we refer to as user-defined labels. There are no specific rules for defining these labels~\cite{Izadi}. As a result, they can be heterogeneous and diverse. Different labels may represent a similar concept. For instance, labels like \emph{Type: Enhancement}, \emph{Enhancement/Feature}, \emph{Feature-Request} are used in many repositories to represent the same concept as the default label \emph{Enhancement}. Due to this variability, Izadi et al.~\cite{Izadi} manually analyzed the issue labels according to their goals.
In the previous phase, we collected 318K labeled issues, comprising 5,641 distinct labels. \figref{TopLabels} displays the 20 most frequently occurring labels in our preliminary dataset. 
Similar labels may vary in form (e.g., bug, Bug) or constituent words (e.g., enhancement, feature request). To address this, we first preprocess the labels to unify their forms and remove noisy characters. Then, we manually identify the semantically relevant labels aligned with our review intentions.

\textbf{Syntactically Preprocessing the Labels:}
As mentioned above, we first unify the syntactical form of the labels. To achieve this, we follow these steps:

\begin{enumerate}
    \item Transforming words to lowercase.
    \item Removing numbers and special characters (such as \#\$ -\_ and emojis) using Regular Expression.
    \item Removing single letters that remain after applying step 2: For example, after removing numbers, label \emph{P1} convert to letter \emph{p}, which does not provide useful information. Our manual analysis also shows that these labels are irrelevant to our selected intentions. They are usually used to indicate an issue's priority or project phase.
    \item Stemming using the SnowballStemmer function in the NLTK library~\cite{NLTK}: For example, three labels \emph{reproduced}, \emph{reproduction}, and \emph{reproducible}, which refer to the same concept, are converted to a single term \emph{reproduc}.
    \item Transforming negative modifiers to the single form ``not'': For example, six labels \emph{not reproduced}, \emph{cannot-reproduce}, \emph{non-reproduce}, \emph{could not reproduce}, \emph{cant-reproduce}, \emph{can't reproduce} will be converted to the unified form \emph{not reproduce} (We provide a detailed description of this step in \secref{preprocess}).
\end{enumerate}

After applying the above steps, we obtained 4,144 syntactically preprocessed labels. Almost half of these labels are attached to at most 11 issues in our dataset, and our manual analysis reveals that they often do not provide sufficient helpful information. In the next step, we manually analyze the labels to identify those related to our intentions. The presence of a large number of repository-specific rare labels without accurate descriptions can reduce the accuracy of our analysis. Therefore, to focus on more common and valuable labels and increase the accuracy of the auxiliary dataset, we select labels that are used at least in 11 issues in our collected dataset. As a result, we manually analyze 2,098 syntactically preprocessed labels in the next step.

\textbf{Semantically Preprocessing the Labels: }
We conducted a manual analysis of issue labels to identify those semantically related to the bug report, feature request, and other review intentions. For this purpose, one author and two graduate students, with an average programming experience of 8 years, independently analyzed the 2,098 labels. Each label was reviewed by at least two analysts separately. Since user-defined labels are subjective and often require clearer and more precise definitions, it is challenging for analysts to find relevant ones. To facilitate a detailed analysis, we provided the necessary label information, including two samples of the original forms, user definitions, and associated issues, which were randomly selected. \tabref{table-sample} presents an example of the information provided to the analysts for three labels. Considering this information, the analysts aimed to identify labels related to our target intentions. They excluded labels that were associated with issues across multiple classes (such as the ``request'' label that was associated with both bug report and feature request issues) or labels that were not relevant to our review intentions (such as priority labels). In addition, some labels refer to code changes or problems from the developer's perspective, such as ``build error'' or ``refactor'', which are related to bugs and improvements in the app's implementation. Since end users typically do not discuss these technical aspects in their reviews, the analysts also removed such labels from consideration.

\begin{table*}
    \caption{Samples of the Provided Information for Each Preprocessed (Pre) Label}
    \renewcommand{\arraystretch}{1.3}
    \begin{tabular}{|c|p{.2\linewidth}|p{.2\linewidth}|p{.25\linewidth}|c|}
        \hline
        \textbf{Pre Labels} & \textbf{Original Labels} & \textbf{Descriptions} & \textbf{Related Issues} & \textbf{Final Labels}\\
        \hline
        type enhanc &
        type/enhancement; Type: enhancement
        & 
        New feature or request; A proposed enhancement to the api or behavior
        & 
        When we are on a screen with a FAB, and the app shows a SnackBar message, currently, the SnackBar is shown on top of the FAB, preventing the user from interacting with the FAB. It would be nice if when the SnackBar bar shows, the user could still be able to interact with the FAB.
        & Feature Request\\
        \hline
        crash & 
        b: crash; [Crash]
        & 
        Crashes Fenix: should link to Sentry, Crash-Stats or GPlay info;
        & **Describe the bug**
        Rotating after creating a new list ends up crashing & Bug Report \\ & & The app crashes, either the foreground or   background.
        & 
        **To Reproduce**
        Steps to reproduce the behavior:
        \begin{enumerate}
            \item Go to Create a new list
            \item Click on Ok in the list name dialog, important step
            \item Rotate the app
        \end{enumerate}

        & 
        \\
        \hline
        type question & 
        Type: Question; [type] question
        & 
        Further information is requested; Request for information or clarification. Not an issue.
        & 
        How can I add objects to maps ? For example, I want add the location of people and do filtering.
        & 
        Other\\
        \hline
    \end{tabular}
\label{table:table-sample}
\end{table*}

As a result, two analysts independently identified related review intention for each label, with an agreement percentage of 76\%. They discussed disagreements in identifying semantically related labels by examining more issues with these labels. Finally, the analysts reached a consensus on most of the labels. The first author made the final decision for the remaining four disputed labels. \tabref{Related_Labels} provides several identified related labels for each intention. As a result of this analysis, labeled issues are reduced to 217.4K, which includes 117.3K (54\%) bug reports, 76.6K (35\%) feature requests, and 24.9K (11\%) other issues. Additionally, 1.4K issues are related to both the bug report and feature request intentions.

It is worth noting that our manual analysis revealed a limitation in the \emph{other} intention class. In the user reviews, the \emph{other} class encompasses a wide range of reviews, including expressing feelings, praises, dislikes, questions, user experiences, etc. However, developers usually tend to express GitHub issues more formally and do not provide this type of emotional expression. During our analysis, we discovered that all 18 issue labels related to the \emph{other} class were about asking user questions. As a result, we only included issues specifically related to user questions in this class. This observation reveals that the labeled issues may not fully encompass the entire spectrum of the \emph{other} class in our target review intentions.

\begin{table*}
    \caption{Some Preprocessed Labels Related to Target Review Intentions}
    \renewcommand{\arraystretch}{1.3}
    \begin{tabular}{lp{0.55\textwidth}cc}
        \hline
        \textbf{Intentions} & \textbf{Some Related Labels} & \textbf{\#Labels} & \textbf{\#Issues}\\
        \hline
        Bug Report & 
        bug, type bug, crash, type confirm bug, is possibl bug, render bug, bug general, type possibl bug, is bug, bug crash, problem, bug beta, need reproduct, type not reproduc, bug minor, defect. & 77 & 117K\\
        \hline
        Feature Request & 
        enhancement, cat enhanc, improvement, feature, type feature, featur enhanc, user stori, propos, idea, suggest, nice to have. & 54 & 77K\\
        \hline
        Other & 
        question, question answer, categori question, faq, is faq candid & 
        18 & 25K\\
        \hline
    \end{tabular}
\label{table:Related_Labels}
\end{table*}

\subsubsection{Extracting Target Information}
\label{sec:extract_info}
The title and body of the issues provide helpful information for describing user intentions and preparing an auxiliary dataset. The title gives a brief textual summary of an issue. However, the body contains various more detailed information, including textual descriptions, error messages, code fragments, etc., some of which are relevant to our purpose. Since our focus is on augmenting labeled review datasets, we need information that describes problems, requests, and the current behavior of an app from the user's perspective. Therefore, considering information specific to developers (such as implementation details or app testing) or irrelevant to the content of user reviews (such as screenshots) increases the error of the predictive model. 

For example, \figref{issue_example} illustrates a bug issue with a specific template comprising several main sections, namely the \emph{First Occurred}, \emph{Steps to Reproduce}, \emph{Expected Behaviour}, and \emph{Actual Behaviour}. The first section contains the date of occurrence of the bug, aiding developers in locating it. The second section lists the detailed steps for reproducing the bug. Typically, users only state an overview of encountered bugs and refer to the last steps in most reviews~\cite{Li}. As we see in \figref{review_example}, a user has mentioned the last three steps without providing many details for replicating the bug. The third section also specifies the desired behavior of the user, which we found to often have the same content as the \emph{Actual Behaviour} section, albeit using a positive verb. This section can lead to errors in the final model. So, the content of the first three sections in the issue shown in \figref{issue_example} provides the details for developers to reproduce and fix this bug, while user reviews do not include this type of information.
Therefore, these sections cannot be helpful for us, and we remove them. On the other hand, the content of the \emph{Actual Behaviour} section is suitable for our purpose. It describes the current behavior of the app and provides an overview of the bug, similar to a user review (\figref{review_example}). 
As a result, we need to choose sections that provide a general description of an issue, similar to user reviews, to create an auxiliary dataset. During this phase, we manually analyze the issue bodies and templates to identify the necessary information. Subsequently, we define 19 linguistic patterns that allow us to extract this relevant information accurately and efficiently.

Developers have the option to either follow a specific template for submitting issues or write them freely and informally. Issue templates usually contain certain sections that provide valuable information for a particular purpose. These templates aid in creating structured content and assist issue authors in recalling the necessary information~\cite{Li_2}. Issue templates are stored in the ``.github/ISSUE\_TEMPLATE'' folder as ``.md'', ``.yaml'', or ``.yml'' files~\cite{Nikeghbal}. These files contain instructions for preparing appropriate content for different sections of a specific issue.  By utilizing these templates, we can identify sections that offer target information compatible with user reviews. In our research, we collected a total of 645 available issue templates from 291 repositories using the GitHub REST API.

\begin{figure}
\centering
  \includegraphics[width=0.4\textwidth]{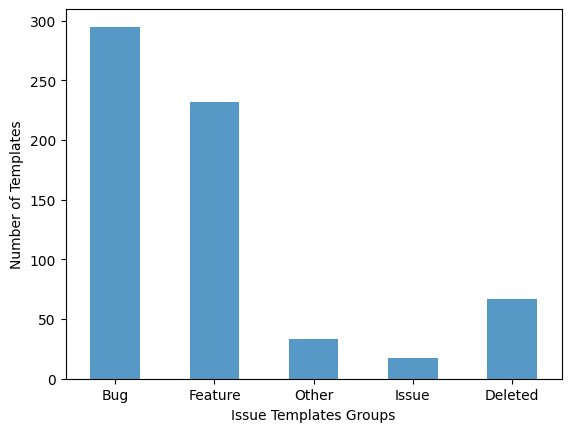}
\caption{The Number of Issue Templates in Five Groups: Bug, Feature, Other, Issue, and Deleted}
\label{fig:temp_group}  
\end{figure}

\begin{figure*}[!t]
    \centering
    \captionsetup[subfloat]{labelfont=footnotesize ,textfont=footnotesize}
    \subfloat[A Bug Template in Md format from the AdAway Repository~\cite{AdAway}]{\includegraphics[width=2.6in]{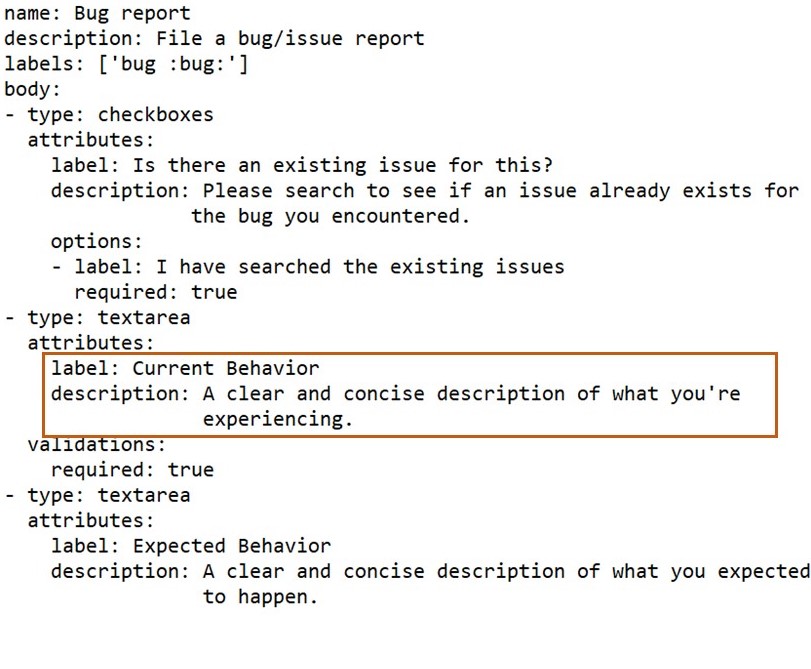}
    \label{temp1}}
    \hfil
    \subfloat[A Bug Template in YAML format from the VocableTrainer-Android Repository~\cite{VocableTrainer}]{\includegraphics[width=2.6in]{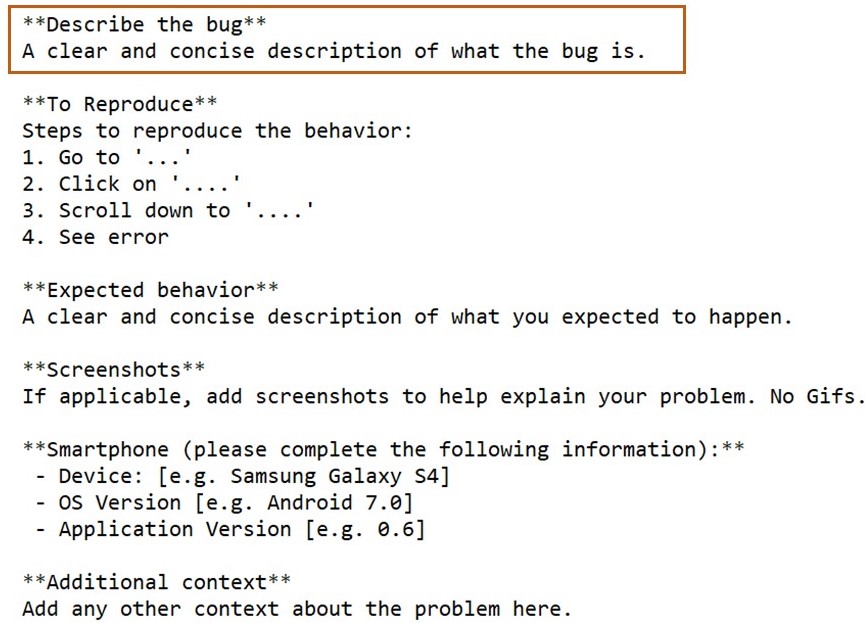}
    \label{temp2}}
    \caption{Two Samples of Bug Template in Yaml and Md Format.}
    \label{fig:two_template}
\end{figure*}

An issue template can be customized for a specific or general goal. We put our collected templates into five groups, \emph{Bug}, \emph{Feature}, \emph{Other}, \emph{Issue}, and \emph{Deleted}, according to the description provided in their definition. The first three groups consist of templates that are relevant to our target intentions, such as ``crash\_report.md'', ``feature\_request.yml'', and ``usage\_question.md''. These templates align with the types of issues we aim to address in our research. The fourth group, \emph{Issue}, consists of general templates that can be used for any purpose, such as ``general\_issue.md'', ``issue-template.md''. Finally, the \emph{Deleted} group contains templates unrelated to our target intentions, such as ``task.md'', ``tech\_debt.md''. As shown in \figref{temp_group}, the number of issue templates in each intended group is depicted. As expected, the \emph{Bug} and \emph{Feature} groups contain the highest number of defined templates, possibly due to the importance of the information provided to developers. We manually analyzed the main sections in 577 templates related to the four groups of Bug, Feature, Other, and Issue to find section titles that contain our target information. For example, \figref{two_template} shows two bug templates in both ``.md'' and ``.yml'' formats. According to the description provided in these templates, the ``Current Behavior'' and ``Describe the Bug'' sections are intended to express the encountered problem. Thus, we considered these sections our target sections, as they provide valuable information for our research purposes.

In our collected data, we found that 708 out of 999 (71\%) repositories have no issue templates. Their issues can be structured in desired formats without following standard definitions. As a result, issue bodies include the arbitrary sections that authors considered necessary when creating issues according to their intentions. So, in addition to analyzing the main sections from related issue templates, we also examined frequent sections that appear in the issue bodies to complete the identified target sections. To this end, we extracted a total of 31K section titles from the body of our collected issues. 
These additional sections from the issue bodies help us capture more diverse information that the standard templates might not cover.
We apply the three standard steps of converting to lowercase, stemming, and removing stop words and special characters to unify them. For stop words, we utilize NLTK's stop word list, but we retain three specific words: \emph{``What,''} \emph{``About,''} and \emph{``Should''}. These words are not removed because our analysis shows that they significantly impact the meaning of the section titles. For example, if we consider these words as stop words the titles ``what happened'' and ``what should happen'' would both be converted to the single term ``happen'' after preprocessing. However, these two titles convey different meanings. The first title usually refers to the app's current behavior, which provides valuable information. The second title, includes the description of the correct or expected behavior, which is differ from what we need for our analysis. Therefore, keeping these words in preprocessed titles is important to preserve their meaning.

After applying the preprocessing steps, we received 21K distinct section titles. It is worth noting that more than 90\% of these titles are used less than four times in the issue bodies. Due to the constraints of our manual analysis, we focus only on the most frequent section titles to enhance the accuracy of identifying target information. As a result, we analyze the 2,098 section titles that have been used at least four times in the issue bodies. Through this investigation, we aim to understand their use cases and determine which sections contain the appropriate information for our research purposes.

After analyzing the target sections from both the issue templates and bodies, we established 19 linguistic patterns to facilitate the extraction of these target section titles from the issue bodies. \tabref{my-pattern} shows our patterns in Regular Expression formats and their associated review intentions. Some patterns, like P1, P4, and P6, are specific to one particular intention, while others, such as P11, have broader applicability. Additionally, we grouped all the one-word section titles under the P19 template. In \tabref{my-pattern-exmpl},  we provide examples to facilitate a better understanding of the defined patterns. Additionally, we carefully examine all the section titles resulting from each of them in our dataset to ensure the accuracy of these patterns.

It is essential to note that the body of an issue may contain multiple sections that can be extracted using defined linguistic patterns. Our investigations show that the first extracted section generally provides more related information than the subsequent ones, often repeating the same information with minor differences. To optimize our performance and avoid redundancies, we consider only the first target section in the body of the issue.
On the other hand, there are some issues in which the body does not adhere to any specific structure. In these cases, our investigations show that issues with several paragraphs contain various information, including implementation details, making it more challenging to identify valuable parts. In contrast, single-paragraph issues usually provide an overview of the intended content. Therefore, to increase the accuracy of the auxiliary dataset, we choose to consider only the unstructured issues that consist of one paragraph and ignore the multi-paragraph issues. By focusing on single-paragraph issues, we can more effectively extract relevant and meaningful information, contributing to the overall quality of the final auxiliary dataset for our research purposes. 

Our goal is to create an auxiliary dataset that minimizes redundant information and enhances the accuracy of the predictive model. This Section proposes a method to identify and extract target information from the body of issues. To ensure the accuracy and reliability of this method for extracting target information, we conducted a manual analysis on a sample of 384 randomly selected issues from our dataset, with a confidence level of 95\% and a confidence interval of 5\%. Two programmers, with an average experience of 8 years, independently analyzed the target information extracted from these issues to determine if it contained the desired information. Also, they discussed their conflicts in order to reach a single decision.
The results of their investigation indicate that in 82.8\% of the cases, the extracted information was as expected. In the remaining cases, the target sections were not identified for various reasons, such as unstructured issues and the use of uncommon titles, resulting in empty output strings.
Since our primary goal is to increase the accuracy of the auxiliary dataset, we can disregard cases that complicate the linguistic patterns and potentially reduce output accuracy. Additionally, as these cases represent a small percentage of the preliminary dataset, ignoring them does not significantly reduce the size of the final auxiliary dataset.
As a result, following the mentioned method in this Section, we can extract the necessary information to form a reliable auxiliary dataset for augmenting primary datasets. Also, for better reader comprehension, we provided samples of information extracted from the issue bodies using our approach in \tabref{extracted_info}.

\begin{table*}
    \caption{Samples of the Target Information Extracted from Issue Bodies}
    \renewcommand{\arraystretch}{1.3}
    \begin{tabular}{|p{.6\linewidth}|p{.2\linewidth}|c|}
        \hline
        \textbf{Raw Issue} & \textbf{Taregt Information} & \textbf{Label}\\
        \hline
        **Describe the bug** & Quick mode UI breaks & Bug Report\\
        Quick mode UI breaks after switching from classic mode. 
        & after switching from classic mode. & \\
        **To Reproduce**
        Steps to reproduce the behavior:
        1. Start training
        2. Answer vocable in classic mode
        3. Switch to quick mode
        4. UI shows all possible buttons of quick mode (correct/wrong/resolve)
        & & \\
        **Expected behavior**
        Only shows buttons currently to be shown on new vocable
        & & 
        \\
        \hline
        **Is your feature request related to a problem? Please describe.**
        & 
        Since the LIVE streaming app is very popular
        & Feature Request \\
        Since the LIVE streaming app is very popular now, hopefully \{N\} may support this feature soon.
        & 
        now, hopefully \{N\} may support this feature soon.
        & \\
        **Describe the solution you'd like**
        & & \\
        May refer to React Native, plenty of plugins for building the LIVE streaming app. & & \\
        **Describe alternatives you've considered** & & \\
        No alternatives, FOREVER SUPPORT NATIVESCRIPT!!! 
        & & \\
        \hline
        \#\#\# Question & 
        Hello, I have a question,
        & Other \\
        Hello, I have a question, is there a way to get comparison results as XML?, thanks
        & is there a way to get comparison results as XML?, thanks
        &
        \\
        \hline
    \end{tabular}
\label{table:extracted_info}
\end{table*}

\begin{table*}
    \caption{Linguistic Patterns (constituent words represented in lemmatized format. B: Bug Report, F: Feature Request, O: Other)}
    \centering
    \renewcommand{\arraystretch}{1.3}
    \begin{tabular}{clccc}
        \hline
        \textbf{Name} & \textbf{Regex Pattern} & \textbf{B} & \textbf{F} & \textbf{O}\\
        \hline
        P1 & r``.*actual.*(?:result\textbar behavior\textbar behaviour
        \textbar outcom\textbar output)'' & \textbullet & &\\
        \hline
        P2 & r``.*(?:current\textbar observ).*(?:result\textbar behavior\textbar behavior\textbar problem)'' &\textbullet & \textbullet & \\
        \hline
        P3 & r``.*(?:describe\textbar descript\textbar explan\textbar explain).*(?:question\textbar bug\textbar problem\textbar featur\textbar crash\textbar use case\textbar usecas)\textbar & & & \\ & (?:question\textbar bug\textbar problem\textbar featur\textbar crash\textbar content\textbar use case\textbar usecas) .*(?:descript\textbar explan\textbar explain)'' & \textbullet & \textbullet & \textbullet\\
        \hline
        P4 & r``.*(?:ask\textbar state\textbar what) question'' & & & \textbullet\\
        \hline
        P5 & r``problem (?:statement\$\textbar address\$)|(?:bug\textbar tell us about) (?:problem\$)'' & \textbullet & \textbullet & \\
        \hline
        P6 & r``.*problem.*(?:tri\textbar attempt).*solv\textbar what.*problem.*solv'' & & \textbullet &\\
        \hline
        P7 & r``.*short descript'' & \textbullet & \textbullet & \\
        \hline
        P8 & r``.*(?: featur want\textbar want featur)'' & & \textbullet & \\
        \hline
        P9        
        & r``.*featur (?:request\textbar suggest\$)|.*suggest (?:featur\$)'' & & \textbullet & \\
        \hline
        P10
        & r``.*what (?:featur\textbar requir)\textbar what.*(?:featur\$)'' & & \textbullet & \\
        \hline
        P11
        & r``what (?:issu\$|about\$)\textbar what issu about'' & \textbullet & \textbullet & \textbullet \\
        \hline
        P12
        & r``.*what(?:(?!expect\textbar should).)*happen'' & \textbullet & & \\
        \hline
        P13    
        & r``(?:what )?(?:user )?(?:problem\$)'' & \textbullet & \textbullet & \\
        \hline
        P14      
        & r``(?:bug\textbar issu) (?:summari\textbar explan)\textbar (?:summari\textbar explan) (?:bug\textbar issu)'' & \textbullet & \textbullet & \\
        \hline
        P15
        & r``(?:subject\textbar featur\textbar bug\textbar describ) (?:issu\$)\textbar issu (?:bug\$\textbar featur\$\textbar request\$\textbar question\$\textbar detail\$)'' & \textbullet & \textbullet & \\
        \hline
        P16
        & r``user (?:benefit\textbar experi)'' & & \textbullet & \\
        \hline
        P17 & r``.*(?:see instead)''
        & & \textbullet & \\
        \hline
        P18 & r``.*featur request relat (?:problem\textbar issu).*'' & & \textbullet & \\
        \hline
        P19 &
        r``(?:overview\$\textbar summari\$\textbar descript\$\textbar issu\$\textbar result\$\textbar problem\$\textbar bug\$\textbar feature\$\textbar usecas\$\textbar use case\$\textbar& & &\\ & usr stori\$\textbar stori\$\textbar question\$\textbar actual\$\textbar observ\$)''
        & \textbullet & \textbullet & \textbullet\\
        \hline
    \end{tabular}
    \label{table:my-pattern}
\end{table*}

\begin{table*}[!t]
    \caption{Examples of Linguistic Patterns}
    \centering
    \renewcommand{\arraystretch}{1.3}
    \begin{tabular}{|c|p{.4\linewidth}|c|p{.4\linewidth}|}
        \hline
        \textbf{Name} & \textbf{Examples} & \textbf{Name} & \textbf{Examples}\\
        \hline
        P1 & -	Actual/Expected behavior & P11 & - What is this issue about?\\
           & -	Actual behaviour after performing these steps &  & - What's the issue\\
        \hline
        P2 & -  Case description and the observed              behavior & P12 & - What would you like to happen?\\
           & -	What is the current behavior? & & - Describe what actually happened.\\
        \hline
        P3 & - Describe the bug in a sentence or two. & P13 & - What is the problem?\\
           & - Describe your question in detail. & & - What/User problem\\
           & - Describe your suggested feature. & & \\
        \hline
        P4 & - What is/are your question(s)? & P14 & - Summary of issue\\
           & - Ask your question. & & - Bug explanation\\
        \hline
        P5 & - Problem statement & P15 & - Issue/Question\\
           & - Tell us about the problem. &  & - Describe the issue\\
        \hline
        P6 & - Problem you are trying to solve & P16 & - Why/User benefit\\
           & - What is the user problem or growth opportunity you want to see solved? & & - User experience\\
        \hline
        P7 & - Provide a short description of the feature. & P17 & - What is the expected output? what do you see instead?\\
           & - Bug - short description & & - What did you see instead?\\
        \hline
        P8 & - Problem you may be having, or feature you want & P18 & - Is your feature request related to a problem? please describe it.\\
           & - Describe the feature you want. & & - Is your feature request related to an issue?\\
           & - Why do you want this feature?  & & \\
        \hline
        P9 & - Step 3: feature request & P19 & - Description\\
           & - Describe your suggested feature. & & - Motivation\\
        \hline
        P10 & - What feature would you like to see?& - & -\\
            & - What is the need and use case of this feature? & &\\
        \hline
    \end{tabular}
    \label{table:my-pattern-exmpl}
\end{table*}

\subsubsection{Preprocessing Textual Data}
\label{sec:preprocess}
We have two types of textual data that should be preprocessed for creating the auxiliary dataset: (i) target information extracted from issue bodies and (ii) issue titles. Due to their nature, we apply several steps to remove noisy and error-prone data. So, we clean textual data using standard preprocessing steps in NLP. These steps include the following:

\begin{enumerate}
    \item Filtering textual data: The body of issues is usually processed by removing additional information that has less value for our purpose, which increases the accuracy of the proposed approach~\cite{Izadi}. We remove noisy contents such as checklists, tags, links, code fragments, variable names, error messages, mention terms, and issue numbers. Additionally, we exclude special phrases listed in \tabref{pre_words}, as well as any phrases that start and end with an underline symbol (such as \emph{``\_created by X\_}''). These phrases do not provide helpful information and usually refer to the author or source of the issue.
    \item Tokenizing the data into constituent words that will serve as the features for the learning model.
    \item Transforming words to lower to unify their appearance.
    \item Removing numbers, punctuations, and special characters (such as \$, \#, and emojis).
    \item Transforming negative modifiers to a single form \emph{``not''}: Negative terms play a crucial role in determining the intention of a review~\cite{Scalabrino}. So, we identify 44 negative modifiers (\tabref{pre_words}) and replace them with ``not'' to reduce the number of features. 
    \item Removing stop words: we use the NLTK stop words, with exceptions for negative modifiers listed in \tabref{pre_words} and three specific words \emph{``Could''}, \emph{``Would''}, and \emph{``Should''}. The exclusion of the former is explained in the previous step. Also, the latter represent important keywords for indicating the review intention~\cite{Maalej_1}. Additionally, the term \emph{``have to''} (similar to \emph{``must''}) can be effective in predicting a feature request review. As this term consists of two words, ``have'' and ``to,'' they are identified as stop words. Consequently, we convert it to the term \emph{``have-to''} to prevent its deletion during preprocessing.    
    \item Transformation to lemma: We apply the WordNet lemmatizer from the NLTK library to unify different word forms.
    \item Removing non-informative items: After applying the above steps to the titles and bodies, we remove issues that contain less than three words because they provide little information~\cite{Assi}. Additionally, we filter out issues with variable names in the title because our investigation shows they are often related to implementation topics.
\end{enumerate}

After applying the preprocessing steps, we have a total of 62.7K final processed labeled issues, including 32.6K Bug Reports, 26.6K Feature Requests, and 3.9K Others. We use a subset of these issues as the auxiliary dataset to augment the primary labeled dataset.

\begin{table*}
    \caption{Special Words Utilized in Preprocessing Steps}
    \centering
    \renewcommand{\arraystretch}{1.3}
    \begin{tabular}{lp{.8\linewidth}}
        \hline
        \textbf{Type} & \textbf{Intended List} \\
        \hline
        Negative Words & no, not, nor, non, none, never, neither, without, isn, isnt, aren, arent, cant, cannot, couldn, couldnt, wasn, wasnt, weren, werent, don, dont, didn, didnt, doesn, doesnt, wouldn, wouldnt, shouldn, shouldnt, haven, havent, hasn, hasnt, hadn, hadnt, won, wont, mustn, mightn, shan, shant, needn, neednt\\
        \hline
        Special Phrases & created by, reported by, reported on, posted by, issue is synchronized with, original issue by
        \\
        \hline
    \end{tabular}
\label{table:pre_words}
\end{table*}

\subsection{Preparing the Primary Dataset}
As mentioned above, previous studies classified user reviews from various aspects, with the most frequent being selected intentions such as bug reports and feature requests. They often labeled a dataset manually through a standard process to train and test their proposed model. In this paper, we utilize five well-known labeled review datasets provided by Guzman et al.~\cite{Guzman}, Maalej et al.~\cite{Maalej_Journal}, Jha et al.~\cite{Jha}, Gu et al.~\cite{Gu}, and  Scalabrino et al.~\cite{Scalabrino} as primary datasets. It is worth noting that the Jha et al. dataset contains some reviews from Maalej et al. We removed these duplicates from the Jha et al. dataset to avoid redundancy, as we use both datasets separately.

There are several reasons for choosing these five datasets. First, they are publicly available and have been labeled through a suitable process by human experts. Second, these datasets have been used in similar research studies~\cite{Devin, Hadi}, making them well-established and reliable sources for comparison and evaluation. Therefore, studying their generalizability improvements through augmentation with the help of an auxiliary dataset from another source can be highly beneficial. Third, the review classes in these datasets can be adapted to our intentions. Since these datasets use different taxonomies or label names, we unified related classes to the \emph{bug report}, \emph{feature request}, and \emph{other} intentions according to their definitions. Additionally, some classes were inconsistent with our intentions or could not be assigned to just one class, so we excluded them. For example, reviews related to the security label in the Scalarbino et al. dataset could include bug reports or requests for improving security-related app features. We present the adapted classes in \tabref{Adopt_Labels}. Forth, selected datasets were collected from various apps with different data sizes. This way, we can analyze the feasibility of augmenting review datasets with an auxiliary dataset derived from GitHub issues under different conditions.

Finally, We preprocess user reviews by applying steps 2 to 8 in \secref{preprocess}. \tabref{Dataset_Spec} provides details of the primary datasets after applying the preprocessing steps.

\begin{table*}
    \caption{Adapted Classes in Primary Datasets}
    \centering
    \renewcommand{\arraystretch}{1.3}
    \begin{tabular}{lllllp{.2\linewidth}}
        \hline
        \textbf{Dataset} & \textbf{Research} & \textbf{Bug Report} & \textbf{Feature Request} & \textbf{Other} & \textbf{Deleted}\\
        \hline
        PD1 & Guzman~\cite{Guzman} & Bug report & User request & Praise, Usage scenario & Feature shortcoming, Feature strength, Complaint, Noise\\
        \hline
        PD2 & Maalej~\cite{Maalej_Journal} & Bug & Feature & UserExperience, Rating & -\\
        \hline
        PD3 & Jha~\cite{Jha} & BugReport & FeatureRequest & Other & -\\
        \hline
        PD4 & Gu~\cite{Gu} & bug reports & feature requests & praise, others & aspect evaluation\\
        \hline
        PD5 & Scalabrino~\cite{Scalabrino} & BUG & FEATURE & OTHER & PERFORMANCE, SECURITY, ENERGY, USABILITY\\
        \hline
    \end{tabular}
\label{table:Adopt_Labels}
\end{table*}

\begin{table*}
    \caption{Details of Primary Datasets Used in Our Approach}
    \centering
    \renewcommand{\arraystretch}{1.3}
    \begin{tabular}{llccccc}
        \hline
        \textbf{Dataset} & \textbf{Platform} & \textbf{\#Apps} & \textbf{\#Reviews} & \textbf{\#Bug Report} & \textbf{\#Feature Request} & \textbf{\#Other}\\
        \hline
        PD1 & Google Play & 7 & 2,794 & 985 & 403 & 1,763\\
        \hline
        PD2 & Google Play, App Store & 458 & 1,218 & 134 & 178 & 958\\
        \hline
        PD3 & Google Play, App Store & - & 3,186 & 1,593 & 799 & 812\\
        \hline
        PD4 & Google Play & 17 & 13,930 & 1,774 & 2,085 & 10,076\\
        \hline
        PD5 & Google Play & 595 & 3,083 & 1,239 & 491 & 1,355\\
        \hline
    \end{tabular}
\label{table:Dataset_Spec}
\end{table*}

\subsection{Dataset Augmentation}
\label{sec:data_aug}
As mentioned earlier, our goal is to enhance the performance of review classification models by augmenting the training dataset with labeled data extracted from GitHub issues. To this end, we prepared an auxiliary dataset of GitHub issues by applying various preprocessing steps. Also, we utilized five manually labeled review datasets as primary datasets and adapted their labels to align with our intentions (bug report, feature request, other). During this phase, we integrate these two types of datasets based on the three methods of \emph{Between-Apps}, \emph{Within-Apps}, and \emph{Within-Context Analysis}. Maalej et al.~\cite{Maalej_Journal} introduced the first two, and we incorporate these methods in our approach based on their definition.

In Between-Apps Analysis, labeled items are randomly selected from various apps to create the training dataset. Therefore, a general model is trained to classify reviews regardless of their apps. Previous approaches are often based on this method due to its simplicity and the availability of labeled datasets. However, in this method, we randomly select a subset of processed issues as an auxiliary dataset to integrate with a primary dataset, thereby creating an augmented dataset. While this method can be effective, it may not fully capture the specific characteristics of individual apps, which could impact the model's performance in app-specific review classification task.

Another method is Within-Apps Analysis, which has received less attention because of the labeling challenges it presents. 
Based on this method, review classifiers are trained for each app using a specifically labeled dataset. In the Within-Apps Analysis, we use the processed issues of the same app to prepare an auxiliary dataset. This method allows us to leverage app-specific information and characteristics, tailoring the classifier to each app's unique review requirements. Indeed, obtaining a sufficient amount of labeled data within each app presents a significant challenge that needs to be addressed to ensure the effectiveness of the Within-Apps Analysis approach. This is particularly true for many Android apps, especially those that are not popular and may have limited labeled issues available. Hassan et al.~\cite{Hassan} demonstrated that analyzing apps while considering the data provided by similar apps can lead to helpful information and reveal other aspects. Therefore, we propose the Within-Context Analysis as an alternative method to overcome the limitation of app-specific labelled datasets. This method leverages the context data of an app to prepare its auxiliary dataset. Instead of relying solely on app-specific labeled issues, we utilize GitHub issues from similar apps with similar characteristics and functionalities. By doing so, we can gather the required information for the target app from a broader range of sources, enhancing the potential for a more comprehensive and contextually relevant review classification.
Additionally, we use the repository descriptions in GitHub to find similarities among apps, as detailed in \secref{find_similars}. 

Since the auxiliary dataset is obtained from another source and has some limitations (such as in the other class), we carefully consider its size compared to the primary dataset. To avoid biasing the results towards the auxiliary dataset and to maintain the integrity of the primary dataset, we ensure that the auxiliary dataset has a smaller size. By managing the size of the auxiliary dataset in this manner, we aim to increase the overall performance.

\subsubsection{Finding Similar Repositories}

\label{sec:find_similars}
\begin{figure}[!t]
\centering
\includegraphics[width=4in]{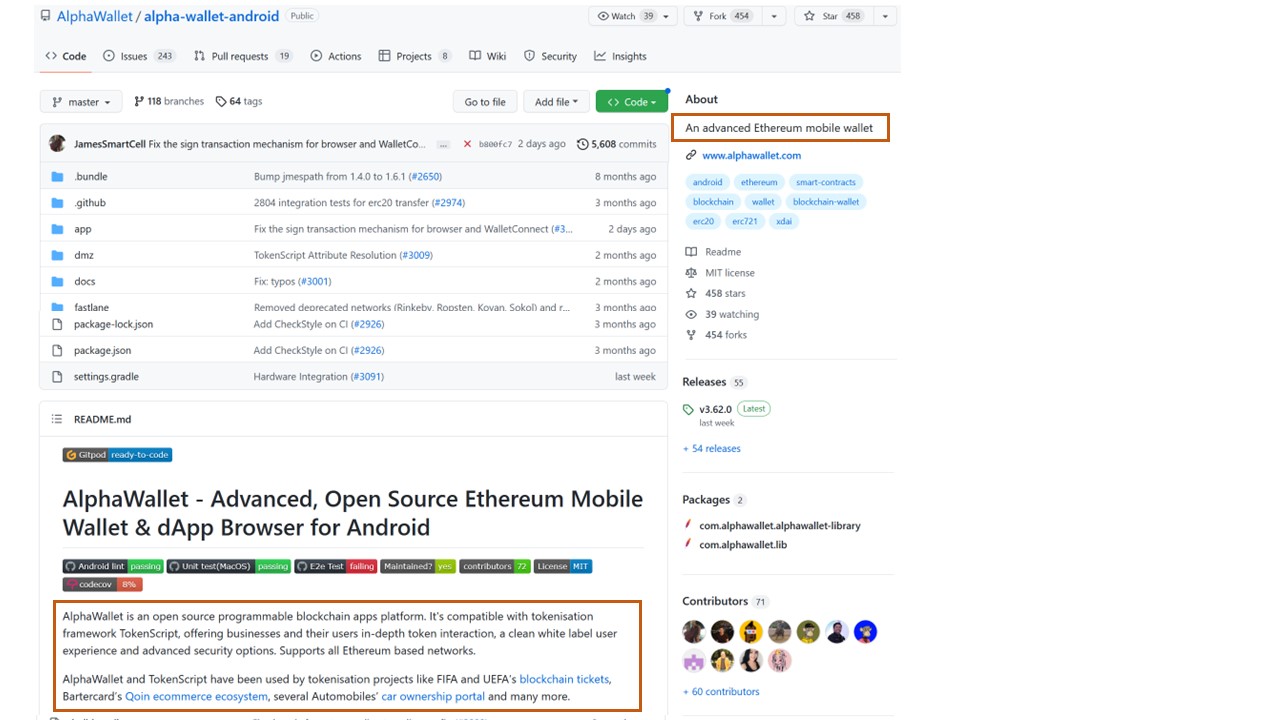}
\caption{An Example of a GitHub Repository}
\label{fig:repo_example}  
\end{figure}

In this step, we create a profile for each repository to find similar ones~\cite{Li_sim}. We utilize the information available in the \emph{ReadMe} file and the \emph{About} section of repositories (as shown in \figref{repo_example}). The ReadMe file contains descriptions of the repository, including its purpose, features, usage instructions, contribution guidelines, etc. We manually analyzed 200 randomly selected ReadMe files to extract the necessary information. Our findings revealed that the initial section of the ReadMe file and sections titled \emph{introduction}, \emph{description}, \emph{features}, \emph{what it does}, \emph{about}, \emph{ about the project}, \emph{overview}, \emph{summary}, \emph{todo} often describe the repository's purpose and features. Furthermore, the About section provides a concise app description. We extract the content of these sections to construct the app profiles and preprocess them using steps 2 to 8 in \secref{preprocess}. We employ tf-idf~\cite{tfidf} to transform each app profile into a vector and calculate cosine similarities to find similar apps.

\subsection{Review Classification}
\label{sec:rev_classify}
As previously mentioned, we utilize processed issues, including issue titles and content of specific sections within issue bodies, to augment manually labeled reviews. 
These are integrated to generate augmented datasets, resulting in more accurate predictions of user review intentions. Recent studies have demonstrated transformer-based models performed well in classifying user reviews~\cite{Hadi, Henao, Araujo}. 
In this paper, we employ our augmented datasets to fine-tune four pre-trained models, BERT, RoBERTa, DistilBERT, and ALBERT.

We aim to classify user reviews into three intention classes: bug reports, feature requests, and others. Consequently, we face a multi-class classification problem with two alternatives (i) training a binary classifier for each class or (ii) training a multi-class classifier~\cite{Maalej_Journal}. Since a review could fall simultaneously into the bug report and feature request categories, we choose the former in training classifiers. Also, prior studies~\cite{Maalej_Journal, Hadi} found that training binary classifiers provide better accuracy than multi-class classifiers. On the other hand, considering that the \emph{others} class comprises reviews that do not pertain to bug reports or feature requests, it becomes apparent that training a distinct classifier for this category is unnecessary. As a result, we train two classifiers specifically for predicting the bug report and feature request classes.

\section{Experiment Settings}
\label{sec:experiment_setting}
\subsection{Research Questions}
This study investigates how information on GitHub can assist developers in training more generalized and accurate models for classifying user reviews. To this end, we try to answer the following research questions:

\begin{enumerate}
    \item \textbf{RQ1}: How does using datasets augmented with GitHub issues using the three methods of Between-Apps, Within-Apps, and Within-Context Analysis increase the performance of review classification models?
    \item \textbf{RQ2}: How does using augmented datasets enhance the generalizability of review classification models when predicting new datasets?
    \item \textbf{RQ3}: How does the volume of auxiliary datasets used for augmenting training datasets impact the performance of review classification models?     
    \item \textbf{RQ4}: How does the augmented dataset improve review classifiers based on different pre-trained models?
\end{enumerate}

To Answer these questions, we designed four experiments to evaluate our proposed approach and answer research questions. First, we evaluated three augmentation methods that enrich the primary labeled datasets. We analyzed the performance of the review classifiers trained using augmented datasets and showed how effectively each method can improve the accuracy of the classification model. Second, we explored augmentation's impact on increasing the generalizability of primary datasets in predicting new ones. Third, We demonstrated how the volume of auxiliary datasets influences the classification model's performance. We identified a suitable volume range for effectiveness, ensuring the model remains unbiased toward auxiliary datasets while yielding optimal results.
Finally, we compared the performance of various transformer-based models trained on augmented datasets for review classification.
The results of these experiments can offer valuable insights when applying our proposed approach to other datasets.

\subsection{Dataset and Models}
we create a manually labeled truth dataset containing reviews from five apps to evaluate and compare models trained on various forms of augmented data in RQ1. Since we use the truth dataset to compare three augmentation methods, we select these apps in such a way that for each one, (i) the associated GitHub repository includes labeled issues from all three target classes; (ii) multiple similar apps are present within our dataset; and (iii) the volume of Google Play reviews is substantial enough to create a proper truth dataset.
\tabref{Truthset_Spec} provides an overview of the selected apps. We selected the apps that closely align with the mentioned conditions in our collected data. Also, there are only a limited number of open-source apps with repositories containing a suitable quantity of labeled bug report and feature request issues. It is noticeable that two of our selected apps, \emph{AntennaPod} and \emph{MetaMask}, have fewer issues within their GitHub repositories compared to the other three. However, we have retained them to ensure diversity within our truth dataset and to investigate the effectiveness of three augmentation methods in less developed applications.

Additionally, in the Within-Apps method, if issues refer to concepts that have already been raised in the reviews, they can be identified by developers under the influence of that review. Therefore, we have a potential information leakage that needs to be managed for result reliability. To prevent this problem, we carefully select reviews where no issues with similar concepts have been posted in the associated GitHub repository prior to them. Therefore, we first vectorized the reviews and issues of the same app using the tf-idf method. Then, for each review in the truth dataset, we calculated its cosine similarity with the issues of the same app. If we find a review with issues having a similarity greater than 0.7, we manually inspect it to determine whether the review and the identified similar issues refer to the same concept. Finally, we retain comments that have not been followed by issues with similar concepts.

To create a truth dataset, we followed the labeling process steps used by Guzman et al.~\cite{Guzman}. First, we prepared a guideline containing the definitions of each class along with several examples. This guideline aided the annotators in comprehending the process better and minimizing disagreements. Second, we randomly selected 200 reviews from each app, including reviews from the bug report, feature request, and other classes. Four students, with an average programming experience of 9.5 years, participated as annotators in the labeling process. We assigned the selected reviews to them so that two annotators assessed each review. They manually analyzed these reviews and selected proper labels for them. Overall, they agreed on 91\% of the labels and achieved a 0.79 Cohen's kappa~\cite{cohen}, providing reliable results. Finally, we investigated disputed reviews and held a meeting to make final decisions about them. Annotators discussed each other during this meeting and agreed on the final labels. \tabref{Truthset_Spec} shows the specifications of the labeled truth dataset.

\begin{table*}
    \caption{Details of Truth Datasets (Revs=Reviews)}
    \centering
    \setlength{\tabcolsep}{1pt}
    \renewcommand{\arraystretch}{1.3}
    \makebox[\textwidth][c]{
    \begin{tabular}{lcccccccc}
        \hline
        \textbf{Apps} & \textbf{\#Total Revs} & \textbf{\#Bug Revs} & \textbf{\#Feature Revs} & \textbf{\#Other Revs} & \textbf{\#Total Issues} &
        \textbf{\#Bug Issues} &
        \textbf{\#Feature Issues} & \textbf{\#Other Issues}\\
        \hline
        ownCloud~\cite{ownCloud, ownCloud_Git} & 200 & 90 & 63 & 55 & 3,086 & 2,560 & 527 & 293\\
        \hline
        AntennaPod~\cite{AtennaPod, AtennaPod_Git} & 200 & 33 & 58 & 109 & 160 & 45 & 107 & 55\\
        \hline
        MetaMask~\cite{MetaMask, MetaMask_Git} & 200 & 115 & 49 & 40 & 159 & 91 & 70 & 30\\
        \hline
        Firefox Nightly~\cite{Firefox_Night, Firefox_Night_Git} & 200 & 38 & 84 & 82 & 6,077 & 5,243 & 854 & 346\\
        \hline
        Firefox Focus~\cite{Firefox_Focus, Firefox_Focus_Git} & 200 & 37 & 78 & 87 & 735 & 597 & 139 & 44 \\
        \hline
        \textbf{SUM} & \textbf{1000} & \textbf{313} & \textbf{332} & \textbf{373} & \textbf{10,217} & \textbf{8,536} & \textbf{1,697} & \textbf{768}\\
        \hline
    \end{tabular}
    }
\label{table:Truthset_Spec}
\end{table*}

We employed the TFAutoModelForSequenceClassification class~\cite{TFAutoModel} from Huggingface's Transformers library to fine-tune pre-trained models. We trained the classification models for 3 epochs using a batch size of 16 and a learning rate of 5e-5 for the Adam optimizer~\cite{adam}. Furthermore, we set the maximum input string length to 50, consistent with our datasets. All models were executed on Google Colab, which offers a free GPU with 12.7 GB of system RAM, 15.0 GB of GPU RAM, and 78.2 GB of disk storage.
Also, our investigation, which corresponds to RQ4, 
demonstrates that the DistilBERT model achieves the highest F1-score in the average of the bug report and feature request results with the shortest time for training and prediction phases. So, we used it to train different review classifiers in preceding experiments.

\subsection{Evaluation Measures}
We use the standard measures of Precision, Recall, and F1-score to evaluate and compare the results of the different models. These measures are calculated using the following formulas:

\begin{equation}
precision = TP/(TP+FP)
\end{equation}

\begin{equation}
recall = TP/(TP+FN)
\end{equation}

\begin{equation}
F1-score = 2*precision*recall/(precision+recall)
\end{equation}

\begin{itemize}[label={-}]
     \item TP: The count of reviews with the target (bug report/feature request) label that are correctly classified.
     \item FP: The count of reviews with the target (bug report/feature request) label that are incorrectly classified.
     \item TN: The count of reviews with a non-target label that are correctly classified.
     \item FN: The count of reviews with a non-target label that are incorrectly classified.
\end{itemize}

As our selected primary datasets exhibit an imbalanced distribution, we employed a 5-fold cross-validation approach to mitigate potential result bias. Consequently, the results obtained represent the average metric values across each fold. To ensure consistent distribution within each fold, we used the StratifiedKFold function~\cite{StratifiedKFold} from the sklearn library for fold selection~\cite{Hadi}.

\section{Experiment Design and Implications}
\label{sec:eval}
\subsection{RQ1: Influences of Augmentation Methods on the Model Performance}

In RQ1, we aim to investigate the effects of augmenting labeled datasets through three proposed methods (Within-App, Within-Context, and Between-App analysis) in improving the performance of bug report and feature request classifiers. To address this, we compared the accuracy of the augmented models, trained based on the augmented datasets, with the baseline models, trained based on the primary datasets, using our truth dataset. To this end, we trained the following five models for each app:
\begin{enumerate}
    \item \textbf{Review-Based}: This is the baseline model in which we only use the manually labeled reviews (primary datasets) as the training dataset for the review classification models.
    \item \textbf{Review + Same Issues}: In this model, we augment the primary datasets through the Within-Apps Analysis method. We create auxiliary datasets using labeled issues from the same app.
    \item \textbf{Reviews + Similar Issues}: In this model, we augment the primary datasets through the Within-Context Analysis method. We create auxiliary datasets using labeled issues from apps similar to the target app. As discussed in \secref{data_aug}, we first identified repositories associated with similar apps in our dataset. Then, we create an auxiliary dataset using their properly labeled issues. We maintained the auxiliary dataset volume at 0.3 of the primary dataset volume, a ratio in the effective range for augmentation volume, as discussed in RQ3.
    \item \textbf{Reviews + (Same + Similar) Issues}: Similar to the previous model, we augment the primary datasets through Within-Context Analysis, with the distinction that we employed issues from both the same and similar apps. Also, we maintained the effective volume of the auxiliary datasets at 0.3. So, we consider the context issues of a target app to include issues related to the same and similar apps.
    \item \textbf{Reviews + Random Issues}: In this model, we augment the primary datasets through the Between-Apps Analysis method. We created auxiliary datasets using randomly selected labeled issues. As with the previous models, the ratio of auxiliary dataset volume to primary dataset volume was maintained at 0.3.
\end{enumerate}

\subsubsection{Results}
In this experiment, we evaluate three proposed methods that augment the training dataset using labeled issues in the review classification task. Tables \ref{table:RQ1_Bug} and \ref{table:RQ1_Feature} show the results of five models trained for each app in the truth dataset. In these tables, underlined numbers indicate the maximum value within each respective column. As can be seen, classification models trained on an augmented dataset can improve F1-score, precision, and recall compared to the baseline models trained on the primary dataset. In the bug report, augmented data led to performance improvements in F1-score by up to 6.3, precision by up to 2.0, and recall by up to 9.2 on average. Additionally, in the feature request, our augmentation approach enhanced F1-score by up to 7.2, precision by up to 6.7, and recall by up to 2.3 on average.

On the other hand, data augmentation can reduces model precision or recall in some cases. In the bug report, the average precision decrease between 1.4 to 5 for Rev+(Sim)I and Rev+(Rand)I, respectively. In the feature request, the average recall decrease between 0.2 to 2.4 for three augmented models. 
This result was predictable because auxiliary datasets have limitations, including (i) they differ from user reviews in the expression. Usually, issues convey concepts more technically with fewer emotional phrases, while user reviews exhibit the opposite tendency; (ii) they do not include all types of user reviews. In the other class, only user questions are usually present in issues.

Also, the analysis of the results for each app shows that our approach typically leads to an increase in model performance, with only a few exceptions. In the bug report, we can see F1-score reduction for some augmented models in \emph{AntennaPod}, \emph{Firefox Nightly}, and \emph{Firefox Focus}. Nevertheless, the maximum value is still associated with an augmented model that utilizes data from relevant apps.
Additionally, in the feature request, the F1-score value for only one augmented model in \emph{MetaMask} reduces with a slight difference of 1.5 compared to the baseline model.
These reductions could be due to the lack of sufficient and appropriate information in the auxiliary dataset. In some apps, the GitHub repository corresponds to the same, or similar apps may have low activity. Therefore, a few relevant issues remain after applying various steps to prepare an auxiliary dataset (\secref{aux_dataset}). For example, apps \emph{AntennaPod} and \emph{MetaMask} have 160 and 159 appropriately labeled issues for our task, respectively.
On the other hand, selected issues may limit to some case of user requirement types in target classes. This problem is more likely in the Between-App Analysis method, in which the issues are randomly selected because the selected issues may belong to apps with less functional similarity with the target app. Therefore, although we perform filters to select the appropriate issues for creating the auxiliary dataset, they may not represent all the various app requirements from the user's point of view.

\begin{table*}
    \caption{Results of Bug Report Classification on Truth Dataset Using Primary (Rev) and Augmented (Rev+(S)I, Rev+(Sim)I, Rev+(S+Sim)I, and Rev+(Rand)I) Datasets in RQ1 (Rev=Reviews, I=Issues, S=Same, Sim=Similar, Rand=Random).}
    \centering
    \setlength{\tabcolsep}{2pt}
    \renewcommand{\arraystretch}{1.3}
    \makebox[\textwidth][c]{
    \begin{tabular}{|c|ccc|ccc|ccc|ccc|ccc|ccc|}
        \hline
            \multicolumn{19}{|c|}{\textbf{Bug Report}} \\
        \hline
            \multirow{2}{*}{Datasets} & 
            \multicolumn{3}{|c|}{ownCloud} & 
            \multicolumn{3}{|c|}{AntennaPod } & 
            \multicolumn{3}{|c|}{MetaMask} &
            \multicolumn{3}{|c|}{Firefox Nightly} &
            \multicolumn{3}{|c|}{Firefox Focus} &
            \multicolumn{3}{|c|}{Average} \\\cline{2-19}
            & P & R & F1 & P & R & F1 & P & R & F1 & P & R & F1 & P & R & F1 & P & R & F1\\
        \hline
            Rev & 76.0 & 65.3 & 62.5 & 73.2 & \underline{66.6} & 65.1 & 78.6 & 57.3 & 61.5 & \underline{87.6} & 50.4 & 63.7 & 72.0 & 67.2 & 63.8 & 77.5 & 61.4 & 63.3\\
        \hline
            Rev+(S)I & 86.1 & 70.7 & 74.7 & \underline{84.9} & 58.2 & \underline{68.2} & \underline{94.6} & 53.4 & 66.0 & 59.7 & \underline{71.6} & 61.5 & 72.0 & 53.0 & 60.1 & \underline{79.5} & 61.4 & 66.1\\
        \hline
            Rev+(Sim)I & 82.9 & 77.6 & 79.6 & 78.9 & 52.7 & 62.6 & 92.9 & \underline{72.0} & \underline{80.3} & 66.3 & 67.4 & \underline{64.9 }& 59.3 & 71.9 & 60.3 & 76.1 & 68.3 & 69.5\\
        \hline
            Rev+(S+Sim)I & \underline{87.8} & 73.1 & 79.0 & 76.5 & 61.2 & 67.0 & 91.7 & 65.9 & 75.4 & 81.2 & 51.6 & 62.9 & 57.6 & 75.6 & \underline{63.9} & 79.0 & 65.5 & \underline{69.6}\\
        \hline
            Rev+(Rand)I & 81.5 & \underline{83.3} & \underline{82.1} & 83.5 & 47.9 & 58.5 & 90.6 & 71.3 & 79.0 & 62.8 & 70.0 & 63.8 & 42.7 & \underline{80.5} & 53.6 & 72.2 & \underline{70.6} & 67.4\\
        \hline
    \end{tabular}
    }
\label{table:RQ1_Bug}
\end{table*}

\begin{table*}
    \caption{Results of Feature Request Classification on Truth Dataset Using Primary (Rev) and Augmented (Rev+(S)I, Rev+(Sim)I, Rev+(S+Sim)I, and Rev+(Rand)I) Datasets in RQ1 (Rev=Reviews, I=Issues, S=Same, Sim=Similar, Rand=Random).}
    \centering
    \setlength{\tabcolsep}{2pt}
    \renewcommand{\arraystretch}{1.3}
    \makebox[\textwidth][c]{
    \begin{tabular}{|c|ccc|ccc|ccc|ccc|ccc|ccc|}
        \hline
            \multicolumn{19}{|c|}{\textbf{Feature Request}} \\
        \hline
            \multirow{2}{*}{Datasets} & 
            \multicolumn{3}{|c|}{ownCloud} & 
            \multicolumn{3}{|c|}{AntennaPod } & 
            \multicolumn{3}{|c|}{MetaMask} &
            \multicolumn{3}{|c|}{Firefox Nightly} &
            \multicolumn{3}{|c|}{Firefox Focus} &
            \multicolumn{3}{|c|}{Average} \\\cline{2-19}
            & P & R & F1 & P & R & F1 & P & R & F1 & P & R & F1 & P & R & F1 & P & R & F1\\
        \hline
            Rev & 54.0 & 79.4 & 57.7 & 51.1 & 82.3 & 56.6 & 57.5 & 83.8 & 63.9 & 60.4 & 75.8 & 61.2 & 51.3 & 80.2 & 57.1 & 54.9 & 80.3 & 59.3\\
        \hline
            Rev+(S)I & 52.5 & \underline{81.6} & 62.0 & \underline{62.9} & 81.4 & \underline{69.5} & 58.7 & 81.2 & 64.7 & 60.9 & 87.9 & 69.3 & \underline{73.0} & 62.6 & \underline{65.6} & \underline{61.6} & 78.9 & 66.2\\
        \hline
            Rev+(Sim)I & 57.9 & 74.9 & 64.0 & 59.7 & 69.7 & 62.6 & 50.5 & \underline{89.8} & 62.3 & 67.2 & 79.3 & 69.7 & 53.4 & \underline{86.7} & 62.9 & 57.7 & 80.1 & 64.3\\
        \hline
            Rev+(S+Sim)I & \underline{62.7} & 71.7 & \underline{65.8} & 49.1 & \underline{87.2} & 60.4 & \underline{64.6} & 84.9 & \underline{72.7} & 61.4 & \underline{89.1} & 72.2 & 53.3 & 80.0 & 61.3 & 58.2 & \underline{82.6} & \underline{66.5}\\
        \hline
            Rev+(Rand)I & 54.9 & 74.3 & 59.0 & 47.0 & 83.1 & 58.7 & 57.1 & 85.3 & 66.1 & \underline{67.6} & 81.9 & \underline{73.5} & 64.7 & 65.1 & 62.7 & 58.3 & 77.9 & 64.0\\
        \hline
    \end{tabular}
    }
\label{table:RQ1_Feature}
\end{table*}

Since reviews represent the user requirements, analysis of them can affect software maintenance decisions to promote the user's satisfaction and the app's success~\cite{Palomba}. Recall can be more important than precision from the developer's point of view in the review classification model related to certain situations, especially for identifying bug reports, as developers probably prefer not to miss them~\cite{Maalej_Journal}.
This is because a high recall ensures that the model identifies most, if not all, actual bug reports. While a high precision ensures that the model mainly identifies true bugs and avoids irrelevant reviews, missing even a single bug report can be more detrimental for developers. Unfixed bugs can lead to crashes, errors, and user frustration, whereas developers can quickly dismiss false positives.
Generally, obtained results show that although the average precision decreases in some augmented models related to bug reports, the F1-score increases due to the significant recall improvement. Therefore, models trained on augmented datasets identify a higher percentage of actual bug report reviews that are helpful for developers. For example, in the bug report, the recall of the baseline model is between 50.4 to 67.2 and, on average, 61.4. 
In our models, recall is between 47.9 to 83.3; on average, it is between 61.4 to 70.6. Additionally, the data augmentation makes the training dataset more diverse and provides more samples for different types of reviews. Hence, trained models on augmented datasets are less biased to specific types, which increases recall. 

Comparing evaluation results of augmentation methods reveals exciting information. In the bug report, models that used contextual issues achieved the best results in the F1-score. On average, F1-score values for Rev+(S+Sim)I and Rev+(Sim)I are 69.6 and 69.5, respectively. Also, the average of F1-score values for the Reviews+(Rand)I is 67.4. So, a randomly selected auxiliary dataset has a lower measure than contextual augmented models. In the feature request, we have almost the same findings. 
Models based on contextual information, with an average F1-score of 66.5 and 64.3 for Rev+(S+Sim)I and Rev+(Sim)I, respectively, had better scores than the random-based model, with an average F1-score of 64.0. Therefore, augmenting labeled reviews using issues of apps that have similar functionality to a target app makes review classification models more robust.
Since the auxiliary dataset randomly selected issues from various apps with different contexts in Between-App Analysis, it can be reasonable that this method achieves less improvements. 

On the other hand, in the bug report, the average F1-score associated with Rev+(Sim)I, 69.5, is better than Rev+(S)I, 66.1. However, in the feature request, we observe the opposite results, where Rev+(S)I outperforms Rev+(Sim)I with an average F1-score of 66.2 compared to 64.3. 
The reason for this difference could be that bug reports for similar apps exhibit a higher degree of similarity to each other~\cite{Tan}. Encountered bugs usually include common problems such as crashing, making errors, freezing screens, etc., which can be found in similar apps. Meanwhile,
due to the wide range of available apps in Google Play, app features are highly varied, which makes it challenging to identify requests for them using our limited dataset. Each application has specific features that are mentioned in issues or reviews. So, bug reports in the similar apps can have more similarity to each other than feature requests.
Therefore, the better performance of the Rev+(Sim)I compared to the Rev+(S)I for bug reports and the opposite results in the feature requests can be rationalized.
 
\emph{\textbf{Answer to RQ1: }Using the training dataset augmented through the three proposed methods can improve F1-score, precision, and recall for review classifiers. The Within-App and Within-Context methods provide better results than the Between-App method.}

\subsection{RQ2: Influences of Dataset Augmentation on Increasing the Generalizability of Model}
In RQ2, we investigate the impact of augmenting primary datasets on increasing the generalizability of the trained classifiers. Devine et al.~\cite{Devin} demonstrated that models trained on seven labeled datasets do not perform well in predicting unseen ones. Therefore, this experiment demonstrates that augmenting review datasets using auxiliary datasets extracted from labeled issues can contribute to the enhancement of performance in trained classifiers. To this end, we use five primary datasets (\tabref{Dataset_Spec}) for training baseline and augmented models.
Devine et al.~\cite{Devin} introduced two types of models, namely \emph{``Single-Dataset''} and \emph{``Leave-One-Out (LOO)''}, for comparison purposes. Their study revealed that the LOO models yield superior results. In this experiment, we also train these two model types using the primary and augmented datasets and compare their results:

\begin{enumerate}
    \item \emph{Single-Dataset}: one dataset (A) as the training dataset, the rest of datasets (PD-{A}) as the test datasets (PD denotes the set of primary datasets).
    \item \emph{Leave-one-out (LOO)}: one dataset (A) as the test dataset, the union of the rest of datasets (union(PD-{A})) as the training dataset.
\end{enumerate}

Additionally, due to the absence of the associated app name for some user reviews within certain primary datasets, it becomes unfeasible to identify the repository of the same app or similar apps. As a result, the only viable augmentation approach for this experiment is the between-app analysis.

On the other hand, in this research question, we analyze how the augmentation of primary datasets increases the generalizability of review classifiers. Although Within-App and Within-Context methods provide better performance for review classifiers, we use the Between-App method in this experiment for two reasons. First, some primary datasets lack available app names, rendering the application of the other two methods unfeasible. Second, we have resource limitations for applying another two methods. For instance, the primary dataset PD1 contains user reviews related to 595 apps, with an average of five reviews for each app. So, we must train 1190 classification models (two binary classifiers for each app) only for this dataset in Within-App and Within-Context methods. Since our classifiers are transformer-based models, we have resource limitations for training them. So, we apply the Between-App method, in which the auxiliary datasets are selected randomly to augment primary datasets.  

\subsubsection{Results}
Tables \ref{table:RQ2-Bug} and \ref{table:RQ2-Feature} present Single-Dataset and LOO classifiers trained using primary and augmented datasets. The top values for each dataset are bolded in tables. Also, underlined numbers show the maximum value of each column. In the bug report, F1-scores increase between 4.1 to 26.0 and recall between 2.2 to 21.2. In the feature request, F1-scores increase between 6.8 to 11.6 and recall between 0.5 to 15.3. As can be seen, the average F1-score and recall of models using the augmented dataset for predicting bug report/feature request reviews improve in all Single-Datasets except two cases in the feature request: recall score of PD3 and F1-score of PD4. In the first one, although the recall is reduced, precision and F1-score have increased. For the latter, the augmentation of the PD4 dataset does not improve the F1-score for predicting feature requests. It could be due to the relatively short length of reviews in this dataset. The average length of feature request reviews in PD4 with 6.5 terms, which is lower than the average length of 12.4 to 25.4 terms in other primary datasets. So, the baseline model trained on the PD4 dataset might face challenges when predicting longer reviews.
Therefore, augmenting this dataset using the auxiliary dataset with an average length of 15.9 makes the model more biased towards issues to predict longer reviews. The average length of FN reviews (feature request reviews that the classifier does not predict) increases from 19 in the baseline model to 21.1 in the augmented model. Also, the average length of TP reviews (feature request reviews that the classifier correctly predicts) decreases from 18.4 in the baseline model to 17.6 in the augmented model.

\begin{table*}
\centering
\caption{Results of Single-Dataset and LOO Models for Bug Report Classification Using Primary and Augmented Datasets in RQ2 (PD=Primary Dataset, AD=Auxiliary Dataset).}
    \renewcommand{\arraystretch}{1.3}
    \setlength{\tabcolsep}{2pt}
    \makebox[\textwidth][c]{
    \begin{tabular}{|c|ccc|ccc|ccc|ccc|ccc|ccc|}
        \hline
            \multicolumn{19}{|c|}{\textbf{Bug Report}} \\
        \hline
            \multirow{2}{*}{Datasets} & 
            \multicolumn{3}{|c|}{PD1} & \multicolumn{3}{|c|}{PD2} & \multicolumn{3}{|c|}{PD3} &
            \multicolumn{3}{|c|}{PD4} &
            \multicolumn{3}{|c|}{PD5} &
            \multicolumn{3}{|c|}{Average} \\\cline{2-19}
            & P & R & F1 & P & R & F1 & P & R & F1 & P & R & F1 & P & R & F1 & P & R & F1 \\
        \thickhline
            PD1 & \cellcolor{lightgray} & \cellcolor{lightgray} & \cellcolor{lightgray} & \underline{\textbf{42.2}} & 55.8 & 47.4 & \textbf{91.0} & 53.0 & 66.1 & \underline{\textbf{60.6}} & 46.6 & 50.7 & \textbf{93.4} & 71.7 & 80.5 
            & \textbf{71.8} & 56.8 & 61.2\\
        \hline
            PD1 + 0.3\textbar\emph{PD1}\textbar AD & \cellcolor{lightgray} & \cellcolor{lightgray} & \cellcolor{lightgray} & 35.2 & \textbf{63.9} & \textbf{44.9} &
            85.3 & \underline{\textbf{73.5}} & \underline{\textbf{78.4}} & 
            41.5 & \textbf{77.1} & \textbf{52.3} &
            89.9 & \textbf{82.3} & \textbf{85.7} & 
            62.9 & \textbf{74.2} & \textbf{65.3}\\
        \thickhline
            PD2 & 60.6 & 19.2 & 25.3 & \cellcolor{lightgray} & \cellcolor{lightgray} & \cellcolor{lightgray} &
            64.3 & 9.9 & 15.9 &
            24.0 & 1.1 & 2.1 & 
            62.4 & 14.7 & 21.4 & 
            52.8 & 11.2 & 16.2\\
        \hline
            PD2 + 0.3\textbar\emph{PD2}\textbar AD & \textbf{87.0} & \textbf{45.5} & \textbf{57.6} & \cellcolor{lightgray} & \cellcolor{lightgray} & \cellcolor{lightgray} & \textbf{86.4} & \textbf{29.4} & \textbf{41.0} &
            \textbf{51.1} & \textbf{22.4} & \textbf{24.8} &
            \textbf{89.6} & \textbf{32.3} & \textbf{45.2} & 
            \textbf{78.5} & \textbf{32.4} & \textbf{42.2}
            \\
        \thickhline
            PD3 & \textbf{87.2} & 76.9 & 81.3 &
            \textbf{40.5} & 64.0 & \textbf{49.0} & \cellcolor{lightgray} & \cellcolor{lightgray} & \cellcolor{lightgray} & 
            \textbf{59.3} & 57.0 & \textbf{56.0} & 
            \textbf{92.6} & 75.7 & 83.0 & 
            \textbf{69.9} & 68.4 & 67.3\\
        \hline
            PD3 + 0.3\textbar\emph{PD3}\textbar AD & 80.0 & \underline{\textbf{89.4}} & \underline{\textbf{84.3}} &
            32.9 & \underline{\textbf{73.6}} & 45.4 & \cellcolor{lightgray} & \cellcolor{lightgray} & \cellcolor{lightgray} & 
            40.8 & \underline{\textbf{82.4}} & 54.3 & 
            85.6 & \underline{\textbf{88.9}} & \underline{\textbf{87.1}} & 
            59.8 & \underline{\textbf{83.6}} & \underline{\textbf{67.8}}\\
        \thickhline
            PD4 & \underline{\textbf{92.4}} & 61.4 & 73.2 &
            \textbf{39.1} & 43.2 & \textbf{40.7} &
            \underline{\textbf{94.7}} & 44.5 & 60.0 &
            \cellcolor{lightgray} & \cellcolor{lightgray} & \cellcolor{lightgray} & 
            \underline{\textbf{95.1}} & 70.4 & 80.5 & 
            \underline{\textbf{80.3}} & 54.9 & 63.6\\
        \hline
            PD4 + 0.3\textbar\emph{PD4}\textbar AD & 87.4 & \textbf{70.9} & \textbf{77.4} &
            33.2 & \textbf{49.5} & 39.1 &
            88.4 & \textbf{60.7} & \textbf{71.2} &
            \cellcolor{lightgray} & \cellcolor{lightgray} & \cellcolor{lightgray} & 
            88.9 & \textbf{77.9} & \textbf{82.5}& 
            74.5 & \textbf{64.8} & \textbf{67.6}\\
        \thickhline
            PD5 & 84.5 & \textbf{77.5} & 79.6 &
            \textbf{38.1} & \textbf{61.2} & \textbf{45.9} &
            87.9 & 55.1 & 65.8 &
            \textbf{59.2} & 52.1 & 51.2 &
            \cellcolor{lightgray} & \cellcolor{lightgray} & \cellcolor{lightgray} & 
            \textbf{67.4} & 61.5 & 60.7\\
        \hline
            PD5 + 0.3\textbar\emph{PD5}\textbar AD & \textbf{88.6} & 77.4 & \textbf{82.2} &
            36.9 & 56.6 & 44.2 &
            \textbf{89.0} & \textbf{59.7} & \textbf{70.9} &
            55.3 & \textbf{61.2} & \underline{\textbf{57.1}}&
            \cellcolor{lightgray} & \cellcolor{lightgray} & \cellcolor{lightgray} & 
            \textbf{67.4} & \textbf{63.7} & \textbf{63.6}\\
        \thickhline
            LOO & \textbf{87.1} & 75.8 & 80.9 &
            \textbf{40.6} & 67.0 & \underline{\textbf{50.3}} &
            \textbf{92.4} & 53.5 & 66.8 &
            \textbf{59.3} & 61.1 & \textbf{55.2} &
            \textbf{90.1} & 77.2 & \textbf{82.2} & 
            \textbf{73.9} & 66.9 & \textbf{67.1}\\
        \hline
            LOO + 0.3\textbar\emph{LOO}\textbar AD & 84.5 & \textbf{79.4} & \textbf{81.2} &
            37.1 & \textbf{68.6} & 48.0 &
            87.3 & \textbf{63.8} & \textbf{73.4} &
            41.5 & \textbf{80.5} & 53.0 & 
            84.1 & \textbf{78.9} & 79.0 & 
            66.9 & \textbf{74.2} & 66.9\\
        \hline
    \end{tabular}
    }
\label{table:RQ2-Bug}
\end{table*}


\begin{table*}
\centering
\caption{Results of Single-Dataset and LOO Models for Feature Request Classification Using Primary and Augmented Datasets in RQ2 (PD=Primary Dataset, AD=Auxiliary Dataset).}
    \setlength{\tabcolsep}{2pt}
    \renewcommand{\arraystretch}{1.3}
    \makebox[\textwidth][c]{
    \begin{tabular}{|c|ccc|ccc|ccc|ccc|ccc|ccc|}
        \hline
            \multicolumn{19}{|c|}{\textbf{Feature Request}} \\
        \hline
            \multirow{2}{*}{Data sets} & 
            \multicolumn{3}{|c|}{PD1} & \multicolumn{3}{|c|}{PD2} & \multicolumn{3}{|c|}{PD3} &
            \multicolumn{3}{|c|}{PD4} &
            \multicolumn{3}{|c|}{PD5} &
            \multicolumn{3}{|c|}{Average} \\\cline{2-19}
            & P & R & F1 & P & R & F1 & P & R & F1 & P & R & F1 & P & R & F1 & P & R & F1 \\
        \thickhline
            PD1 & \cellcolor{lightgray} & \cellcolor{lightgray} & \cellcolor{lightgray} & \underline{\textbf{39.4}} & 12.0 & 17.0 &
            \underline{\textbf{80.8}} & 27.0 & 37.0 & \underline{\textbf{76.0}} & 21.7 & 30.1 &
            \underline{\textbf{81.7}} & 44.1 & 54.1 & 
            \underline{\textbf{69.5}} & 26.2 & 34.6\\
        \hline
            PD1 + 0.3\textbar\emph{PD1}\textbar AD & \cellcolor{lightgray} & \cellcolor{lightgray} & \cellcolor{lightgray} & 27.9 & \textbf{16.2} & \textbf{19.9} &
            69.4 & \textbf{46.5} & \textbf{54.7} & 
            53.7 & \textbf{46.3} & \textbf{48.0} &
            72.2 & \textbf{57.0} & \underline{\textbf{62.3}} & 
            55.8 & \textbf{41.5} & \textbf{46.2}\\
        \thickhline
            PD2 & 20.7 & 18.4 & 17.3 & 
            \cellcolor{lightgray} & \cellcolor{lightgray} & \cellcolor{lightgray} &
            21.9 & 10.4 & 11.4 &
            12.8 & 4.1 & 4.2 & 
            22.0 & 22.0 & 19.8 & 
            19.4 & 13.7 & 13.2\\
        \hline
            PD2 + 0.3\textbar\emph{PD2}\textbar AD & 
            \textbf{37.5} & \textbf{22.9} & \textbf{20.5} & \cellcolor{lightgray} & \cellcolor{lightgray} & \cellcolor{lightgray} & 
            \textbf{48.1} & \textbf{21.9} & \textbf{21.7} &
            \textbf{43.3} & \textbf{21.3} & \textbf{18.3} &
            \textbf{39.1} & \textbf{23.5} & \textbf{21.3} & 
            \textbf{42.0} & \textbf{22.4} & \textbf{20.4}
            \\
        \thickhline
            PD3 & 
            33.5 & \underline{\textbf{81.1}} & 45.6 &
            14.5 & \textbf{40.2} & 19.8 & \cellcolor{lightgray} & \cellcolor{lightgray} & \cellcolor{lightgray} & 
            31.4 & \underline{\textbf{82.0}} & 41.0 & 
            37.0 & \underline{\textbf{81.8}} & 47.8 & 
            29.1 & \underline{\textbf{71.3}} & 38.5\\
        \hline
            PD3 + 0.3\textbar\emph{PD3}\textbar AD &
            \textbf{53.8} & 67.5 & \underline{\textbf{59.3}} &
            \textbf{24.2} & 26.8 & \textbf{24.8} & 
            \cellcolor{lightgray} & \cellcolor{lightgray} & \cellcolor{lightgray} & 
            \textbf{47.3} & 64.7 & \textbf{52.1} & 
            \textbf{58.8} & 61.0 & \textbf{58.9} &
            \textbf{46.0} & 55.0 & \textbf{48.8}\\
        \thickhline
            PD4 & \underline{\textbf{61.5}} & 56.5 & \textbf{57.8} &
            \textbf{29.3} & 19.4 & \textbf{22.8} &
            \underline{\textbf{80.8}} & 52.7 & \textbf{62.2} &
            \cellcolor{lightgray} & \cellcolor{lightgray} & \cellcolor{lightgray} & 
            \textbf{71.6} & 55.2 & \textbf{61.0} & 
            \textbf{60.8} & 46.0 & \underline{\textbf{50.9}}\\
        \hline
            PD4 + 0.3\textbar\emph{PD4}\textbar AD & 
            39.5 & \textbf{72.8} & 49.5 &
            15.4 & \textbf{28.6} & 18.7 &
            56.4 & \underline{\textbf{73.4}} & 62.1 &
            \cellcolor{lightgray} & \cellcolor{lightgray} & \cellcolor{lightgray} & 
            42.6 & \textbf{74.3} & 52.3 & 
            38.5 & \textbf{62.3} & 45.7\\
        \thickhline
            PD5 & 
            57.6 & 60.5 & 50.4 &
            25.2 & \textbf{23.7} & 19.3 &
            \textbf{65.5} & \textbf{51.1} & 47.4 &
            \textbf{55.7} & 48.2 & 37.2 &
            \cellcolor{lightgray} & \cellcolor{lightgray} & \cellcolor{lightgray} & 
            \textbf{51.0} & 45.9 & 38.5\\
        \hline
            PD5 + 0.3\textbar\emph{PD5}\textbar AD & 
            \textbf{58.7} & \textbf{63.6} & \textbf{58.2} &
            \textbf{27.1} & 21.7 & \textbf{22.4} &
            64.1 & 50.8 & \textbf{53.8} &
            50.9 & \textbf{49.8} & \textbf{46.6}&
            \cellcolor{lightgray} & \cellcolor{lightgray} & \cellcolor{lightgray} & 
            50.2 & \textbf{46.4} & \textbf{45.3}\\
        \thickhline
            LOO & 
            \textbf{50.0} & 69.9 & 52.5 &
            \textbf{31.6} & 28.9 & \underline{\textbf{26.8}} &
            \textbf{62.5} & 67.3 & 55.2 &
            \textbf{56.0} & 58.0 & 51.0 &
            \textbf{57.4} & 72.5 & \textbf{60.1} & 
            \textbf{51.5} & 59.3 & \textbf{49.1}\\
        \hline
            LOO + 0.3\textbar\emph{LOO}\textbar AD & 
            48.7 & \textbf{74.7} & \textbf{58.8} &
            15.8 & \underline{\textbf{40.4}} & 19.3 &
            63.6 & \textbf{67.7} & \underline{\textbf{64.0}} &
            51.9 & \textbf{60.0} & \underline{\textbf{54.5}} & 
            36 & \textbf{74.9} & 47.5 & 
            43 & \textbf{63.5} & 48.8\\
        \hline
    \end{tabular}
    }
\label{table:RQ2-Feature}
\end{table*}

Also, the baseline model trained on PD4 provides a maximum score of 50.9. The reason can be related to the volume of labeled dataset in PD4. We have 2,085 reviews with feature request labels, providing a richer training dataset than others whose count of feature request reviews is 178 to 799 (\tabref{Dataset_Spec}). 
The average recall in LOO models increases with data augmentation in both classes. However, the average F1-score has a slight reduction 0f 0.2 in the bug report and 0.3 in the feature request. Since primary datasets include reviews from various apps and auxiliary datasets selected randomly, this low reduction of F1-score in LOO is probable. As shown in RQ1, two other augmentation methods which cannot use them in this experiment usually perform better than the Between-App method. Nevertheless, the recall increase of 7.3 in the bug report and 4.4 in the feature request shows that augmenting the training dataset can provide more confidence in finding bug reports/feature requests for developers. So, if developers aim to find user feedback and miss fewer ones for app maintenance, this negligible reduction in F1-score can be ignored.

Devine et al.~\cite{Devin} investigate the performance of review classifiers in predicting new datasets that have not seen any review from them in the training dataset. They demonstrate that models trained on LOO (combined multiple datasets) outperformed Single-Dataset models. In this experiment, LOO models do not necessarily provide the best results but are close to maximum scores. In the bug report, models with the highest F1-scores in order related to augmented PD3 (67.8), augmented PD4 (67.6), baseline PD3 (67.3), and baseline LOO (67.1). In the feature request, models with the highest F1-score in order related to baseline PD4 (50.9), baseline LOO (49.1), and jointly augmented LOO and PD3 (48.8). So, models that augment with the Between-App method can outperform LOO models.   

Generally, the maximum value of average F1-score and recall for the bug report belongs to the augmented PD3. Also, the baseline model of PD3 achieves the best results compared to other baselines in these two measures. In the feature request, the maximum value of average F1-score and recall are related to LOO and PD3 datasets. So, models trained on the PD3 dataset have more generalizability for prediction of review intentions than other primary datasets, with the best F1-score results in the bug report and a slight difference from the best scores in the feature request. 
In addition, PD1 and PD4 datasets have the highest precision scores in the bug report and feature request models. Still, they identify almost half of the bug reports and less than half of the feature requests in user reviews. These datasets can be helpful for developers that need more precision in finding specific types of reviews.
On the other hand, the augmentation of the primary dataset increases the precision score in some models related to: the PD2 dataset in predicting all other primary datasets, the PD5 dataset in some cases, and the PD3 dataset only in the feature request. So, according to the richness of the selected issues for creating an auxiliary dataset, the precision of classifiers may increase, as mentioned in RQ1.

Using the PD2 dataset as the training or testing dataset provides different results and needs more analysis. As can be seen, bug report/feature request classifiers trained on this dataset have the lowest values for evaluation measures in predicting other primary datasets. On the other hand, models based on augmented datasets often do not outperform baselines in predicting the PD2 dataset. One reason for this low generalizability can be caused by the low volume of this dataset (\tabref{Dataset_Spec}). Hence, the augmentation of PD2 makes a significant improvement in the results. Another reason can be its labeling policy, which differs from other primary datasets. In this dataset, user complaints about app features were considered feature requests in some cases, while they were identified as bug reports in other datasets. For example, review \emph{``Not working properly. wont download emails to inbox.''} in PD2 has the feature request label. Nevertheless, similar reviews like \emph{``After lollipop's update my calendar doesn't work properly.''} in PD4 and \emph{``Have turned off as note it does not work properly.''} in PD3 have labeled bug reports.

\emph{\textbf{Answer to RQ2: } Overall, augmenting primary datasets through the Between-App method increases the generalizability of trained models and improves the results in F1-score and recall. The relatively shorter length of samples within the primary dataset can potentially lead to model bias towards the auxiliary dataset when making predictions for reviews with longer lengths. It can reduce the generalizability of the augmented model. The PD3 dataset is more generalizable in predicting other datsets.}

\subsection{RQ3: Influences of Volume of Auxiliary Dataset on Model Performance}
Our results show that the volume of the auxiliary dataset used in augmenting the primary dataset affects the model performance.
We aim to discuss the volume of auxiliary datasets as an important parameter for improving review classifiers in data augmentation. For this reason, we compare the results of models trained on primary datasets augmented by auxiliary datasets with different volumes.
Therefore, in RQ3, we focus on finding the effective range for the volume of the auxiliary dataset.
We use the LOO models that combine multiple primary datasets for training review classifiers. Similar to RQ2, we employed the Between-App method for data augmentation in this research question because of our limitations.

The auxiliary dataset consists of processed issues that differ slightly from user reviews in their expressions and do not include all types of them (e.g., lack of some cases in the other class). Therefore, we consider the volume of the auxiliary dataset smaller than the volume of the primary dataset to avoid biasing the trained model to it. 
We train bug report and feature request classification models on varying augmented datasets. In these datasets, the aforementioned ratio is adjusted incrementally in steps of 0.1 within the range of [0, 1]. Due to the constraints of resources and the time-intensive nature of training transformer-based models, this experiment exclusively involves the training of LOO models, encompassing all available datasets.

Therefore, we analyze obtained results and investigate how increasing the volume ratio improves model performance. We determine an effective range for the ratio of auxiliary dataset volume to primary dataset volume, which provides the best classification performance. Such a range could be efficiently applied to other datasets as well. 

\subsubsection{Results}
Tables \ref{table:RQ3_Bug} and \ref{table:RQ3_Feature} detail the results of ten models trained on datasets augmented by different volumes of auxiliary dataset for bug reports and feature requests, respectively. Bold numbers represent the maximum value of each column. On average, recall scores of all augmented models increase from 2.5 to 11 in the bug report and 2.2 to 9.5 in the feature request. We often observe a slight decrease of F1-score in augmented models, which we discussed in RQ2. This reduction can be significant in some volume ratios. For example, in the bug report, the volume ratio of 0.9 obtained an F1-score of 62.5, with a 4.5 difference compared to the baseline. 

\begin{table}
    \caption{Results of Bug Report Classification Using Augmented Datasets with Varied Volumes in RQ3 (AD=Auxiliary Dataset).}
    \centering
    \renewcommand{\arraystretch}{1.3}
    \setlength{\tabcolsep}{2pt}
    \makebox[\textwidth][c]{
    \begin{tabular}{|c|ccc|ccc|ccc|ccc|ccc|ccc|}
        \hline
            \multicolumn{19}{|c|}{Bug Report} \\
        \hline
            \multirow{2}{*}{Datasets} & 
            \multicolumn{3}{|c|}{PD1} & 
            \multicolumn{3}{|c|}{PD2} & 
            \multicolumn{3}{|c|}{PD3} &
            \multicolumn{3}{|c|}{PD4} &
            \multicolumn{3}{|c|}{PD5} &
            \multicolumn{3}{|c|}{Average} \\\cline{2-19}
            & P & R & F1 & P & R & F1 & P & R & F1 & P & R & F1 & P & R & F1 & P & R & F1\\
        \hline
            LOO & 87.1 & 75.8 & 80.9 & \textbf{40.6} & 67.0 & \textbf{50.3} & 92.4 & 53.5 & 66.8 & \textbf{59.3} & 61.1 & 55.2 & \textbf{90.1} & 77.2 & 82.2 & \textbf{73.9} & 66.9 & 67.1\\
        \hline
            LOO + 0.1\textbar\emph{LOO}\textbar AD  & 86.3 & 76.3 & 79.9 & 33.9 & 63.9 & 41.2 & \textbf{92.8} & 53.7 & 67.3 & 51.4 & 70.1 & \textbf{56.9} & 88.5 & 82.9 & 84.9 & 70.6 & 69.4 & 66.0\\
        \hline
        \rowcolor{lightapricot}
            LOO + 0.2\textbar\emph{LOO}\textbar AD & \textbf{88.5} & 78.1 & 82.3 & 36.5 & 66.9 & 45.6 & 79.0 & 75.0 & 75.0 & 52.7 & 68.5 & 54.8 & 84.8 & 83.4 & 82.7 & 68.3 & 74.4 & 68.1\\
            
        \hline
        \rowcolor{lightapricot}
            LOO + 0.3\textbar\emph{LOO}\textbar AD & 84.5 & 79.4 & 81.2 & 37.1 & 68.6 & 48.0 & 87.3 & 63.8 & 73.4 & 41.5 & 80.5 & 53.0 & 84.1 & 78.9 & 70.9 & 66.9 & 74.2 & 66.9\\
        \hline
        \rowcolor{lightapricot}
            LOO + 0.4\textbar\emph{LOO}\textbar AD & 84.9 & 79.6 & 81.5 & 38.9 & 61.0 & 47.1 & 82.6 & \textbf{75.1} & \textbf{78.2} & 43.0 & 78.3 & 53.4 & 81.6 & 84.0 & 81.2 & 66.2 & 75.6 & \textbf{68.3}\\
        \hline
        \rowcolor{lightapricot}
            LOO + 0.5\textbar\emph{LOO}\textbar AD & 87.8 & 78.3 & 82.4 & 38.9 & 69.5 & 47.7 & 88.2 & 56.5 & 67.6 & 39.8 & \textbf{82.4} & 52.5 & 86.8 & 79.0 & 81.3 & 68.3 & 73.1 & 66.3\\
        \hline
        \rowcolor{lightapricot}
           LOO + 0.6\textbar\emph{LOO}\textbar AD & \textbf{88.5} & 73.5 & 79.7 & 38.3 & 60.9 & 46.7 & 87.2 & 63.6 & 72.1 & 41.4 & 77.0 & 53.0 & 85.8 & 84.6 & 84.4 & 68.2 & 71.9 & 67.2\\
        \hline
        \rowcolor{lightapricot}
           LOO + 0.7\textbar\emph{LOO}\textbar AD & 80.7 & \textbf{87.9} & \textbf{84.1} & 35.1 & 70.6 & 46.6 & 84.5 & 69.5 & 74.4 & 40.2 & 78.2 & 52.5 & 83.1 & 83.5 & 81.5 & 64.7 & \textbf{77.9} & 67.8\\
        \hline
           LOO + 0.8\textbar\emph{LOO}\textbar AD & 85.1 & 75.6 & 79.0 & 29.4 & \textbf{73.0} & 41.4 & 87.3 & 62.4 & 71.9 & 55.0 & 55.5 & 44.8 & 84.7 & 86.5 & \textbf{85.2} & 68.3 & 70.6 & 64.5\\
        \hline
           LOO + 0.9\textbar\emph{LOO}\textbar AD & 77.2 & 77.9 & 74.0 & 28.5 & 72.4 & 39.3 & 89.7 & 54.6 & 66.9 & 37.4 & 77.9 & 49.9 & 79.9 & \textbf{87.8} & 82.7 & 62.5 & 74.1 & 62.6\\
        \hline
           LOO + 1.0\textbar\emph{LOO}\textbar AD & 84.7 & 83.0 & 83.8 & 34.6 & 64.9 & 44.2 & 86.5 & 68.8 & 76.6 & 37.2 & 81.8 & 48.5 & 89 & 72.9 & 78.8 & 66.4 & 74.3 & 66.4\\
        \hline
    \end{tabular}
    }
\label{table:RQ3_Bug}

\vspace*{1 cm}

    \caption{Results of Feature Request Classification Using Augmented Datasets with Varied Volumes in RQ3 (AD=Auxiliary Dataset).}
    \centering
    \renewcommand{\arraystretch}{1.3}
    \setlength{\tabcolsep}{2pt}
    \makebox[\textwidth][c]{
    \begin{tabular}{|c|ccc|ccc|ccc|ccc|ccc|ccc|}
        \hline
            \multicolumn{19}{|c|}{Feature Request} \\
        \hline
            \multirow{2}{*}{Datasets} & 
            \multicolumn{3}{|c|}{PD1} & 
            \multicolumn{3}{|c|}{PD2} & 
            \multicolumn{3}{|c|}{PD3} &
            \multicolumn{3}{|c|}{PD4} &
            \multicolumn{3}{|c|}{PD5} &
            \multicolumn{3}{|c|}{Average} \\\cline{2-19}
            & P & R & F1 & P & R & F1 & P & R & F1 & P & R & F1 & P & R & F1 & P & R & F1\\
        \hline
            LOO & 50.0 & 69.9 & 52.5 & \textbf{31.6} & 28.9 & \textbf{26.8} & 52.5 & 67.3 & 55.2 & \textbf{56.0} & 58.0 & 51.0 & 57.4 & 52.5 & \textbf{60.1} & \textbf{51.5} & 59.3 & \textbf{49.1}\\
        \hline
            LOO + 0.1\textbar\emph{LOO}\textbar AD & 48.9 & 70.4 & 50.8 & 17.7 & \textbf{46.4} & 22.9 & 60.6 & 72.8 & 63.3 & 45.0 & 68.8 & 48.0 & \textbf{59.7} & 63.1 & 52.9 & 46.4 & 64.3 & 47.6\\
        \hline
            LOO + 0.2\textbar\emph{LOO}\textbar AD & 32.0 & \textbf{81.9} & 44.0 & 14.5 & 40.2 & 21.7 & 52.7 & 76.4 & 61.1 & 38.0 & 72.7 & 45.0 & 43.8 & 72.9 & 48.8 & 36.2 & \textbf{68.8} & 44.1\\
        \hline
        \rowcolor{lightapricot}
            LOO + 0.3\textbar\emph{LOO}\textbar AD &
            48.7 & 74.7 & \textbf{58.8} &
            15.8 & 40.4 & 19.3 &
            63.6 & 67.7 & 64.0 &
            51.9 & 60.0 & \textbf{54.5} & 
            36.0 & 74.9 & 47.5 & 
            43.0 & 63.5 & 48.8\\
        \hline
        \rowcolor{lightapricot}
            LOO + 0.4\textbar\emph{LOO}\textbar AD & 46.4 & 73.0 & 54.7 & 14.5 & 37.7 & 19.5 & \textbf{66.7} & 67.0 & \textbf{64.3} & 42.4 & 67.5 & 51.5 & 28.5 & \textbf{84.8} & 39.9 & 39.7 & 66.0 & 46.0\\
        \hline
        \rowcolor{lightapricot}
           LOO + 0.5\textbar\emph{LOO}\textbar AD & 43.7 & 70.7 & 52.0 & 12.9 & 39.0 & 18.6 & 62.8 & 69.0 & 63.7 & 47.0 & 58.9 & 51.5 & 40.1 & 76.4 & 49.7 & 41.3 & 62.8 & 47.1\\
        \hline
        \rowcolor{lightapricot}
           LOO + 0.6\textbar\emph{LOO}\textbar AD & 45.2 & 69.2 & 45.5 & 16.4 & 27.4 & 18.5 & 55.7 & 73.6 & 61.9 & 37.7 & 69.2 & 44.9 & 45.4 & 74.2 & 53.3 & 40.1 & 62.7 & 44.8\\
        \hline
        \rowcolor{lightapricot}
           LOO + 0.7\textbar\emph{LOO}\textbar AD & 46.2 & 75.2 & 55.8 & 15.5 & 32.9 & 19.2 & 53.4 & 74.3 & 59.2 & 39.8 & 63.4 & 47.5 & 53.5 & 70.5 & 57.7 & 41.7 & 63.3 & 47.9\\
        \hline
        \rowcolor{lightapricot}      
           LOO + 0.8\textbar\emph{LOO}\textbar AD & \textbf{51.8} & 70.4 & 55.7 & 23.3 & 40.4 & 25.5 & 51.1 & \textbf{76.2} & 58.2 & 36.6 & 68.4 & 45.4 & 49.4 & 70.1 & 54.5 & 42.4 & 65.1 & 47.9\\
        \hline
           LOO + 0.9\textbar\emph{LOO}\textbar AD & 47.4 & 68.6 & 53.8 & 22.3 & 27.5 & 22.2 & 55.8 & 68.4 & 56.9 & 35.5 & \textbf{73.6} & 44.4 & 47.9 & 74.3 & 55.0 & 41.4 & 62.5 & 46.5\\
        \hline
           LOO + 1.0\textbar\emph{LOO}\textbar AD & 47.8 & 70.6 & 56.6 & 17.8 & 34.9 & 21.5 & 58.9 & 66.0 & 56.2 & 36.9 & 66.7 & 45.8 & 43.4 & 69.2 & 52.2 & 41.0 & 61.5 & 46.5\\
        \hline
    \end{tabular}
    }
\label{table:RQ3_Feature}

\end{table}

\begin{figure*}[!t]
\centering
  \includegraphics[width=4in,height=2.7in]{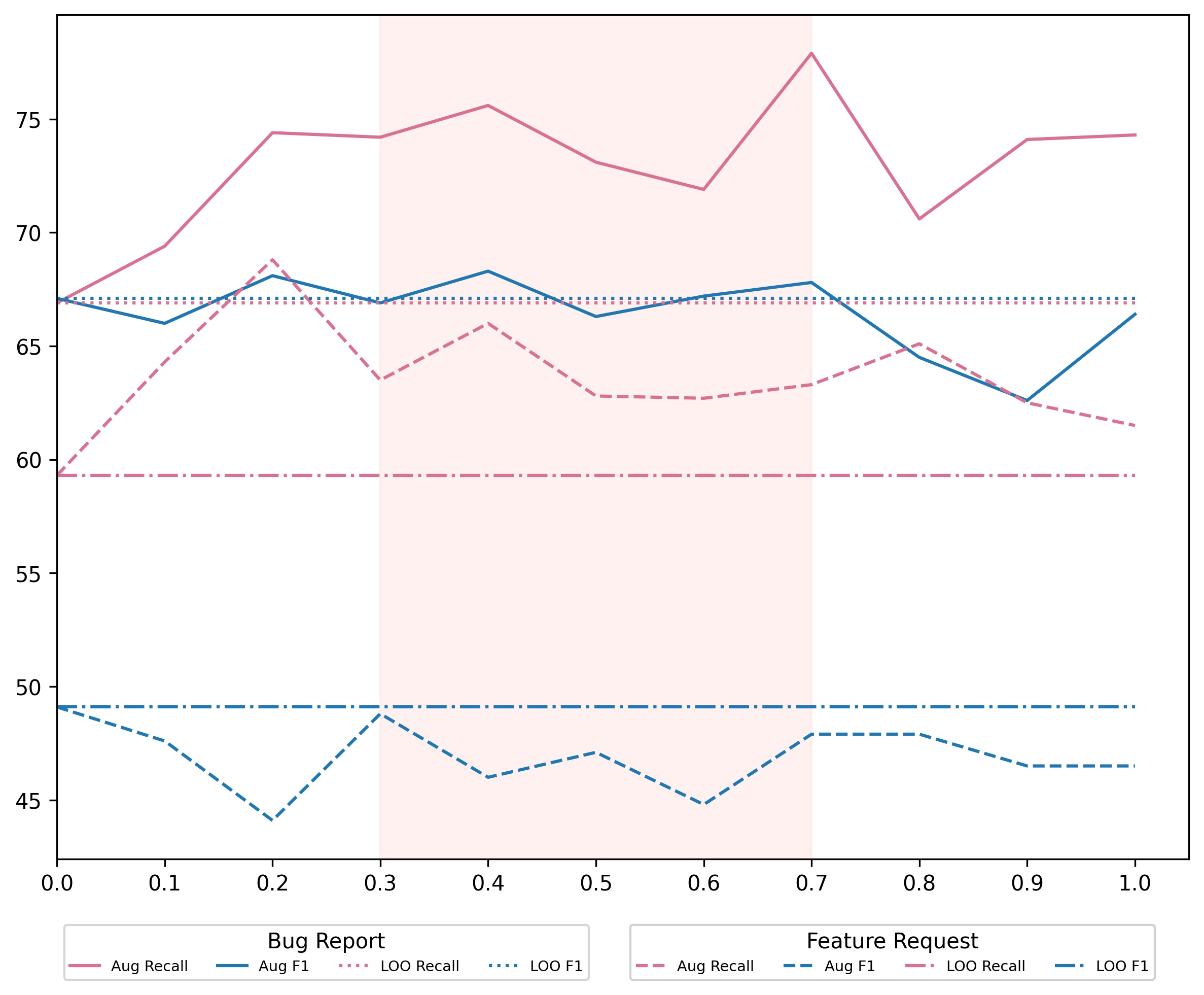}
\caption{The Average F1-score and Recall for the Bug Report/Feature Request Classifiers Trained on Datasets Augmented with Various Volume Ratios.}
\label{fig:trend}  
\end{figure*}

Due to the variety of Android apps and their constituent features, if the volume of the auxiliary dataset is low, it cannot provide enough information to increase the generalizability of the review classifier. We use transformer-based models in our experiments. So, millions of parameters need to be updated to fine-tune classification models. For this reason, low volume can have a negative impact. For example, the volume ratio of 0.1 in the bug report has a 1.1 decrease in the average F1-score and a minimal increase in average recall (2.5) over the other models with augmented dataset. Also, the volume ratio of 0.2 in the feature request has the lowest value of average F1-score (44.1) with a difference of 5 compared to the baseline (49.1). A higher volume for the auxiliary dataset can increase the model's generalizability (e.g., volume ratios 0.2 in the bug report and 0.3 in the feature request). However, this result may differ for top volume ratios that further reduce average F1-score. For example, in the bug report, volume ratios of 0.9 and 0.8 have the most reductions of 4.5 and 2.6 for average F1-score, respectively. Also, in the feature request, volume ratios of 0.9 and 1.0 have one of the most reductions of 2.6 for average F1-score and the slightest increases of 2.2 and 3.2 in the average recall, respectively. The reason is likely that the classification model is biased toward the auxiliary dataset derived from the issues in the high volume. Therefore, we can see these reductions in average F1-score due to the differences between issues and reviews. 

Figure \ref{fig:trend} shows the changing trends of average F1-score and recall. In this figure, we only show our most important measures, recall and F1-score for a more understandable display and better result analysis. We can see that the data augmentation hurts the classifier results in some volume ratio range and cannot help the performance improvement 
In the bug report, models with a volume ratio of 0.2 to 0.7 provide the best results for F1-score and recall. Therefore, this interval can be considered the \emph{Effective Range}, a range that data augmentation caused a positive effect. In this interval, the model performance fluctuates for F1-score between 66.3 to 68.3 with average changes of +0.3 and recall between 71.9 to 77.9 with average changes of +7.0 compared to the baseline. Also, there is a considerable reduction in F1-score for volume ratios 0.1, 0.8 and 0.9. In the feature request, the effective range is 0.3 to 0.8, which has a better overall trend in F1-score and recall. In this interval, the model performance fluctuates for F1-score between 44.8 to 48.8 with an average change of +4.6 and recall between 62.7 to 66 with an average change of -2 compared to the baseline. The volume ratio of 0.6, which is in the effective range of the feature request, has the second smallest value for F1-score, 44.8, with a minimum F1-score of 44.1. However, we consider the overall trend of the range 0.3 to 0.8 to provide a positive effect and ignore this case. Since we select auxiliary datasets randomly in this experiment, they will likely not contain the best samples. In addition, the F1-score reduction in volume ratios of 0.9 and 1.0 is 2.6, which is not very significant compared to others, but the recall measures are the smallest scores among augmented models. Therefore, we consider them outside the effective range of the feature request. 
The F1-score of all augmented models in the feature request are lower than the baseline, which could be due to the augmentation method used. As explained in RQ1, because of the app diversity, we face more challenges for finding feature request reviews, especially in the Between-App method, which the auxiliary dataset is randomly selected.

According to the observed results, we consider the intersection of effective ranges in the bug report and the feature request, 0.3 to 0.7, as the overall effective range for choosing the volume of the auxiliary dataset. It can help to augment primary datasets appropriately. Therefore, we chose volume ratios from the effective range to achieve better results in other experiments. Since many models need to be trained to answer our RQs, we choose a volume ratio of 0.3 to reduce the training and prediction time as much as possible. Generally, models using augmented datasets in the effective range often contain the best results except for predicting the PD2 for the feature request. 
It is worth mentioning two tips: (i) volume ratios within the effective range may not include the best results (such as the ratio of 0.4 in predicting PD5 for the feature request), (ii) volume ratios outside the effective range may also provide good results (such as the ratio of 0.8 in the bug report for PD5 prediction). We determine the effective range based on average values for reflecting overall behaviour that can be applied to other datasets.

\emph{\textbf{Answer to RQ3: } Augmenting the primary dataset using very low or high volumes of the auxiliary dataset can lead to a reduction in model performance. There is an effective range of 0.3 to 0.7 for volume ratio that often causes better results.}

\subsection{RQ4: Influences of Dataset Augmentation on Transformer-Based Models}
In this research question, our goal is to explore the impact of dataset augmentation on the performance of classifiers trained using various transformer-based algorithms. 
We discuss the results of review classifiers based on four models BERT~\cite{bert_model}, RoBERTa~\cite{roberta_model}, ALBERT~\cite{albert_model}, and DistilBERT~\cite{distilbert_model}. RoBERTa~\cite{roberta} stands as \emph{``Robustly-Optimized BERT''} that performs better than BERT~\cite{bert} by applying modifications in training and masking strategies. Also, ALBERT~\cite{albert} and DistilBERT~\cite{distilbert} use techniques to provide faster and lighter models without significant performance degradation~\cite{Izadi-2}. Since transformer-based models outperformed traditional algorithms for our task in previous studies~\cite{Hadi, Henao}, we do not consider them in this experiment.
The results of this experiment will facilitate the selection of the appropriate algorithm for training binary review classifiers.

\subsubsection{Results}
We report the results of trained review classifiers based on four algorithms, BERT, DistilBERT, RoBERTa, and ALBERT, using LOO and augmented LOO datasets in \tabref{RQ4}. The numbers represent the average of five LOO models, where each time one of the primary datasets is used as the test dataset, like the ``Average'' column in \tabref{RQ3_Bug} and \tabref{RQ3_Feature}. Bolded and underlined numbers indicate the maximum values for each algorithm and each column, respectively. As detailed in RQ2, the Between-App method is also employed in this experiment.
The results obtained in various algorithms provide a consistent finding: augmenting the training dataset with processed issues can improves F1-score and recall. On average, using the augmented dataset increases these measures in all models except for a few cases: the F1-score in the DistilBERT models for the bug report and the feature request prediction, which have a slight decrease of 0.2 and 0.3, respectively, the recall score in the RoBERTa model for the feature request prediction, which has increased in precision and F1-scores, and the F1-score in the BERT model for the feature request prediction, which has reduced 2.4. As can be seen, the DistilBERT algorithm has the best performance in predicting the reviews related to target classes, with an average F1-score of 58.1 among the baseline models and 57.1 among augmented models. Also, the trained models based on the BERT algorithm have closer results to it.

\begin{table*}
    \footnotesize
    \centering
    \renewcommand{\arraystretch}{1.3}
    \caption{Results of Review Classification Using Primary (LOO) and Augmented Datasets with Different Transformer-Based Models in RQ4 (*minimum time)}
    \setlength{\tabcolsep}{2pt}
    \makebox[\textwidth][c]{
    \begin{tabular}{|c|c|ccc|cc||ccc|cc||ccc|}
        \hline
            \multirow{2}{*}{Algorithms} &
            \multirow{2}{*}{Datasets} &
            \multicolumn{5}{|c||}{Bug Report} &
            \multicolumn{5}{|c||}{Feature Request} & 
            \multicolumn{3}{|c|}{Average}
            \\\cline{3-15}
            & & P & R & F1 & Training(sec) & Prediction(sec) & P & R & F1 & Training(sec) & Prediction(sec)& P & R & F1 
            \\
        \thickhline
            \multirow{2}{*}{BERT} & LOO & 
            \textbf{72.2} & 67.1 & 66.4 & 629.3 & 20.7 &
            \underline{\textbf{60.7}} & 45.7 & \textbf{46.1} & 620.8 & 20.1 &
            \underline{\textbf{66.4}} & 56.4 & \textbf{56.2} 
            \\
        \cline{2-15}
            & LOO + 0.3\textbar\emph{LOO}\textbar AD & 
            62.9 & \underline{\textbf{76.5}} & \textbf{66.8} & 771.6 & 20.5 &
            41.7 & \underline{\textbf{63.9}} & 43.7 & 801.5 & 20.5 & 
            52.3 & \underline{\textbf{70.2}} & 55.2\\

        \thickhline
            \multirow{2}{*}{RoBERTa} & LOO & 
            42.2 & 39.4 & 37.3 & 639.3 & 20.1 &
            23.9 & \textbf{27.7} & 24.0 & 629.4 & 19.9 & 33.0 & 33.5 & 30.6 \\
        \cline{2-15}
            & LOO + 0.3\textbar\emph{LOO}\textbar AD & 
            \textbf{47.2} & \textbf{50.9} & \textbf{44.4} & 795.4 & 19.8 &
            \textbf{32.1} & 23.5 & \textbf{24.7} & 819.7 & 20.2 & 
            \textbf{39.6} & \textbf{37.2} & \textbf{34.5}\\
            
        \thickhline
            \multirow{2}{*}{DistilBERT} & LOO & 
            \underline{\textbf{73.9}} & 66.9 & \textbf{67.1} & 327.6* & 11.1* &
            \textbf{51.5} & 59.3 & \underline{\textbf{49.1}} & 323.2* & 10.8* & \textbf{62.7} & 63.1 & \underline{\textbf{58.1}} \\
        \cline{2-15}
            & LOO + 0.3\textbar\emph{LOO}\textbar AD & 
            66.9 & \textbf{74.2} & 66.9 & 430.4* & 11.2* &
            43.0 & \textbf{63.5} & 48.8 & 432.5* & 11.6* & 
            54.9 & \textbf{68.8} & 57.8 \\
            
        \thickhline
            \multirow{2}{*}{ALBERT} & LOO & 
            68.0 & 64.0 & 63.1 & 539.5 & 19.7 &
            56.9 & 36.4 & 39.0 & 553.0 & 20.5 & 
            \textbf{62.4} & 50.2 & 51.0 \\
        \cline{2-15}
            & LOO + 0.3\textbar\emph{LOO}\textbar AD & 
            \textbf{68.9} & \textbf{71.6} & \underline{\textbf{68.1}} & 722.7 &
            20.2 &
            \textbf{43.5} & \textbf{54.4} & \textbf{45.2} & 725.6 & 20.2 & 
            56.2 & \textbf{63.0} & \textbf{56.6} \\
        \hline
    \end{tabular}
    }
\label{table:RQ4}
\end{table*}

In addition, all maximum values of each column for the bug report/feature request correspond to classifiers fine-tuned DistilBERT and then BERT model, except for the maximum F1-score in the bug report, which belongs to the classifier trained on ALBERT using an augmented dataset. ALBERT classifiers have close results to DistilBERT classifiers in the bug report. However, they have more differences, especially for the recall score in the feature request. 
On the other hand, RoBERTa models have the lowest results for all evaluation measures in the bug report/feature request. Since RoBERTa is a larger model with more parameters than BERT, a larger training dataset is needed to fine-tune it, while our LOO dataset contains 10,000 to 23,000 reviews. Therefore, RoBERTa is not a suitable model for training review classifiers. Considering that the volume of the training dataset for fine-tuning the pre-trained models in our task is small, choosing a model with fewer parameters is better. On the other hand, we observe that DistilBERT models have the lowest training and prediction time by a significant difference with others. For this reason, we used DistilBERT to train classifiers in other experiments.

\emph{\textbf{Answer to RQ4:} Data augmentation in various transformer-based models provides consistent results and often improves F1-score and recall. DistilBERT models provide the best results among baseline and augmented models and require the minimum time for training and prediction. Then, BERT and ALBERT models have better performance, respectively. RoBERTa is a large model unsuitable for our purpose as we have some limitations in preparing a large training dataset.}

\subsection{Discussion}

In this paper, our goal is to extract valuable information from the labeled issues and use them to enrich the training dataset. Our experiments show that using the processed issues as an auxiliary dataset result in an improved review classifier, notably in terms of F1-score and recall metrics. User reviews, especially those related to the bug report, can help developers in maintenance decisions~\cite{Gu}. Therefore, losing a smaller percentage of user reviews related to app maintenance enhances the reliability of the review classifier.
Also, since reviews may change over time~\cite{Dabrowski}, using processed issues as part of the training dataset makes it easier and faster to update it.

In addition, the appropriate selection of an auxiliary dataset plays a significant role in improving the accuracy of review classifiers. The auxiliary dataset should be carefully chosen to contain rich information and be well-representative of the various user requirements related to the target applications to enhance the results. Since we use labeled issues to prepare the auxiliary dataset, we face limitations such as different expressions and the absence of certain types of reviews, which can lead to a slight decrease in precision or recall in specific cases, as detailed in RQ1. In our proposed approach, we tried to manage these problems as much as possible through issue processing and filtering. Additionally, since the ratio of auxiliary dataset volume to primary dataset volume can affect the accuracy of the results, it is important to consider the recommended range of 0.3 to 0.7. On the other hand, different labeling policies of manually labeled datasets can also affect the expected results. Therefore, selecting appropriate datasets for training classification models can help provide outputs that align with developers' goals.

Generally, our proposed approach can be employed using the Within-App, Within-Context, and Between-App Analysis methods. The first two methods are applicable for training an App-Specific classifier, while the last method facilitates training a General-Purpose classifier.

\subsubsection{Training App-Specific Classifiers}
Considering the diversity of available apps, training classifiers specific to each app can provide more useful results because they can learn more accurately app related features. However, creating app-specific classifiers is challenging because of difficulties in preparing labeled datasets. 
To our knowledge, previous studies often  do  not  focus on it.
Our proposed approach makes it possible to use GitHub issues for this purpose. If the developers and users  are actively engaged within the GitHub repository of an app, issues can effectively represent the requirements of the app's users. Therefore, the app repository can can serve as the best source for augmenting a primary dataset. However, it is typically necessary to incorporate a larger volume of issues to create an effective auxiliary dataset. The volume of issues significantly impacts the quality of results obtained. Augmenting with either a very small or large volume of data not only fails to be beneficial but can also negatively impact prediction results. So, we identify the similar app(s) that contain sufficient information on the labeled issues related to the target classes to benefit from them for overcoming the lack of information.

\subsubsection{Training General-Purpose Classifiers}
Data augmentation can also help train general-purpose classifiers. The results of our experiments in RQ2 show that using our approach can increase the generalizability of available manually labeled datasets. Some datasets are limited to a small number of apps. For example, PD1 only includes user reviews of seven apps. The labeled processed issues in our dataset are selected from 999 apps, offering substantial diversity. Therefore, the data augmentation becomes the training dataset more general and not limited to specific domains or apps. As explained in RQ2, models based on LOO and PD3 datasets are among the best results. So, using PD3 or a combination of available datasets provide a better dataset to train the general-purpose classifiers.

\section{Related Work}
\label{sec:rel_work}
In this section, we discuss previous research focusing on two categories of studies that are based on app store reviews and other sources of data.
\subsection{Approaches Based on App Store Reviews}
Some previous research analyzes user reviews based on Natural Language Processing (NLP) and Information Retrieval (IR) techniques.
As one of the first research in user review analysis, Iacob et al.~\cite{Iacob} propose the MARA tool. They conduct manual analysis and define 237 linguistic rules for extracting feature requests. They rank identified requests by their length and frequency.
Guzman et al.~\cite{Guzman_2} automatically analyze user reviews and identify app features at two levels. They filter out the extracted collocations to discover low-level features and compute sentiment scores for them. These features are then organized into groups using topic modeling techniques.
Gao et al.~\cite{Gao} propose an approach that automatically extracts and prioritizes the main topics derived from user reviews. In addition to the information provided by users, they also use the relevance of the topics to rank them. Also, they visualize and analyze topic dynamics over time.
While these unsupervised approaches aid in extracting valuable insights from user reviews, it's important to note that supervised algorithms can provide better performance in identifying reviews related to predefined classes. 

Many studies used traditional learning algorithms to train user review classifiers.
Chen et al.~\cite{Chen} present AR-Miner, an approach to identify informative user reviews using the EMNB algorithm and group them by applying the topic modeling. They provide a scheme for ranking generated groups and their instances.
Villarroel et al.~\cite{Villarroel}~\cite{Scalabrino} present the CLAP tool that helps developers plan the next release. They classify user reviews in the bug report, suggestion for new feature, and other classes using their constituent words, N-grams, and ratings. Also, they cluster the reviews related to the two first classes, which are of developer interest. Then, clusters were prioritized to recommend the most important ones by defining several independent features.
Maalej et al.~\cite{Maalej_1}\cite{Maalej_Journal} classify reviews into four classes bug report, feature request, user experience, and rating. They employ this approach based on constituent words, collocations, general metadata, and sentiment scores derived from user reviews.
Panichella et al.~\cite{Panichella} classify review sentences using a combination of three features: constituent terms, grammatical patterns, and sentiment scores. They analyze developer discussions about maintenance tasks and adapt them with user review categories provided by Pagano et al.~\cite{Pagano}. As a result, they select helpful categories from the developer's perspective relevant to the app evolution.
Guzman et al.~\cite{Guzman} identify a taxonomy of seven categories (bug report, feature strength, feature shortcoming, user request, praise, complaint, and usage scenario) that provide helpful information for the app evolution. They represent the vectorized form of a user review by the tf-idf weighted vector and several numerical features (such as the word and character count). Also, they use four traditional algorithms and their ensembles to classify user reviews based on their taxonomy.
Jha et al.~\cite{Jha} present an approach to classify and summarize helpful reviews. In this approach, reviews represent by semantic frames that are automatically extracted and more efficient than linguistic patterns obtained by manual analysis.
Gu et al.~\cite{Gu} aim to summarize user reviews. First, they classify reviews into five categories, aspect evaluation, bug report, feature request, praise, and other, and use two types of textual features, lexical and structural. Also, they summarize the sentences of the aspect evaluation category by identifying and clustering the aspects and their related opinions and sentiments. This approach employs defined patterns to extract app aspects.

Recent studies show that deep learning algorithms can outperform traditional algorithms in review classification. Aslam et al.~\cite{Aslam} use a CNN-based network that considers the textual, non-textual and sentiment features extracted from app reviews as classifier inputs. Also, Stanik et al.~\cite{Stanik} present an approach to classify user reviews and tweets in two languages, English and Italian. In addition to traditional models, they use a CNN-based network for classifier training. Due to the relatively small volume of labeled datasets, transfer learning can be helpful in review classification task~\cite{Henao}. Henao et al.~\cite{Henao} argue that BERT-based models outperform models based on traditional machine learning algorithms. Also, Araujo et al.~\cite{Araujo} compare the various techniques available for review representation and demonstrate that the pre-trained transformer-based models achieve better results in review analysis tasks.
Hadi et al.~\cite{Hadi} conduct a more comprehensive study comparing pre-trained models and previous approaches for the review classification task in different settings. They focused their research on six labeled datasets from prior works and integrated them and four popular approaches. In addition, they pre-train new models specific to this domain using collected user reviews and show that using them results in better performance. So, we also use pre-trained models in our approach.
In most previous works, predictive models have been trained and tested on one dataset. Devine et al.~\cite{Devin} discuss how a review classifier trained on one dataset performs well in predicting the intention of reviews related to other apps or datasets. This study uses seven datasets labeled in previous studies and shows that the results of model evaluations on unseen datasets cannot perform as well as the model evaluations on the same dataset. Also, they argue that merging labeled datasets can help overcome this generalizability problem. Our paper provides another solution by augmenting the labeled review dataset with processed issues that can result in better performance than the merged dataset.

\subsection{Approaches Based on Other Sources of data}
Different sources, such as issue tracker systems, developer emails, and social platforms, contain valuable information for development and maintenance tasks~\cite{Devin}. Twitter is one of the sources that has been the focus of researchers. Guzman et al.~\cite{Guzman_3} manually analyze tweets related to software applications (including desktop, mobile, or both) and identify 26 categories for their content. Also, they train traditional classifiers to predict the relevance of these categories with the various types of stakeholders.
Nayebi et al.~\cite{Nayebi} classify tweets and user reviews into three categories: bug reports, feature requests, and others. Then, they extract the topics of each category and analyze their similarity in reviews and tweets. They found that analyzing tweets can provide developers with worthy information besides information extracted from reviews. 
Sorbo et al.~\cite{Sorbo} classify the information provided in the developer emails into six categories with manual analysis. They define linguistic patterns to identify the corresponding category.
Shi et al.~\cite{Shi} present an approach to identify feature requests in developer emails. They categorize sentences using 81 fuzzy rules and find feature requests using sequential pattern mining. In these approaches, training and testing datasets are selected from the same source.

Some studies use available data from one source to help analyze and extract information from another source. Huang et al.~\cite{Huang} examine whether the taxonomy introduced by Sorbo et al.~\cite{Sorbo}, provided by analyzing the email contents, can be useful for categorizing GitHub Comments. They customize it and make it suitable for the analysis of GitHub comments. Also, they use a CNN-based neural network to categorize them automatically. Their results show that using classifiers trained on a dataset related to one source (developer email/issues) to predict a dataset related to another source (issues/developer email) does not provide better results than the classifiers that use data from the same source for training and prediction. Therefore, since data from different sources usually have a different nature, we consider the size of the auxiliary datasets to be smaller than the primary datasets in our approach to avoid the excessive biasing model to them.
Stanik~\cite{Stanik}, Henao~\cite{Henao}, and Devine~\cite{Devin} also use Twitter datasets in their studies. However, only the last one examines merging them with review datasets to improve the performance of trained models. 

Al-Safoury et al.~\cite{Al-Safoury} aggregate the information provided in user reviews and GitHub issues. They extract bug reports and feature requests from them and merge similar ones to create a complete change feature list by computing textual similarity. In our proposed approach, we process issues to adapt with app store reviews. Then, we use them to augment the labeled review dataset. As far as we know, GitHub issues have not been used in previous approaches to improve the performance of the review classification.
Zhang et al.~\cite{Zhang} utilized the similarity between user reviews and GitHub issues to categorize unlabeled issues. They applied the SURF tool~\cite{Sorbo_1} to identify app reviews related to problem discovery and feature requests and incorporated their similarity with unlabeled issues in the final similarity measure. In this study, our objective differs from their research, as we utilize labeled issues to assist in categorizing user reviews.

Additionally, some studies augment the text of reviews to improve the accuracy of review prediction. Wang et al.~\cite{Wang} emphasize that using other sources can help prepare accurate classifiers. They extract feature words related to each category from app changelogs and use them to augment reviews. Liu et al.~\cite{Liu} also utilize app descriptions for this purpose. Our proposed approach differs from this research because we aim to augment the labeled dataset and not add additional information to the text of reviews.

\section{Threats to Validity}
\label{sec:threat}

In this section, we examine internal, external, and construct threats to the validity of our proposed approach and the decisions made to address them.

\textbf{\emph{Internal Validity}} One of the threats that could impact the accuracy of the classifiers is the inaccurate or insufficient information provided by GitHub repositories. We identify issues related to target review intentions, which include bug reports, feature requests, and other categories, using the assigned labels. Wrong samples could be included in the auxiliary datasets if these  labels are erroneously assigned due to human mistakes. It can reduce the accuracy of the models trained on augmented datasets. For this reason, we try to select repositories that enhances our confidence level by applying two conditions: 1) having a minimum of two contributors capable of reviewing labels, and 2) ensuring the presence of more than 30 labeled issues, which indicates a certain degree of experience in labeling and managing issues. 
Additionally, we should adapt issue labels with target review intentions. First, we syntactically and semantically preprocess the labels and unify similar ones. Then, an author and two students manually analyze them to find related issue labels to the bug report, feature request, and other intentions. Therefore, another threat is the misidentification of related labels. To avoid this, we provide some required information (such as samples of original formats, definitions, and illustrative examples for each label) which helps them to identify the relevant intention correctly. Also, analysts perform investigations independently to reduce subjective biases. To decide on disagreement labels that do not have a clear definition, they review more samples during a meeting and determine the final class. In addition, due to the difficulty of this process and to avoid the boredom of analysts, which could potentially reduces the accuracy, we consider more commonly occurring labels (used in a minimum of 11 issues) in the analysis process. These frequently employed labels are highly likely to align with our intended purpose.

Issues contain various detailed information that is not suitable for our purpose. Therefore, we define 19 language patterns to extract target information. Other threats can be the completeness and correctness of our defined patterns. To ensure the former, we tried to thoroughly analyze all issues in addition to the templates for identifying the target information because most repositories do not have defined templates. We manually analyze the most frequent sections of the issues repeated at least four times to increase accuracy. We identify sections related to our goals and define linguistic patterns based on them. Also, we meet some conditions in our analysis to ensure the correctness of the information extracted by our defined patterns. First, we categorize the templates into five groups and analyze only four groups that suit our purposes to remove unrelated cases. We identify the main sections that contain an overall textual description of an issue. Second, we randomly select 384 GitHub issues, with a confidence level of 95\% and a confidence interval of 5\%. These selected issues were then examined for the accuracy of the identified target information, involving the collaboration of two programmers. The results revealed that in 82.8\% of instances, the target information was accurately extracted.
Third, we manually check all extracted sections by applying our patterns to all issues.

The completeness and correctness of the information extracted from associated data of GitHub repositories, utilized in creating their profiles, are other threats we face. For completeness, we examine the available data in GitHub repositories to describe their primary objectives or features. We use ReadMe files to create profiles and identify sections containing this information type. However, the constituent sections within ReadMe files can exhibit greater diversity compared to issues, but the first section often provides relevant information. Since some of the ReadMe files have limited content, the ``About'' section of repositories can also be intriguing. For this reason, in addition to identifying the most frequent target sections, we incorporate ``About'' sections in preparing repository profiles. To verify the correctness of the extracted information, we examine the ReadMe files of 200 repositories which constitute 20\% of our repository count. 

To prepare the truth dataset, we performed a human analysis that individual mistakes and biases of labelers can be threats to our study. We followed the standard process outlined by Gazman et al.~\cite{Guzman} to address these concerns. Therefore, we prepare a guideline that includes class definitions, examples, and potential ambiguities to establish consistent decision-making rules. Also, we tried to avoid subjective bias through (i) independent labeling and (ii) convening decision meetings to address contested labels, with the labelers' involvement. We reached a Cohen's kappa of 79.5, which indicates good agreement among the labelers. Another potential concern involves errors within our implementation code. We reviewed the code multiple times and shared it with other researchers to ensure reproducibility.

\textbf{\emph{External Validity}}
One potential concern is that our findings might not apply to all types of reviews. Our research relies on the three most common review intentions found in the available labeled review datasets, which have corresponding issue labels. However, for other review categories, considerably fewer related issues are available. As a result, our chosen intentions are better suited for conducting a thorough investigation into the feasibility of enhancing review datasets. Also, we adapt the labels of the selected review datasets to each other to create consistent primary datasets.

To ensure the generalizability of results in our proposed approach, we collect a relatively complete dataset of Android issues and labeled reviews to prepare the auxiliary and primary datasets, respectively. To achieve this, we leverage AndroZooOpen, the largest available collection of Android apps on GitHub,  to identify appropriate repositories for our experimental purposes. After filtering repositories and issues in different phases, we obtained a dataset of 62.7K processed issues, which includes 32.6K bug reports, 26.6K feature requests, and 3.9K others. Also, we selected five datasets from the available labeled review datasets that best align with our purposes. We create a truth dataset most compatible with the required conditions for applying various augmented methods. In addition, The selected reviews encompass all three intention classes, facilitating a comprehensive analysis of the results.
We compared the performance of the baseline and augmented classifiers in predicting the truth dataset to assess the generalizability of the proposed approach. In addition, we investigate the feasibility of predicting each primary dataset using models trained on other datasets. Also, we maintained consistent hyperparameters across all experiments and refrained from adjusting them for individual datasets to prevent overfitting the results to specific data.

\textbf{\emph{Construct Validity}}
We employed common evaluation measures, precision, recall, and F1-score, used in previous studies, to compare the baseline and augmented review classifiers. One concern could be the selection of baseline models for our experiments. Previous research has demonstrated that pre-trained models outperform other approaches based on traditional machine learning algorithms~\cite{Hadi}. Consequently, we did not consider the latter as baselines and instead concentrated on state-of-the-art models.

\section{Conclusions and Future Work}
\label{sec:concusion}
Users convey their requirements to developers by sharing reviews. As a result, analyzing user reviews can provide an understanding of the encountered problems and their expectations. Many studies have been conducted to classify user reviews automatically. Researchers prepare truth datasets through manual labeling, which are used to train and evaluate ML-based classifiers. Due to the challenges of the labeling process, these datasets are often small in size and restricted to specific apps or domains. Hence, they may not effectively represent all user reviews on the app stores. Devin et al. demonstrated that models trained on available labeled datasets exhibit diminished generalization in predicting new cases than expected.
In addition to app stores, specific platforms like GitHub's issue-tracking system can incorporate user feedback. GitHub issues usually consist of similar content to user reviews, accompanied by assigned labels that clarify their intent. Although developers often write them with more technical details, we can identify and extract relevant information for review classification tasks. Therefore, we aim to augment the labeled dataset by combining the processed labeled issues with manually labeled reviews to improve app review classifiers. For this reason, we proposed an approach comprising five phases: collecting the required data, processing labeled issues to prepare an auxiliary dataset, adapting five labeled review datasets to prepare the primary datasets, dataset augmentation through three methods of Within-App, Within-Context, and Between-App Analysis, and classifying unseen reviews using state-of-the-art pre-trained models.

Then, we designed several experiments to investigate the effectiveness of the proposed approach in improving the performance of review classifiers. First, we examined the impact of each augmentation method on training app-specific and general-purpose models. We prepare a truth dataset of 1000 user reviews from five apps, involving four developers in its labeling process. These reviews were augmented using three proposed methods to compare their accuracy with the baselines.
Our results showed that augmenting the review datasets using various methods can increase the F1-score between 2.8 to 6.3 for the bug report and 5.0 to 7.2 for the feature request. We demonstrated that augmenting manually labeled datasets increases the generalizability of publicly available datasets in predicting new cases. Additionally, we found that the effective range of auxiliary dataset volume is a critical parameter in the process of review augmentation.
After analyzing the results across various ranges, we identified the effective range of 0.3 and 0.7. Finally, we compared results from different transformer-based pre-trained models, revealing that the DistilBERT model provides the best results. In general, our findings provide valuable insights for training more accurate review classifiers by leveraging labeled GitHub issues while minimizing the loss of information. 

We intend to extend our approach to encompass additional sources, such as Twitter, in our future endeavors. It involves employing GitHub issues to augment their labeled datasets and subsequently analyzing the outcomes achieved. We have plans to delve into the generalizability of the augmented reviews, assessing their predictive capabilities on datasets originating from other sources. Also, manually labeled datasets do not use uniform guidelines. Their labeling policies may lead to different decisions in some cases. We can investigate how the diversity of the primary datasets and differences in labeling policies can impact the results of their augmentation. Moreover, we extract useful information from issues using language patterns identified through manual inspection. This extracted data holds the potential to be used in the development of an automated model capable of pinpointing the valuable sections within issues. Prompt-based models, such as GPT-4, have recently provided good results in various domains. However, they do not necessarily yield optimal performance in more specialized areas. Therefore, a more extensive analysis is required to fully evaluate this model, which can be addressed in future research.

\bibliographystyle{ACM-Reference-Format}
\bibliography{Main}


\begin{thebibliography}{76}


\ifx \showCODEN    \undefined \def \showCODEN     #1{\unskip}     \fi
\ifx \showDOI      \undefined \def \showDOI       #1{#1}\fi
\ifx \showISBNx    \undefined \def \showISBNx     #1{\unskip}     \fi
\ifx \showISBNxiii \undefined \def \showISBNxiii  #1{\unskip}     \fi
\ifx \showISSN     \undefined \def \showISSN      #1{\unskip}     \fi
\ifx \showLCCN     \undefined \def \showLCCN      #1{\unskip}     \fi
\ifx \shownote     \undefined \def \shownote      #1{#1}          \fi
\ifx \showarticletitle \undefined \def \showarticletitle #1{#1}   \fi
\ifx \showURL      \undefined \def \showURL       {\relax}        \fi
\providecommand\bibfield[2]{#2}
\providecommand\bibinfo[2]{#2}
\providecommand\natexlab[1]{#1}
\providecommand\showeprint[2][]{arXiv:#2}

\bibitem[AdA(2023)]%
        {AdAway}
 \bibinfo{year}{2023}\natexlab{}.
\newblock \bibinfo{title}{AdAway}.
\newblock
\newblock
\urldef\tempurl%
\url{https://github.com/AdAway/AdAway}
\showURL{%
Retrieved May 2023 from \tempurl}


\bibitem[alb(2023)]%
        {albert_model}
 \bibinfo{year}{2023}\natexlab{}.
\newblock \bibinfo{title}{albert-base-v2}.
\newblock
\newblock
\urldef\tempurl%
\url{https://huggingface.co/albert/albert-base-v2}
\showURL{%
Retrieved Jan 2023 from \tempurl}


\bibitem[Ate(2023a)]%
        {AtennaPod_Git}
 \bibinfo{year}{2023}\natexlab{a}.
\newblock \bibinfo{title}{AtennaPod}.
\newblock
\newblock
\urldef\tempurl%
\url{https://github.com/AntennaPod/AntennaPod}
\showURL{%
Retrieved May 2023 from \tempurl}


\bibitem[Ate(2023b)]%
        {AtennaPod}
 \bibinfo{year}{2023}\natexlab{b}.
\newblock \bibinfo{title}{AtennaPod}.
\newblock
\newblock
\urldef\tempurl%
\url{https://play.google.com/store/apps/details?id=de.danoeh.antennapod}
\showURL{%
Retrieved May 2023 from \tempurl}


\bibitem[ber(2023)]%
        {bert_model}
 \bibinfo{year}{2023}\natexlab{}.
\newblock \bibinfo{title}{bert-base-uncased}.
\newblock
\newblock
\urldef\tempurl%
\url{https://huggingface.co/google-bert/bert-base-uncased}
\showURL{%
Retrieved Jan 2023 from \tempurl}


\bibitem[dis(2023)]%
        {distilbert_model}
 \bibinfo{year}{2023}\natexlab{}.
\newblock \bibinfo{title}{distilbert-base-uncased}.
\newblock
\newblock
\urldef\tempurl%
\url{https://huggingface.co/distilbert/distilbert-base-uncased}
\showURL{%
Retrieved Jan 2023 from \tempurl}


\bibitem[Fir(2023a)]%
        {Firefox_Focus_Git}
 \bibinfo{year}{2023}\natexlab{a}.
\newblock \bibinfo{title}{Firefox Focus for Android}.
\newblock
\newblock
\urldef\tempurl%
\url{https://github.com/mozilla-mobile/focus-android}
\showURL{%
Retrieved Jan 2023 from \tempurl}


\bibitem[Fir(2023b)]%
        {Firefox_Focus}
 \bibinfo{year}{2023}\natexlab{b}.
\newblock \bibinfo{title}{Firefox Focus: No Fuss Browser}.
\newblock
\newblock
\urldef\tempurl%
\url{https://play.google.com/store/apps/details?id=org.mozilla.focus}
\showURL{%
Retrieved Jan 2023 from \tempurl}


\bibitem[Fir(2023c)]%
        {Firefox_Night_Git}
 \bibinfo{year}{2023}\natexlab{c}.
\newblock \bibinfo{title}{Firefox for Android}.
\newblock
\newblock
\urldef\tempurl%
\url{https://github.com/mozilla-mobile/fenix}
\showURL{%
Retrieved Jan 2023 from \tempurl}


\bibitem[Fir(2023d)]%
        {Firefox_Night}
 \bibinfo{year}{2023}\natexlab{d}.
\newblock \bibinfo{title}{Firefox Nightly for Developers}.
\newblock
\newblock
\urldef\tempurl%
\url{https://play.google.com/store/apps/details?id=org.mozilla.fenix}
\showURL{%
Retrieved Jan 2023 from \tempurl}


\bibitem[RES(2023)]%
        {RESTAPI}
 \bibinfo{year}{2023}\natexlab{}.
\newblock \bibinfo{title}{GitHub REST API documentation}.
\newblock
\newblock
\urldef\tempurl%
\url{https://docs.github.com/en/rest?apiVersion=2022-11-28}
\showURL{%
Retrieved Jan 2023 from \tempurl}


\bibitem[Met(2023a)]%
        {MetaMask_Git}
 \bibinfo{year}{2023}\natexlab{a}.
\newblock \bibinfo{title}{MetaMask}.
\newblock
\newblock
\urldef\tempurl%
\url{https://github.com/MetaMask/metamask-mobile}
\showURL{%
Retrieved Jan 2023 from \tempurl}


\bibitem[Met(2023b)]%
        {MetaMask}
 \bibinfo{year}{2023}\natexlab{b}.
\newblock \bibinfo{title}{MetaMask-Blockchain Wallet}.
\newblock
\newblock
\urldef\tempurl%
\url{https://play.google.com/store/apps/details?id=io.metamask}
\showURL{%
Retrieved Jan 2023 from \tempurl}


\bibitem[own(2023a)]%
        {ownCloud_Git}
 \bibinfo{year}{2023}\natexlab{a}.
\newblock \bibinfo{title}{Nextcloud Android app}.
\newblock
\newblock
\urldef\tempurl%
\url{https://github.com/nextcloud/android}
\showURL{%
Retrieved Jan 2023 from \tempurl}


\bibitem[NLT(2023)]%
        {NLTK}
 \bibinfo{year}{2023}\natexlab{}.
\newblock \bibinfo{title}{NLTK}.
\newblock
\newblock
\urldef\tempurl%
\url{https://www.nltk.org}
\showURL{%
Retrieved Jan 2023 from \tempurl}


\bibitem[own(2023b)]%
        {ownCloud}
 \bibinfo{year}{2023}\natexlab{b}.
\newblock \bibinfo{title}{ownCloud}.
\newblock
\newblock
\urldef\tempurl%
\url{https://play.google.com/store/apps/details?id=com.owncloud.android}
\showURL{%
Retrieved Jan 2023 from \tempurl}


\bibitem[rob(2023)]%
        {roberta_model}
 \bibinfo{year}{2023}\natexlab{}.
\newblock \bibinfo{title}{roberta-base}.
\newblock
\newblock
\urldef\tempurl%
\url{https://huggingface.co/FacebookAI/roberta-base}
\showURL{%
Retrieved Jan 2023 from \tempurl}


\bibitem[Str(2023)]%
        {StratifiedKFold}
 \bibinfo{year}{2023}\natexlab{}.
\newblock \bibinfo{title}{StratifiedKFold}.
\newblock
\newblock
\urldef\tempurl%
\url{https://scikit-learn.org/stable/modules/generated/sklearn.model_selection.StratifiedKFold.html}
\showURL{%
Retrieved Jan 2023 from \tempurl}


\bibitem[TFA(2023)]%
        {TFAutoModel}
 \bibinfo{year}{2023}\natexlab{}.
\newblock \bibinfo{title}{TFAutoModelforSequenceClassification}.
\newblock
\newblock
\urldef\tempurl%
\url{https://huggingface.co/transformers/v3.0.2/model_doc/auto.html#tfautomodelforsequenceclassification}
\showURL{%
Retrieved Jan 2023 from \tempurl}


\bibitem[Voc(2023)]%
        {VocableTrainer}
 \bibinfo{year}{2023}\natexlab{}.
\newblock \bibinfo{title}{VocableTrainer-Android}.
\newblock
\newblock
\urldef\tempurl%
\url{https://github.com/0xpr03/VocableTrainer-Android}
\showURL{%
Retrieved May 2023 from \tempurl}


\bibitem[Abedini et~al\mbox{.}(2024)]%
        {DATAR}
\bibfield{author}{\bibinfo{person}{Yasaman Abedini},
  \bibinfo{person}{Mohammad~Hadi Hajihosseini}, {and} \bibinfo{person}{Abbas
  Heydarnoori}.} \bibinfo{year}{2024}\natexlab{}.
\newblock \showarticletitle{{DATAR}: A Dataset for Tracking App Releases}. In
  \bibinfo{booktitle}{\emph{Proceedings of the 21st IEEE/ACM International
  Conference on Mining Software Repositories}}.
\newblock


\bibitem[Al-Safoury et~al\mbox{.}(2022)]%
        {Al-Safoury}
\bibfield{author}{\bibinfo{person}{Laila Al-Safoury}, \bibinfo{person}{Akram
  Salah}, {and} \bibinfo{person}{Soha Makady}.}
  \bibinfo{year}{2022}\natexlab{}.
\newblock \showarticletitle{Integrating user reviews and issue reports of
  mobile apps for change requests detection}.
\newblock \bibinfo{journal}{\emph{International Journal of Advanced Computer
  Science and Applications}} \bibinfo{volume}{13}, \bibinfo{number}{12}
  (\bibinfo{year}{2022}).
\newblock


\bibitem[Araujo et~al\mbox{.}(2022)]%
        {Araujo}
\bibfield{author}{\bibinfo{person}{Adailton~F. Araujo},
  \bibinfo{person}{Marcos~P.S. Golo}, {and} \bibinfo{person}{Ricardo~M.
  Marcacini}.} \bibinfo{year}{2022}\natexlab{}.
\newblock \showarticletitle{Opinion mining for app reviews: an analysis of
  textual representation and predictive models}.
\newblock \bibinfo{journal}{\emph{Automated Software Engineering}}
  \bibinfo{volume}{29}, \bibinfo{number}{1} (\bibinfo{year}{2022}),
  \bibinfo{pages}{1--30}.
\newblock


\bibitem[Aslam et~al\mbox{.}(2020)]%
        {Aslam}
\bibfield{author}{\bibinfo{person}{Naila Aslam}, \bibinfo{person}{Waheed~Yousuf
  Ramay}, \bibinfo{person}{X~I~A Kewen}, {and} \bibinfo{person}{Nadeem
  Sarwar}.} \bibinfo{year}{2020}\natexlab{}.
\newblock \showarticletitle{Convolutional neural network based classification
  of app reviews}.
\newblock \bibinfo{journal}{\emph{IEEE Access}}  \bibinfo{volume}{8}
  (\bibinfo{year}{2020}), \bibinfo{pages}{185619--185628}.
\newblock


\bibitem[Assi et~al\mbox{.}(2021)]%
        {Assi}
\bibfield{author}{\bibinfo{person}{Maram Assi}, \bibinfo{person}{Safwat
  Hassan}, \bibinfo{person}{Yuan Tian}, {and} \bibinfo{person}{Ying Zou}.}
  \bibinfo{year}{2021}\natexlab{}.
\newblock \showarticletitle{FeatCompare: Feature comparison for competing
  mobile apps leveraging user reviews}.
\newblock \bibinfo{journal}{\emph{Empirical Software Engineering}}
  \bibinfo{volume}{26}, \bibinfo{number}{5} (\bibinfo{year}{2021}),
  \bibinfo{pages}{1--38}.
\newblock


\bibitem[Borges et~al\mbox{.}(2016)]%
        {Borges}
\bibfield{author}{\bibinfo{person}{Hudson Borges}, \bibinfo{person}{Andre
  Hora}, {and} \bibinfo{person}{Marco~Tulio Valente}.}
  \bibinfo{year}{2016}\natexlab{}.
\newblock \showarticletitle{Predicting the popularity of github repositories}.
  In \bibinfo{booktitle}{\emph{Proceedings of the 12th ACM SIGSOFT
  International Conference on Predictive Models and Data Analytics in Software
  Engineering}}. \bibinfo{pages}{1--10}.
\newblock


\bibitem[Borges et~al\mbox{.}(2020)]%
        {Li}
\bibfield{author}{\bibinfo{person}{Hudson Borges}, \bibinfo{person}{Andre
  Hora}, {and} \bibinfo{person}{Marco~Tulio Valente}.}
  \bibinfo{year}{2020}\natexlab{}.
\newblock \showarticletitle{Automated bug reproduction from user reviews for
  Android applications}. In \bibinfo{booktitle}{\emph{Proceedings of the 42nd
  IEEE/ACM International Conference on Software Engineering: Software
  Engineering in Practice}}. \bibinfo{pages}{51--60}.
\newblock


\bibitem[Chen et~al\mbox{.}(2014)]%
        {Chen}
\bibfield{author}{\bibinfo{person}{Ning Chen}, \bibinfo{person}{Jialiu Lin},
  \bibinfo{person}{Steven C~H Hoi}, \bibinfo{person}{Xiaokui Xiao}, {and}
  \bibinfo{person}{Boshen Zhang}.} \bibinfo{year}{2014}\natexlab{}.
\newblock \showarticletitle{AR-miner: mining informative reviews for developers
  from mobile app marketplace}. In \bibinfo{booktitle}{\emph{Proceedings of the
  36th IEEE/ACM International Conference on Software Engineering}}.
  \bibinfo{pages}{767--778}.
\newblock


\bibitem[Ciurumelea et~al\mbox{.}(2017)]%
        {Ciurumelea}
\bibfield{author}{\bibinfo{person}{Adelina Ciurumelea},
  \bibinfo{person}{Andreas Schaufelbuhl}, \bibinfo{person}{Sebastiano
  Panichella}, {and} \bibinfo{person}{Harald~C. Gall}.}
  \bibinfo{year}{2017}\natexlab{}.
\newblock \showarticletitle{Analyzing reviews and code of mobile apps for
  better release planning}. In \bibinfo{booktitle}{\emph{Proceedings of the
  24th IEEE International Conference on Software Analysis, Evolution and
  Reengineering}}. \bibinfo{pages}{91--102}.
\newblock


\bibitem[Cohen(1960)]%
        {cohen}
\bibfield{author}{\bibinfo{person}{Jacob Cohen}.}
  \bibinfo{year}{1960}\natexlab{}.
\newblock \showarticletitle{A coefficient of agreement for nominal scales.}
\newblock \bibinfo{journal}{\emph{Educational and psychological measurement}}
  \bibinfo{volume}{20}, \bibinfo{number}{1} (\bibinfo{year}{1960}).
\newblock


\bibitem[Deocadez et~al\mbox{.}(2017)]%
        {Deocadez}
\bibfield{author}{\bibinfo{person}{Roger Deocadez}, \bibinfo{person}{Rachel
  Harrison}, {and} \bibinfo{person}{Daniel Rodriguez}.}
  \bibinfo{year}{2017}\natexlab{}.
\newblock \showarticletitle{Preliminary study on applying Semi-Supervised
  Learning to app store analysis}. In \bibinfo{booktitle}{\emph{Proceedings of
  the 21st International Conference on Evaluation and Assessment in Software
  Engineering}}. \bibinfo{pages}{320--323}.
\newblock


\bibitem[Devine et~al\mbox{.}(2023)]%
        {Devin}
\bibfield{author}{\bibinfo{person}{Peter Devine}, \bibinfo{person}{Yun~Sing
  Koh}, {and} \bibinfo{person}{Kelly Blincoe}.}
  \bibinfo{year}{2023}\natexlab{}.
\newblock \showarticletitle{Evaluating software user feedback classifier
  performance on unseen apps, datasets, and metadata}.
\newblock \bibinfo{journal}{\emph{Empirical Software Engineering}}
  \bibinfo{volume}{28}, \bibinfo{number}{1} (\bibinfo{year}{2023}).
\newblock


\bibitem[Devlin et~al\mbox{.}(2019)]%
        {bert}
\bibfield{author}{\bibinfo{person}{Jacob Devlin}, \bibinfo{person}{Ming-Wei
  Chang}, \bibinfo{person}{Kenton Lee}, {and} \bibinfo{person}{Kristina
  Toutanova}.} \bibinfo{year}{2019}\natexlab{}.
\newblock \showarticletitle{BERT: pre-training of deep bidirectional
  transformers for language understanding}. In
  \bibinfo{booktitle}{\emph{Proceedings of the Annual Conference of the North
  American Chapter of the Association for Computational Linguistics: Human
  Language Technologies}}. \bibinfo{pages}{4171--4186}.
\newblock


\bibitem[Dhinakaran et~al\mbox{.}(2018)]%
        {Dhinakaran}
\bibfield{author}{\bibinfo{person}{Venkatesh~T Dhinakaran},
  \bibinfo{person}{Raseshwari Pulle}, \bibinfo{person}{Nirav Ajmeri}, {and}
  \bibinfo{person}{Pradeep~K Murukannaiah}.} \bibinfo{year}{2018}\natexlab{}.
\newblock \showarticletitle{App review analysis via active learning}. In
  \bibinfo{booktitle}{\emph{Proceedings of the 26th IEEE International
  Requirements Engineering Conference}}. \bibinfo{pages}{170--181}.
\newblock


\bibitem[Di~Sorbo et~al\mbox{.}(2015)]%
        {Sorbo}
\bibfield{author}{\bibinfo{person}{Andrea Di~Sorbo},
  \bibinfo{person}{Sebastiano Panichella}, \bibinfo{person}{Corrado~A.
  Visaggio}, \bibinfo{person}{Massimiliano Di~Penta}, \bibinfo{person}{Gerardo
  Canfora}, {and} \bibinfo{person}{Harald~C. Gall}.}
  \bibinfo{year}{2015}\natexlab{}.
\newblock \showarticletitle{Development emails content analyzer: Intention
  mining in developer discussions}. In \bibinfo{booktitle}{\emph{Proceedings of
  the 30th IEEE/ACM International Conference on Automated Software
  Engineering}}. \bibinfo{pages}{12--23}.
\newblock


\bibitem[Dąbrowski et~al\mbox{.}(2022)]%
        {Dabrowski}
\bibfield{author}{\bibinfo{person}{Jacek Dąbrowski}, \bibinfo{person}{Emmanuel
  Letier}, \bibinfo{person}{Anna Perini}, {and} \bibinfo{person}{Angelo Susi}.}
  \bibinfo{year}{2022}\natexlab{}.
\newblock \showarticletitle{Analysing app reviews for software engineering: a
  systematic literature review}.
\newblock \bibinfo{journal}{\emph{Empirical Software Engineering}}
  \bibinfo{volume}{27}, \bibinfo{number}{2} (\bibinfo{year}{2022}),
  \bibinfo{pages}{43}.
\newblock


\bibitem[Gao et~al\mbox{.}(2015)]%
        {Gao}
\bibfield{author}{\bibinfo{person}{Cuiyun Gao}, \bibinfo{person}{Hui Xu},
  \bibinfo{person}{Junjie Hu}, {and} \bibinfo{person}{Yangfan Zhou}.}
  \bibinfo{year}{2015}\natexlab{}.
\newblock \showarticletitle{AR-Tracker: track the dynamics of mobile apps via
  user review mining}. In \bibinfo{booktitle}{\emph{The 9th IEEE Symposium on
  Service-Oriented System Engineering}}. \bibinfo{pages}{284--290}.
\newblock


\bibitem[Gu and Kim(2015)]%
        {Gu}
\bibfield{author}{\bibinfo{person}{Xiaodong Gu} {and} \bibinfo{person}{Sunghun
  Kim}.} \bibinfo{year}{2015}\natexlab{}.
\newblock \showarticletitle{What parts of your apps are loved by users?}. In
  \bibinfo{booktitle}{\emph{Proceedings of the 30th IEEE/ACM International
  Conference on Automated Software Engineering}}. \bibinfo{pages}{760--770}.
\newblock


\bibitem[Guzman et~al\mbox{.}(2017)]%
        {Guzman_3}
\bibfield{author}{\bibinfo{person}{Emitza Guzman}, \bibinfo{person}{Rana
  Alkadhi}, {and} \bibinfo{person}{Norbert Seyff}.}
  \bibinfo{year}{2017}\natexlab{}.
\newblock \showarticletitle{An exploratory study of Twitter messages about
  software applications}.
\newblock \bibinfo{journal}{\emph{Requirements Engineering}}
  \bibinfo{volume}{22}, \bibinfo{number}{3} (\bibinfo{year}{2017}),
  \bibinfo{pages}{387--412}.
\newblock


\bibitem[Guzman et~al\mbox{.}(2015)]%
        {Guzman}
\bibfield{author}{\bibinfo{person}{Emitza Guzman}, \bibinfo{person}{Muhammad
  El-Haliby}, {and} \bibinfo{person}{Bernd Bruegge}.}
  \bibinfo{year}{2015}\natexlab{}.
\newblock \showarticletitle{Ensemble methods for app review classification: an
  approach for software evolution}. In \bibinfo{booktitle}{\emph{Proceedings of
  the 30th IEEE/ACM International Conference on Automated Software
  Engineering}}. \bibinfo{pages}{771--776}.
\newblock


\bibitem[Guzman and Maalej(2014)]%
        {Guzman_2}
\bibfield{author}{\bibinfo{person}{Emitza Guzman} {and} \bibinfo{person}{Walid
  Maalej}.} \bibinfo{year}{2014}\natexlab{}.
\newblock \showarticletitle{How do users like this feature? a fine grained
  sentiment analysis of app reviews}. In \bibinfo{booktitle}{\emph{Proceedings
  of the 22nd IEEE International Requirements Engineering Conference}}.
  \bibinfo{pages}{153--162}.
\newblock


\bibitem[Hadi and Fard(2021)]%
        {Hadi}
\bibfield{author}{\bibinfo{person}{Mohammad~Abdul Hadi} {and}
  \bibinfo{person}{Fatemeh~H. Fard}.} \bibinfo{year}{2021}\natexlab{}.
\newblock \showarticletitle{Evaluating pre-trained models for user feedback
  analysis in software engineering: A study on classification of app-reviews}.
\newblock \bibinfo{journal}{\emph{arXiv preprint arXiv:2104.05861}}
  (\bibinfo{year}{2021}).
\newblock


\bibitem[Hassan et~al\mbox{.}(2022)]%
        {Hassan}
\bibfield{author}{\bibinfo{person}{Safwat Hassan}, \bibinfo{person}{Heng Li},
  {and} \bibinfo{person}{Ahmed~E. Hassan}.} \bibinfo{year}{2022}\natexlab{}.
\newblock \showarticletitle{On the importance of performing app analysis within
  peer groups}. In \bibinfo{booktitle}{\emph{Proceedings of the 29th IEEE
  International Conference on Software Analysis, Evolution and Reengineering}}.
  \bibinfo{pages}{890--901}.
\newblock


\bibitem[Henao et~al\mbox{.}(2021)]%
        {Henao}
\bibfield{author}{\bibinfo{person}{Pablo~Restrepo Henao},
  \bibinfo{person}{Jannik Fischbach}, \bibinfo{person}{Dominik Spies},
  \bibinfo{person}{Julian Frattini}, {and} \bibinfo{person}{Andreas
  Vogelsang}.} \bibinfo{year}{2021}\natexlab{}.
\newblock \showarticletitle{Transfer learning for mining feature requests and
  bug reports from tweets and app store reviews}. In
  \bibinfo{booktitle}{\emph{Proceedings of the 29th IEEE International
  Requirements Engineering Conference Workshops}}. \bibinfo{pages}{80--86}.
\newblock


\bibitem[Huang et~al\mbox{.}(2018)]%
        {Huang}
\bibfield{author}{\bibinfo{person}{Qiao Huang}, \bibinfo{person}{Xin Xia},
  \bibinfo{person}{David Lo}, {and} \bibinfo{person}{Gail~C. Murphy}.}
  \bibinfo{year}{2018}\natexlab{}.
\newblock \showarticletitle{Automating intention mining}.
\newblock \bibinfo{journal}{\emph{IEEE Transactions on Software Engineering}}
  \bibinfo{volume}{46}, \bibinfo{number}{10} (\bibinfo{year}{2018}),
  \bibinfo{pages}{1098--1119}.
\newblock


\bibitem[Iacob and Harrison(2013)]%
        {Iacob}
\bibfield{author}{\bibinfo{person}{Claudia Iacob} {and} \bibinfo{person}{Rachel
  Harrison}.} \bibinfo{year}{2013}\natexlab{}.
\newblock \showarticletitle{Retrieving and analyzing mobile apps feature
  requests from online reviews}. In \bibinfo{booktitle}{\emph{Proceedings of
  the 10th IEEE/ACM International Conference on Mining Software Repositories}}.
  \bibinfo{pages}{41--44}.
\newblock


\bibitem[Izadi et~al\mbox{.}(2022)]%
        {Izadi}
\bibfield{author}{\bibinfo{person}{Maliheh Izadi}, \bibinfo{person}{Kiana
  Akbari}, {and} \bibinfo{person}{Abbas Heydarnoori}.}
  \bibinfo{year}{2022}\natexlab{}.
\newblock \showarticletitle{Predicting the objective and priority of issue
  reports in software repositories.}
\newblock \bibinfo{journal}{\emph{Empirical Software Engineering}}
  \bibinfo{volume}{27}, \bibinfo{number}{2} (\bibinfo{year}{2022}),
  \bibinfo{pages}{50}.
\newblock


\bibitem[Izadi et~al\mbox{.}(2021)]%
        {Izadi-2}
\bibfield{author}{\bibinfo{person}{Maliheh Izadi}, \bibinfo{person}{Abbas
  Heydarnoori}, {and} \bibinfo{person}{Georgios Gousios}.}
  \bibinfo{year}{2021}\natexlab{}.
\newblock \showarticletitle{Topic recommendation for software repositories
  using multi-label classification algorithms}.
\newblock \bibinfo{journal}{\emph{Empirical Software Engineering}}
  \bibinfo{volume}{26}, \bibinfo{number}{5} (\bibinfo{year}{2021}).
\newblock


\bibitem[Jha and Mahmoud(2018)]%
        {Jha}
\bibfield{author}{\bibinfo{person}{Nishant Jha} {and} \bibinfo{person}{Anas
  Mahmoud}.} \bibinfo{year}{2018}\natexlab{}.
\newblock \showarticletitle{Using frame semantics for classifying and
  summarizing application store reviews}.
\newblock \bibinfo{journal}{\emph{Empirical Software Engineering}}
  \bibinfo{volume}{23}, \bibinfo{number}{6} (\bibinfo{year}{2018}),
  \bibinfo{pages}{3734--3767}.
\newblock


\bibitem[Khalajzadeh et~al\mbox{.}(2022)]%
        {Khalajzadeh}
\bibfield{author}{\bibinfo{person}{Hourieh Khalajzadeh},
  \bibinfo{person}{Mojtaba Shahin}, \bibinfo{person}{Humphrey~O. Obie},
  \bibinfo{person}{Pragya Agrawal}, {and} \bibinfo{person}{John Grundy}.}
  \bibinfo{year}{2022}\natexlab{}.
\newblock \showarticletitle{Supporting Developers in Addressing Human-centric
  Issues in Mobile Apps}.
\newblock \bibinfo{journal}{\emph{IEEE Transactions on Software Engineering}}
  \bibinfo{volume}{49}, \bibinfo{number}{4} (\bibinfo{year}{2022}),
  \bibinfo{pages}{2149--2168}.
\newblock


\bibitem[Kingma and Ba(2014)]%
        {adam}
\bibfield{author}{\bibinfo{person}{Diederik~P. Kingma} {and}
  \bibinfo{person}{Jimmy~Lei Ba}.} \bibinfo{year}{2014}\natexlab{}.
\newblock \showarticletitle{Adam: A method for stochastic optimization}.
\newblock \bibinfo{journal}{\emph{arXiv preprint arXiv:1412.6980}}
  (\bibinfo{year}{2014}).
\newblock


\bibitem[Lan et~al\mbox{.}(2019)]%
        {albert}
\bibfield{author}{\bibinfo{person}{Zhenzhong Lan}, \bibinfo{person}{Mingda
  Chen}, \bibinfo{person}{Sebastian Goodman}, \bibinfo{person}{Kevin Gimpel},
  \bibinfo{person}{Piyush Sharma}, {and} \bibinfo{person}{Radu Soricut}.}
  \bibinfo{year}{2019}\natexlab{}.
\newblock \showarticletitle{Albert: A lite bert for self-supervised learning of
  language representations}.
\newblock \bibinfo{journal}{\emph{arXiv preprint arXiv:1909.11942}}
  (\bibinfo{year}{2019}).
\newblock


\bibitem[Li et~al\mbox{.}(2017)]%
        {Li_sim}
\bibfield{author}{\bibinfo{person}{Quanlai Li}, \bibinfo{person}{Yan Li1},
  \bibinfo{person}{Pavneet~Singh Kochhar}, \bibinfo{person}{Xin Xia}, {and}
  \bibinfo{person}{David Lo}.} \bibinfo{year}{2017}\natexlab{}.
\newblock \showarticletitle{Detecting similar repositories on GitHub.}. In
  \bibinfo{booktitle}{\emph{Proceedings of the 24th IEEE International
  Conference on Software Analysis, Evolution and Reengineering}}.
  \bibinfo{pages}{13--23}.
\newblock


\bibitem[Li et~al\mbox{.}(2023)]%
        {Li_2}
\bibfield{author}{\bibinfo{person}{Zhixing Li}, \bibinfo{person}{Yue Yu},
  \bibinfo{person}{Tao Wang}, \bibinfo{person}{Yan Lei}, \bibinfo{person}{Ying
  Wang}, {and} \bibinfo{person}{Huaimin Wang}.}
  \bibinfo{year}{2023}\natexlab{}.
\newblock \showarticletitle{To follow or not to follow: understanding
  issue/pull-request templates on GitHub}.
\newblock \bibinfo{journal}{\emph{IEEE Transactions on Software Engineering}}
  \bibinfo{volume}{49}, \bibinfo{number}{4} (\bibinfo{year}{2023}).
\newblock


\bibitem[Liu et~al\mbox{.}(2020)]%
        {Liu_1}
\bibfield{author}{\bibinfo{person}{Pei Liu}, \bibinfo{person}{Li Li},
  \bibinfo{person}{Yanjie Zhao}, \bibinfo{person}{Xiaoyu Sun}, {and}
  \bibinfo{person}{Johnl Grundy}.} \bibinfo{year}{2020}\natexlab{}.
\newblock \showarticletitle{AndroZooOpen: collecting large-scale open source
  Android apps for the research community}. In
  \bibinfo{booktitle}{\emph{Proceedings of the 17th IEEE/ACM International
  Conference on Mining Software Repositories}}. \bibinfo{pages}{548--552}.
\newblock


\bibitem[Liu(2018)]%
        {Liu}
\bibfield{author}{\bibinfo{person}{Xiaoyu Liu, Yuzhou~Wang}.}
  \bibinfo{year}{2018}\natexlab{}.
\newblock \showarticletitle{Analyzing reviews guided by app descriptions for
  the software development and evolution}.
\newblock \bibinfo{journal}{\emph{Journal of Software: Evolution and Process}}
  \bibinfo{volume}{30}, \bibinfo{number}{12} (\bibinfo{year}{2018}).
\newblock


\bibitem[Liu et~al\mbox{.}(2019)]%
        {roberta}
\bibfield{author}{\bibinfo{person}{Yinhan Liu}, \bibinfo{person}{Myle Ott},
  \bibinfo{person}{Naman Goyal}, \bibinfo{person}{Jingfei Du},
  \bibinfo{person}{Mandar Joshi}, \bibinfo{person}{Danqi Chen},
  \bibinfo{person}{Omer Levy}, \bibinfo{person}{Mike Lewis},
  \bibinfo{person}{Luke Zettlemoyer}, {and} \bibinfo{person}{Veselin
  Stoyanov}.} \bibinfo{year}{2019}\natexlab{}.
\newblock \showarticletitle{Roberta: A robustly optimized bert pretraining
  approach}.
\newblock \bibinfo{journal}{\emph{arXiv preprint arXiv:1907.11692}}
  (\bibinfo{year}{2019}).
\newblock


\bibitem[Maalej et~al\mbox{.}(2016)]%
        {Maalej_Journal}
\bibfield{author}{\bibinfo{person}{Walid Maalej}, \bibinfo{person}{Zijad
  Kurtanović}, \bibinfo{person}{Hadeer Nabil}, {and}
  \bibinfo{person}{Christoph Stanik}.} \bibinfo{year}{2016}\natexlab{}.
\newblock \showarticletitle{On the automatic classification of app reviews}.
\newblock \bibinfo{journal}{\emph{Requirements Engineering}}
  \bibinfo{volume}{21}, \bibinfo{number}{3} (\bibinfo{year}{2016}),
  \bibinfo{pages}{311--331}.
\newblock


\bibitem[Maalej and Nabil(2015)]%
        {Maalej_1}
\bibfield{author}{\bibinfo{person}{Walid Maalej} {and} \bibinfo{person}{Hadeer
  Nabil}.} \bibinfo{year}{2015}\natexlab{}.
\newblock \showarticletitle{Bug report, feature request, or simply praise? On
  automatically classifying app reviews}. In
  \bibinfo{booktitle}{\emph{Proceedings of the 23rd IEEE International
  Requirements Engineering Conference}}. \bibinfo{pages}{116--125}.
\newblock


\bibitem[Mazrae et~al\mbox{.}(2021)]%
        {Rostami}
\bibfield{author}{\bibinfo{person}{Pooya~Rostami Mazrae},
  \bibinfo{person}{Maliheh Izadi}, {and} \bibinfo{person}{Abbas Heydarnoori}.}
  \bibinfo{year}{2021}\natexlab{}.
\newblock \showarticletitle{Automated Recovery of Issue-Commit Links Leveraging
  Both Textual and Non-textual Data}. In \bibinfo{booktitle}{\emph{Proceedings
  of the 37th IEEE International Conference on Software Maintenance and
  Evolution}}.
\newblock


\bibitem[Nayebi et~al\mbox{.}(2018)]%
        {Nayebi}
\bibfield{author}{\bibinfo{person}{Maleknaz Nayebi}, \bibinfo{person}{Henry
  Cho}, {and} \bibinfo{person}{Guenther Ruhe}.}
  \bibinfo{year}{2018}\natexlab{}.
\newblock \showarticletitle{App store mining is not enough for app
  improvement}.
\newblock \bibinfo{journal}{\emph{Empirical Software Engineering}}
  \bibinfo{volume}{23}, \bibinfo{number}{5} (\bibinfo{year}{2018}),
  \bibinfo{pages}{2764--2794}.
\newblock


\bibitem[Nikeghbal et~al\mbox{.}(2024)]%
        {Nikeghbal}
\bibfield{author}{\bibinfo{person}{Nafiseh Nikeghbal},
  \bibinfo{person}{Amir~Hossein Kargaran}, {and} \bibinfo{person}{Abbas
  Heydarnoori}.} \bibinfo{year}{2024}\natexlab{}.
\newblock \showarticletitle{{GIRT-Model}: Automated Generation of Issue Report
  Templates}. In \bibinfo{booktitle}{\emph{Proceedings of the 21st IEEE/ACM
  International Conference on Mining Software Repositories}}.
\newblock


\bibitem[Pagano and Maalej(2013)]%
        {Pagano}
\bibfield{author}{\bibinfo{person}{Dennis Pagano} {and} \bibinfo{person}{Walid
  Maalej}.} \bibinfo{year}{2013}\natexlab{}.
\newblock \showarticletitle{User feedback in the appstore : an empirical
  study}. In \bibinfo{booktitle}{\emph{Proceedings of the 21st IEEE
  International Requirements Engineering Conference}}.
  \bibinfo{pages}{125--134}.
\newblock


\bibitem[Palomba et~al\mbox{.}(2015)]%
        {Palomba}
\bibfield{author}{\bibinfo{person}{Fabio Palomba}, \bibinfo{person}{Mario
  Linares-Vasquez}, \bibinfo{person}{Gabriele Bavota}, \bibinfo{person}{Rocco
  Oliveto}, \bibinfo{person}{Massimiliano Di~Penta}, \bibinfo{person}{Denys
  Poshyvanyk}, {and} \bibinfo{person}{Andrea De~Lucia}.}
  \bibinfo{year}{2015}\natexlab{}.
\newblock \showarticletitle{User reviews matter! Tracking crowdsourced reviews
  to support evolution of successful apps}. In
  \bibinfo{booktitle}{\emph{Proceedings of the 31st IEEE International
  Conference on Software Maintenance and Evolution}}.
  \bibinfo{pages}{291--300}.
\newblock


\bibitem[Panichella et~al\mbox{.}(2015)]%
        {Panichella}
\bibfield{author}{\bibinfo{person}{Sebastiano Panichella}, \bibinfo{person}{Di
  Sorbo}, \bibinfo{person}{Corrado Aaron}, {and} \bibinfo{person}{C Harald}.}
  \bibinfo{year}{2015}\natexlab{}.
\newblock \showarticletitle{How can I improve my app? classifying user reviews
  for software maintenance and evolution}. In
  \bibinfo{booktitle}{\emph{Proceedings of the 31st IEEE International
  Conference on Software Maintenance and Evolution}}.
  \bibinfo{pages}{281--290}.
\newblock


\bibitem[Salton and Buckley(1988)]%
        {tfidf}
\bibfield{author}{\bibinfo{person}{Gerard Salton} {and}
  \bibinfo{person}{Christopher Buckley}.} \bibinfo{year}{1988}\natexlab{}.
\newblock \showarticletitle{Term-weighting approaches in automatic text
  retrieval}.
\newblock \bibinfo{journal}{\emph{Information processing and management}}
  \bibinfo{volume}{24}, \bibinfo{number}{5} (\bibinfo{year}{1988}),
  \bibinfo{pages}{513--523}.
\newblock


\bibitem[Sanh et~al\mbox{.}(2019)]%
        {distilbert}
\bibfield{author}{\bibinfo{person}{Victor Sanh}, \bibinfo{person}{Lysandre
  Debut}, \bibinfo{person}{Julien Chaumond}, {and} \bibinfo{person}{Thomas
  Wolf}.} \bibinfo{year}{2019}\natexlab{}.
\newblock \showarticletitle{DistilBERT, a distilled version of BERT: smaller,
  faster, cheaper and lighter}.
\newblock \bibinfo{journal}{\emph{arXiv preprint arXiv:1910.01108}}
  (\bibinfo{year}{2019}).
\newblock


\bibitem[Scalabrino et~al\mbox{.}(2019)]%
        {Scalabrino}
\bibfield{author}{\bibinfo{person}{Simone Scalabrino},
  \bibinfo{person}{Gabriele Bavota}, \bibinfo{person}{Barbara Russo},
  \bibinfo{person}{Massimiliano~Di Penta}, {and} \bibinfo{person}{Rocco
  Oliveto}.} \bibinfo{year}{2019}\natexlab{}.
\newblock \showarticletitle{Listening to the crowd for the release planning of
  mobile apps}.
\newblock \bibinfo{journal}{\emph{IEEE Transactions on Software Engineering}}
  \bibinfo{volume}{45}, \bibinfo{number}{1} (\bibinfo{year}{2019}),
  \bibinfo{pages}{68--86}.
\newblock


\bibitem[Shi et~al\mbox{.}(2021)]%
        {Shi}
\bibfield{author}{\bibinfo{person}{Lin Shi}, \bibinfo{person}{Celia Chen},
  \bibinfo{person}{Qing Wang}, {and} \bibinfo{person}{Barry Boehm}.}
  \bibinfo{year}{2021}\natexlab{}.
\newblock \showarticletitle{Automatically detecting feature requests from
  development emails by leveraging semantic sequence mining}.
\newblock \bibinfo{journal}{\emph{Requirements Engineering}}
  \bibinfo{volume}{26}, \bibinfo{number}{2} (\bibinfo{year}{2021}),
  \bibinfo{pages}{255--271}.
\newblock


\bibitem[Sorbo et~al\mbox{.}(2016)]%
        {Sorbo_1}
\bibfield{author}{\bibinfo{person}{Andrea~Di Sorbo},
  \bibinfo{person}{Sebastiano Panichella}, \bibinfo{person}{Carol~V Alexandru},
  \bibinfo{person}{Junji Shimagaki}, \bibinfo{person}{Corrado~A Visaggio},
  \bibinfo{person}{Gerardo Canfora}, {and} \bibinfo{person}{Harald. Gall}.}
  \bibinfo{year}{2016}\natexlab{}.
\newblock \showarticletitle{What would users change in my app? summarizing app
  reviews for recommending Software Changes}. In
  \bibinfo{booktitle}{\emph{Proceedings of the 24th ACM SIGSOFT International
  Symposium on Foundations of Software Engineering}}.
  \bibinfo{pages}{499--510}.
\newblock


\bibitem[Stanik et~al\mbox{.}(2019)]%
        {Stanik}
\bibfield{author}{\bibinfo{person}{Christoph Stanik}, \bibinfo{person}{Marlo
  Haering}, {and} \bibinfo{person}{Walid Maalej}.}
  \bibinfo{year}{2019}\natexlab{}.
\newblock \showarticletitle{Classifying multilingual user feedback using
  traditional machine learning and deep learning}. In
  \bibinfo{booktitle}{\emph{Proceedings of the 27th IEEE International
  Requirements Engineering Conference Workshops}}. \bibinfo{pages}{220--226}.
\newblock


\bibitem[Tan and Li(2020)]%
        {Tan}
\bibfield{author}{\bibinfo{person}{Shin~Hwei Tan} {and}
  \bibinfo{person}{Ziqiang Li}.} \bibinfo{year}{2020}\natexlab{}.
\newblock \showarticletitle{Collaborative bug finding for android apps}. In
  \bibinfo{booktitle}{\emph{Proceedings of the 42th IEEE/ACM International
  Conference on Software Engineering}}. \bibinfo{pages}{1335--1347}.
\newblock


\bibitem[Tizard et~al\mbox{.}(2019)]%
        {Tizard}
\bibfield{author}{\bibinfo{person}{James Tizard}, \bibinfo{person}{Hechen
  Wang}, \bibinfo{person}{Lydia Yohannes}, {and} \bibinfo{person}{Kelly
  Blincoe}.} \bibinfo{year}{2019}\natexlab{}.
\newblock \showarticletitle{Can a conversation paint a picture? Mining
  requirements in software forums}. In \bibinfo{booktitle}{\emph{Proceedings of
  the 27th IEEE International Requirements Engineering Conference}}.
  \bibinfo{pages}{17--27}.
\newblock


\bibitem[Villarroel et~al\mbox{.}(2016)]%
        {Villarroel}
\bibfield{author}{\bibinfo{person}{Lorenzo Villarroel},
  \bibinfo{person}{Gabriele Bavota}, \bibinfo{person}{Barbara Russo},
  \bibinfo{person}{Rocco Oliveto}, {and} \bibinfo{person}{Massimiliano~Di
  Penta}.} \bibinfo{year}{2016}\natexlab{}.
\newblock \showarticletitle{Release planning of mobile apps based on user
  reviews}. In \bibinfo{booktitle}{\emph{Proceedings of the 38th IEEE/ACM
  International Conference on Software Engineering}}. \bibinfo{pages}{14--24}.
\newblock


\bibitem[Wang et~al\mbox{.}(2019)]%
        {Wang}
\bibfield{author}{\bibinfo{person}{Chong Wang}, \bibinfo{person}{Tao Wang},
  \bibinfo{person}{Peng Liang}, \bibinfo{person}{Maya Daneva}, {and}
  \bibinfo{person}{Marten Van~Sinderen}.} \bibinfo{year}{2019}\natexlab{}.
\newblock \showarticletitle{Augmenting app reviews with app changelogs: An
  approach for app reviews classification}. In
  \bibinfo{booktitle}{\emph{Proceedings of the 31st International Conference on
  Software Engineering and Knowledge Engineering}}. \bibinfo{pages}{398--403}.
\newblock


\bibitem[Zhang et~al\mbox{.}(2019)]%
        {Zhang}
\bibfield{author}{\bibinfo{person}{Tao Zhang}, \bibinfo{person}{Haoming Li},
  \bibinfo{person}{Zhou Xu}, \bibinfo{person}{Jian Liu},
  \bibinfo{person}{Rubing Huang}, {and} \bibinfo{person}{Yiran Shen}.}
  \bibinfo{year}{2019}\natexlab{}.
\newblock \showarticletitle{Labelling issue reports in mobile apps}.
\newblock \bibinfo{journal}{\emph{IET Software}} \bibinfo{volume}{13},
  \bibinfo{number}{6} (\bibinfo{year}{2019}), \bibinfo{pages}{528--542}.
\newblock


\end{thebibliography}

\end{document}